%% file: main.tex
\documentclass{LMCS}

\def\dOi{10(3:8)2014}
\lmcsheading%
{\dOi}
{1--61}
{}
{}
{Sep.~30, 2012}
{Aug.~21, 2014}
{}

\ACMCCS{[{\bf Software and its engineering}]: Software creation and
  management---Software verification and validation; Software
  organization and properties---Software functional
  properties---Formal methods---Model checking}


\usepackage[utf8]{inputenc}

\usepackage{url}
\usepackage{hyperref}

\usepackage{soul}
\usepackage{color}
\usepackage{latexsym}
\usepackage{amsxtra} 
\usepackage{amssymb}
\usepackage{amsmath}
\usepackage{yhmath}

\usepackage{multirow}

\usepackage{algorithm}
\usepackage{algorithmicx}
\usepackage[noend]{algpseudocode}

\usepackage{subfig}
\usepackage{graphics}
\usepackage{epsfig}
\usepackage{color}

\usepackage{tikz}
\usetikzlibrary{arrows,decorations.pathmorphing,backgrounds,positioning,fit,petri,calc,automata}
\usetikzlibrary{calc,decorations.pathreplacing}
\usepackage{arrayjobx} 

\input{commands}

\input{tikzmacros-period}

\input{tikzmacros-termination}

\begin{document}

\title{Deciding Conditional Termination}

\author[M.~Bozga]{Marius Bozga\rsuper a}	
\address{{\lsuper{a,b}}Univ. Grenoble Alpes/CNRS/VERIMAG, F-38000, Grenoble France}	
\email{\{bozga,iosif\}@imag.fr}  
\thanks{{\lsuper{a,b}}Supported by the French National Research Agency (project
  ANR-09-SEGI-016 VERIDYC)}

\author[R.~Iosif]{Radu Iosif\rsuper b}	
\address{\vspace{-18 pt}}	
\thanks{{\lsuper{b,c}}Supported by the Rich Model Toolkit initiative
  (formally ESF COST action IC0901).}

\author[F.~Kone\v{c}n\'{y}]{Filip Kone\v{c}n\'{y}\rsuper c} 
\address{{\lsuper c}EPFL IC IIF LARA, Station 14, 1015 Lausanne, Switzerland}
\email{filip.konecny@epfl.ch} 
\thanks{{\lsuper c}Supported by the Czech Science Foundation (project
  P103/10/0306).}






\begin{abstract}
We address the problem of conditional termination, which is that of
defining the set of initial configurations from which a given program
always terminates. First we define the dual set, of initial
configurations from which a non-terminating execution exists, as the
greatest fixpoint of the function that maps a set of states into its
pre-image with respect to the transition relation. This definition
allows to compute the weakest non-termination precondition if at least
one of the following holds: (i) the transition relation is
deterministic, (ii) the descending Kleene sequence over-approximating
the greatest fixpoint converges in finitely many steps, or (iii) the
transition relation is well founded. We show that this is the case for
two classes of relations, namely octagonal and finite monoid affine
relations. Moreover, since the closed forms of these relations can be
defined in Presburger arithmetic, we obtain the decidability of the
termination problem for such loops.

We show that the weakest non-termination precondition for octagonal
relations can be computed in time polynomial in the size of the binary
representation of the relation. Furthermore, for every well-founded
octagonal relation, we prove the existence of an effectively
computable well-founded witness relation for which a linear ranking
function exists. For the class of linear affine relations we show that
the weakest non-termination precondition can be defined in Presburger
arithmetic if the relation has the finite monoid property. Otherwise,
for a more general subclass, called polynomially bounded affine
relations, we give a method of under-approximating the termination
preconditions.

Finally, we apply the method of computing weakest non-termination
preconditions for conjunctive relations (octagonal or affine) to
computing termination preconditions for programs with complex
transition relations. We provide algorithms for computing transition
invariants and termination preconditions, and define a class of
programs, whose control structure has no nested loops, for which these
algorithms provide precise results. Moreover, it is shown that, for
programs with no nested control loops, and whose loops are labeled
with octagonal constraints, the dual problem i.e.\ the existence of
infinite runs, is NP-complete.
\end{abstract}

\maketitle\vfill

\section{Introduction}

The termination problem asks whether every computation of a given
program ends in a halting state. The universal termination problem asks
whether a given program always terminates for every possible input
configuration. Both problems are among the first ever to be shown
undecidable, by A. Turing \cite{turing}. In many cases however,
programs will terminate when started in certain configurations, and
may\footnote{If the program is non-deterministic, the existence of a
  single infinite run, among other finite runs, suffices to consider
  an initial configuration non-terminating.} run forever, when started
in other configurations. The problem of determining the set of
configurations from which a program terminates on all paths is called
{\em conditional termination}.

In this paper we focus on programs that handle integer variables,
performing Presburger arithmetic tests and (possibly
non-deterministic) updates. A first observation is that the set of
configurations from which an infinite computation is possible is the
greatest fixpoint of the pre-image $\pre_R$ of the program's
transition relation\footnote{This definition is the dual of the {\em
    reachability set}, needed for checking safety properties: the
  reachability set is the least fixpoint of the post-image of the
  transition relation.}  $R$. This set, called the {\em weakest
  recurrent set}, and denoted $\wrs(R)$ in our paper, is the limit of
the descending sequence $\pre_R^0(\true), \pre_R^1(\true),
\pre_R^2(\true), \ldots$, i.e.\ $\wrs(R) = \bigcap_{i=1}^\infty
pre_R^n(\true)$, if either (i) the pre-image of the transition
relation is continuous (this is the case, for instance, when the
transition relation is deterministic), (ii) the descending Kleene
sequence that over-approximates the greatest fixpoint eventually
stabilizes, or (iii) the relation is well founded, i.e.\ $\wrs(R) =
\emptyset$. If, moreover, the closed form defining the infinite
sequence of precondition sets $\{\pre_R^n(\true)\}_{n \geq 1}$ can be
defined using a decidable fragment of arithmetic, we obtain
decidability proofs for the universal termination problem.

\paragraph{\bf Contributions of this paper}
The main novelty in this paper is of rather theoretical nature: we
show that the non-termination preconditions for integer transition
relations defined as either {\em octagons} or {\em linear affine loops
  with finite monoid property} are definable in quantifier-free
Presburger arithmetic. Thus, the universal termination problem for
such program loops is decidable. However, since quantifier elimination
in Presburger arithmetic is a complex procedure, we have developed
alternative ways of deriving the preconditions for non-termination,
and in particular:
\begin{itemize}
\item for {\em octagonal relations}, we use a result from
  \cite{cav10}, namely that the sequence $\{R^i\}_{i \geq 0}$ is, in
  some sense, periodic. Based on this, we develop an algorithm that
  computes the weakest non-termination precondition of $R$ in time
  polynomial in the size of the binary representation of
  $R$. Moreover, we investigate the existence of linear ranking
  functions and prove that for each well-founded octagonal relation,
  there exists an effectively computable witness relation for $R$,
  i.e.\ a relation that is well-founded if and only if the original
  relation is well-founded and, in this case, it also has a linear
  ranking function.

\item for {\em linear affine relations}, weakest recurrent sets can be
  defined in Presburger arithmetic if we consider several restrictions
  concerning the transformation matrix. If the matrix $A$ defining $R$
  has eigenvalues which are either zeros or roots of unity, all
  non-zero eigenvalues being of multiplicity one (these conditions are
  equivalent to the finite monoid property of
  \cite{boigelot-thesis,finkel-leroux}), then $\wrs(R)$ is Presburger
  definable.  Otherwise, if all non-zero eigenvalues of $A$ are roots
  of unity, of multiplicities greater or equal to one, $\wrs(R)$ can
  be expressed using polynomial terms. In this case, we can
  systematically issue Presburger termination preconditions, which are
  safe under-approximations of the complement of the $\wrs(R)$ set.
\end{itemize}

Unfortunately, in practice, the cases in which the closed form of the
sequence of preconditions $\{\pre^n_R(\true)\}_{n \geq 0}$ is
definable in a decidable fragment of arithmetic, are fairly rare. All
relations considered so far are conjunctive, meaning that they can
represent only simple program loops of the form {\tt
  while(condition)\{body\}} where the loop body contains no further
conditional constructs. Whereas in reality such simple programs are
rare, our results can be used as building blocks of other termination
proof methods \cite{Cook:2006}, which discard {\em lasso-shaped}
non-termination counterexamples one by one. Our method can be used for
proving non-termination as well, by embedding it into general
algorithms, such as \cite{henzinger-rybalchenko-popl08}.

In order to deal with more complicated program loops, we use the
method of {\em transition invariants} \cite{transition-invariants} to
compute safe under-approximations of the strongest termination
preconditions. Concretely, we compute a {\em transition invariant},
which is an over-approximation of the transitive closure of the
transition relation of the program, restricted to the states reachable
from some set of initial configurations. If one can find a finite
union $R^\#_1 \cup \ldots \cup R^\#_m$ of octagonal relations that is
a transition invariant, then we can compute an over-approximation of
the weakest non-termination precondition as $\wrs(R^\#_1) \cup \ldots \cup
\wrs(R^\#_m)$. The required termination precondition is the complement
of this set.

This method can infer non-termination preconditions for programs
without procedure calls. It is moreover shown to be complete, and to
yield the precise result for a class of programs without nested loops,
called {\em flat}. Moreover, we studied a restriction of flat programs
in which all transitions within loops are labeled with octagonal
constraints, and found that, for this restricted class, the problem of
existence of infinite runs is NP-complete.


We have implemented the computation of transition invariants and
procedure summaries in the \textsc{Flata} tool for the analysis of
integer programs. Several experiments on inferring non-termination
preconditions have been performed, and reported.

\paragraph{\bf Roadmap} 
The paper is organized as follows. Section \ref{sec:definitions}
introduces the notation and some basic concepts needed throughout the
paper. Section \ref{sec:termination:fixpoint} defines weakest
recurrent sets as greatest fixpoints of the pre-image of the
transition relation. Sections \ref{sec:termination:octagons} and
\ref{sec:termination:affine} apply this definition to the computation
of weakest recurrent sets for octagonal and linear affine
relations. Section \ref{sec:intprograms} extends the computation of
weakest termination preconditions from simple conjunctive loops to
integer programs, and Section \ref{sec:experiments} reports on the
implementation and experiments performed on several integer
programs. Finally, Section \ref{sec:conclusion} concludes. 

The core results presented in this paper have been reported in
\cite{tacas12}. In addition to the work presented in \cite{tacas12},
here we improve the time complexity upper bound for the computation of
weakest non-termination preconditions for octagonal relations, and
give a polynomial time algorithm. Moreover, we extend the results from
\cite{tacas12} from simple conjunctive program loops to computing
non-termination preconditions for full integer programs (whose
transition rules are defined using quantifier-free Presburger
arithmetic), by giving a decidability result to the universal
termination problem, for a class of {\em flat} programs, i.e.\ without
nested loops, and no branching within loops.

\subsection{Related Work}

The literature on program termination is vast. Most work focuses
however on universal termination, i.e.\ the question if a program will
always terminate on all inputs, such as the techniques for
synthesizing linear ranking functions of Sohn and Van Gelder
\cite{sohn-vangelder91} or Podelski and Rybalchenko
\cite{podelski-rybalchenko}, and the more sophisticated method of
Bradley, Manna and Sipma \cite{bradley-manna-sipma}, which synthesizes
lexicographic polynomial ranking functions, suitable when dealing with
disjunctive loops. However, not every terminating program (loop) has
a~ linear (polynomial) ranking function. In this paper, we show that
for an entire class of non-deterministic linear relations, defined
using octagons, termination is always witnessed by a~computable
octagonal relation that has a~linear ranking function.

A closely related work direction investigates the termination of
programs abstracted using {\em size-change graphs}, i.e.\ graphs in
which nodes are variables and edges indicate the decrease of values in
a well-founded domain. In \cite{amir-ben-amran06} the size-change
termination problem is investigated for graphs annotated with
difference bounds constraints. It is shown that, even if the general
problem is undecidable, the restriction to size-change graphs with at
most one incoming size-change arc per variable is PSPACE-complete. Our
results are incomparable, since we consider multiple incoming
size-change arcs, but restrict the control structure of the decidable
class of programs to be {\em flat}, i.e.\ no nested loops are
allowed. Moreover, we focus on the problem of computing the weakest
non-termination precondition for simple loops labeled with octagonal
relations, and solve it using a PTIME algorithm.

Another line of work considers the decidability of termination for
simple (conjunctive) linear loops. Initially, Tiwari \cite{tiwari04}
showed decidability of termination for affine linear loops interpreted
over {\em reals}, while Braverman \cite{braverman04} refined this
result by showing decidability over {\em rationals} and over {\em
  integers}, for homogeneous relations of the form $C_1\vec{x} > 0
~\wedge~ C_2\vec{x} \geq 0 ~\wedge~ \vec{x'} = A\vec{x}$. The
non-homogeneous integer case seems to be much more difficult as it is
closely related to the open {\em Skolem's Problem} (see,
e.g.\ \cite{ouaknine-worrell12} for a discussion on this problem):
given a~linear recurrence $\{u_i\}_{i \geq 0}$, determine whether $u_i
= 0$ for some $i \geq 0$. The related problem of existence of linear
ranking functions for linear affine loops has been studied in
\cite{ben-amram-genaim-popl13}. This problem has been found to be in
PTIME when the program variables range over mathematical reals, and
coNP-complete when they range over integers. 

To our knowledge, the first work on proving the existence of
non-terminating computations is arguably \cite{payet-mesnard04}, in
the context of Constraint Logic Programming. Another important
contribution, which considers simple imperative loops, is reported in
\cite{henzinger-rybalchenko-popl08}. The notion of {\em recurrent
  sets} occurs in this work, however, without the connection with
fixpoint theory, which is introduced in the present work. Finding
recurrent sets in \cite{henzinger-rybalchenko-popl08} is complete with
respect to a~predefined set of templates, typically linear systems of
rational inequalities.

The work which is closest to ours is probably that of Cook et
al.\ \cite{byron}. In that paper, the authors develop an algorithm for
deriving termination preconditions by first guessing a~ranking
function candidate (typically the linear term from the loop
condition) and then inferring a~supporting assertion which
guarantees that the candidate function decreases with each
iteration. The step of finding a~supporting assertion requires a~
fixpoint iteration in order to find an invariant condition. Unlike
our work, the authors of \cite{byron} do not address issues related
to completeness: the method is not guaranteed to find the weakest
precondition for termination, even in cases when this set can be
computed. On the other hand, it is applicable to a~large range of
programs extracted from real-life software. To compare our method
with theirs, we tried the examples available in \cite{byron}.
For those which are polynomially bounded affine relations, we used our
under-approximation method and have computed termination
preconditions, which turn out to be slightly more general than the
ones reported in \cite{byron}.

\section{Preliminary Definitions}\label{sec:definitions}

We denote by $\zed$, $\nat$ and $\nat_+$ the sets of integers,
positive (including zero) and strictly positive integers,
respectively. We denote by $\zed_\infty$ and $\zed_{-\infty}$ the sets
$\zed \cup \{\infty\}$ and $\zed \cup \{-\infty\}$, respectively. In
this paper we use a set of variables $\vec{x} =
\{x_1,x_2,\ldots,x_N\}$, for a given integer constant $N > 0$. The set
of \emph{primed} variables is $\vec{x'} =
\{x'_1,x'_2,\ldots,x'_N\}$. These variables are assumed to be ranging
over $\zed$. For a set $S \subseteq \zed$ of integers, we denote by
$\min S$ the smallest integer $s \in S$, if one exists, and by $\inf
S$ the largest element $m \in \zed_{-\infty}$ such that $m \leq s$,
for all $s \in S$. If $S = \emptyset$, we convene that $\min S = \inf
S = \infty$.

A \emph{linear term} $t(\x)$ over a set of variables in $\vec{x}$ is a
linear combination of the form $a_0 + \sum_{i=1}^N a_ix_i$, where
$a_0,a_1, \ldots, a_N \in \zed$. \emph{Presburger arithmetic} is the
first-order logic over {\em atomic propositions} of the form~$t(\x)
\leq 0$. Presburger arithmetic has quantifier elimination and is
decidable \cite{presburger29}. Moreover, the satisfiability of its
{\em quantifier-free fragment} is NP-complete in the size of the
binary representation of the formula \cite{Verma05Complexity}. For
simplicity, we consider only formulas in Presburger arithmetic in this
paper.

For a first-order logical formula $\varphi$, let $FV(\varphi)$ denote
the set of its free variables. By writing $\varphi(\vec{x})$ we imply
that $FV(\varphi) \subseteq \vec{x}$. For a formula
$\varphi(\vec{x})$, we denote by $\varphi[t_1/x_1,\ldots,t_N/x_N]$ the
formula obtained from $\varphi$ by syntactically replacing each free
occurrence of $x_1,\ldots,x_N$ with the terms $t_1,\ldots,t_N$,
respectively. For a first-order logical formula $\varphi$, let
$Atom(\varphi)$ denote the set of atomic propositions in $\varphi$.

A \emph{valuation} of $\vec{x}$ is a function $\smash{\nu : \vec{x}
  \arrow{}{} \zed}$. The set of all such valuations is denoted by
$\zed^{\vec{x}}$. If $\nu \in \zed^{\vec{x}}$, we denote by $\nu
\models \varphi$ the fact that the formula obtained from $\varphi$ by
replacing each occurrence of $x_i$ with $\nu(x_i)$ is
valid. Similarly, an arithmetic formula $\phi_R(\vec{x}, \vec{x'})$
defining a relation $R \subseteq \zed^{\vec{x}} \times \zed^{\vec{x}}$
is evaluated with respect to two valuations $\nu_1$ and $\nu_2$, by
replacing each occurrence of $x_i$ with $\nu_1(x_i)$ and each
occurrence of $x'_i$ with $\nu_2(x_i)$. The satisfaction relation is
denoted $(\nu_1,\nu_2) \models \phi_R$. By $\models \varphi$ we denote
the fact that $\varphi$ is \emph{valid}, i.e.\ logically equivalent to
$\true$. We say that an arithmetic formula $\varphi(\vec{x})$ is
\emph{consistent} if there exists a valuation $\nu$ such that $\nu
\models \varphi$. We use the symbols $\Rightarrow, \iff$ to denote
logical implication and equivalence, respectively. The consistency of
a formula $\varphi$ is usually denoted by writing $\varphi \not\iff
\false$. In the following, we will sometimes abuse notation and use
the same symbols for relations (sets) and their defining formulas.

The composition of two relations $R_1, R_2 \subseteq \zed^{\vec{x}}
\times \zed^{\vec{x}}$ is defined as $R_1 \circ R_2 = \{(\nu,\nu') \in
\zed^{\vec{x}} \times \zed^{\vec{x}} ~|~ \exists \nu'' \in
\zed^{\vec{x}} ~.~ (\nu,\nu'')\in R_1 ~\wedge~ (\nu'',\nu')\in
R_2\}$. The {\em identity relation} on $\x$ is defined as
$\mathcal{I}_{\x} = \{(\nu,\nu) ~|~ \nu \in \zed^\x\}$. For any
relation $R \subseteq \zed^{\vec{x}}$, we define $R^0 =
\mathcal{I}_{\x}$ and $R^{i+1} = R^i \circ R$, for all $i \geq 0$. The
relation $R^i$ is called the {\em $i$-th power} of $R$ in the
sequel. With these notations, $R^+ = \bigcup_{i=1}^\infty R^i$ denotes
the \emph{transitive closure} of $R$, and $R^* = R^+ \cup
\mathcal{I}_{\vec{x}}$ denotes the \emph{reflexive and transitive
  closure} of $R$. A relation $R \subseteq \zed^{\vec{x}} \times
\zed^{\vec{x}}$ is said to be {\em deterministic} if and only if
$(\nu,\nu') \in R$ and $(\nu,\nu'') \in R$ implies $\nu'=\nu''$, for
all $\nu,\nu',\nu'' \in \zed^{\vec{x}}$. Let $\pre_R : 2^{\zed^\x}
\rightarrow 2^{\zed^\x}$ be the {\em pre-image} function defined as
$\pre_R(S) = \{\nu ~|~ \exists \nu' \in S ~.~ (\nu,\nu') \in R\}$, for
any $S \subseteq \zed^\x$.

A function $F : 2^{\zed^{\vec{x}}} \rightarrow 2^{\zed^{\vec{x}}}$ is
said to be {\em monotonic} if and only if $S \subseteq T$ implies
$F(S) \subseteq F(T)$, for any two sets $S,T \subseteq
\zed^{\vec{x}}$, and {\em $\cap$-continuous} if and only if
$F(\cap_{i=1}^\infty S_i) = \cap_{i=1}^\infty F(S_i)$, for any
infinite sequence $\{S_i\}_{i=1}^\infty$ of valuation sets, where $S_i
\subseteq \zed^{\vec{x}}$ for all $i \geq 1$. The {\em greatest
  fixpoint} $F$ is the largest set $S$ such that $F(S) = S$, and is
denoted $\gfp(F)$.

\section{Weakest Preconditions for Non-termination}
\label{sec:termination:fixpoint}

This section is concerned with the definition of weakest preconditions
for non-termination, and the characterization of such preconditions as
greatest fixpoints of the pre-image function. We also give certain
conditions under which these fixpoints are computable as limits of
descending Kleene sequences, and finally, define them using
first-order integer arithmetic.

In the rest of this section, let $\vec{x} = \{x_1,\ldots,x_N\}$ be a
set of variables ranging over integers, for some constant $N > 0$. We
start by proving several properties of the pre-image function.

\begin{prop}\label{prop:pre:properties}
Let $R,R' \subseteq \zed^\x \times \zed^\x$ be relations and $S,S'
\subseteq \zed^\x$ be sets of valuations. The following hold:
\begin{enumerate}
 \item If $R \subseteq R'$ and $S \subseteq S'$ then $\pre_R(S)
   \subseteq \pre_{R'}(S')$. Consequently, $\pre_R$ is monotonic.
 \item If $1 \leq n \leq m$ then $\pre^n_R(S) \supseteq \pre^m_R(S)$.
   Consequently, the sequence $\{ \pre^n_R(\zed^\x) \}_{n\geq 1}$ is descending.
\end{enumerate}
\end{prop}
\proof{(1) Let $\nu \in \pre_R(S)$ be a valuation. Hence there exist
  $\nu' \in S \subseteq S'$ such that $(\nu,\nu') \in R \subseteq
  R'$. But then $\nu \in \pre_{R'}(S')$. Monotonicity of $\pre_R$
  follows by taking $R'=R$. (2) We have: 
  \[\begin{array}{rcll}
  \zed^\x & \supseteq & \pre_R(\zed^\x) & \mbox{since $\zed^\x$ is the universal set} \\
  \pre_R(\zed^\x) & \supseteq & \pre^2_R(\zed^\x) & \mbox{by the monotonicity of $\pre_R$ at point (1)} \\
  & \ldots & \\
  \pre^n_R(\zed^\x) & \supseteq & \pre^{n+1}_R(\zed^\x)
  \end{array}\]
  Hence the sequence $\{\pre^n_R\}_{n \geq 1}$ is descending.  \qed}
 
We next define the notions of {\em $*$-consistent} and {\em well-founded} relation.

\begin{defi}
A relation $R \subseteq \zed^{\vec{x}} \times \zed^{\vec{x}}$ is said
to be {\em $*$-consistent} if and only if, for any $m \geq 0$, there
exists a finite sequence of valuations $\{\nu_i\}_{i=1}^m$, where
$\nu_i \in \zed^{\vec{x}}$ for all $i \geq 1$, such that
$(\nu_i,\nu_{i+1}) \in R$, for all $i=1,\dots,m-1$. $R$ is said to be
{\em well founded} if and only if there is no infinite sequence of
valuations $\{\nu_i\}_{i\geq1}$, such that $\nu_i \in \zed^\vec{x}$
and $(\nu_i,\nu_{i+1}) \in R$, for all $i\geq 0$.
\end{defi}
Notice that if a~relation is not $*$-consistent, then it is also
well founded. However the dual is not true. For instance, the relation
$R = \{(n,n-1) ~|~ n > 0\}$ is both $*$-consistent and
well founded. Also notice that a~relation $R$ is $*$-consistent if and
only if $R^i$ is consistent for all $i\geq1$.

\begin{defi}
A set $S \subseteq \zed^{\vec{x}}$ is said to be a~{\em
  non-termination precondition} for a relation $R \subseteq
\zed^{\vec{x}} \times \zed^{\vec{x}}$ if and only if for each $\nu \in
S$ there exists an infinite sequence of valuations $\{\nu_i\}_{i \geq
  0}$ such that $\nu=\nu_0$ and $\nu_i \in \zed^{\vec{x}}$,
$(\nu_i,\nu_{i+1}) \in R$, for all $i\geq 0$.
\end{defi}
If $S_0,S_1,\ldots$ are all non-termination preconditions for $R$,
then the (possibly infinite) union $\bigcup_{i=0,1,\ldots} S_i$ is a~
non-termination precondition for $R$ as well. The set $\wnt(R) =
\bigcup\{S \in \zed^{\vec{x}} ~|~ S \textrm{ is a~non-termination
  precondition for } R\}$ is called the {\em weakest non-termination
  precondition} for $R$. A~relation $R$ is well founded if and only if
$\wnt(R) = \emptyset$. A~set $S$ such that $S \cap \wnt(R) =
\emptyset$ is called a {\em termination precondition}.

\begin{defi}
  A~set $S \subseteq \zed^{\vec{x}}$ is said to be {\em recurrent} for
  a~relation $R \subseteq \zed^{\vec{x}} \times \zed^{\vec{x}}$ if and only
  if $S \subseteq \pre_R(S)$.
\end{defi}
\noindent
Notice that if $S$ is a~recurrent set for a~relation $R$, then for
each $\nu\in S$ there exists $\nu'\in S$ such that $(\nu,\nu')\in R$.

\begin{prop}\label{recurrent-union}
Let $S_0,S_1,\ldots \in \zed^{\vec{x}}$ be a~(possibly infinite)
sequence of sets, all of which are recurrent for a~relation $R \in
\zed^{\vec{x}} \times \zed^{\vec{x}}$. Then their union
$\bigcup_{i=0,1,\ldots} S_i$ is recurrent for $R$ as well.
\end{prop}
\proof{ For each $i$ we have $S_i \subseteq \pre_R(S_i) \subseteq
  \pre_R(\bigcup_{j=0,1,\ldots} S_j)$. The last inclusion is by the
  monotonicity of $\pre_R$. Hence $\bigcup_{j=0,1,\ldots} S_j
  \subseteq \pre_R(\bigcup_{j=0,1,\ldots} S_j)$. \qed} The set
$\wrs(R) = \bigcup\{S \in \zed^{\vec{x}} ~|~ S ~\mbox{is a~recurrent
  set for $R$}\}$ is called the {\em weakest recurrent set} for
$R$. By Proposition \ref{recurrent-union}, $\wrs(R)$ is recurrent for
$R$.  The following lemma shows that in fact, $\wrs(R)$ is exactly the set
of valuations from which an infinite iteration of $R$ is possible and,
equivalently, the greatest fixpoint of the transition relation's 
pre-image.

\begin{lem}\label{wrs-wnt-gfp}
   For every relation $R \subseteq \zed^{\vec{x}} \times \zed^{\vec{x}}$,
   \[ \wrs(R) = \wnt(R) = \gfp(\pre_R)\textrm{.} \]
\end{lem}
\proof{``$\wrs(R) = \gfp(\pre_R)$'' By the {\em Knaster-Tarski Fixpoint 
  Theorem}\footnote{We use the version given as Prop. A.10 in 
  \cite{nielson-nielson-hankin}, pg. 400.}, \[\gfp(\pre_R) = 
  \bigcup \{S ~|~ S \subseteq \pre_R(S)\} = \wrs(R)\textrm{.}\]
  
  ``$\wrs(R) \subseteq \wnt(R)$'' Let $\nu_0 \in \wrs(R)$ be
  a~valuation. Then there exists $\nu_1 \in \wrs(R)$ such that
  $(\nu_0,\nu_1)\in R$. Applying this argument infinitely many times,
  one can construct an infinite sequence $\nu_0,\nu_1,\nu_2,\ldots$
  such that $(\nu_i,\nu_{i+1})\in R$, for all $i \geq 0$. Hence $\nu_0
  \in \wnt(R)$. 
  
  ``$\wnt(R) \subseteq \wrs(R)$'' Let $\nu_0 \in
  \wnt(R)$ be a~valuation and let $\nu_0,\nu_1,\nu_2,\dots$ be an
  arbitrary infinite sequence such that $(\nu_i,\nu_{i+1})\in R$, for
  all $i \geq 0$.  Clearly, $\nu_1 \in \wnt(R)$ too. Consequently,
  $\nu_0\in \pre_R(\wnt(R))$ for each state $\nu_0\in \wnt(R)$ and
  hence, $\wnt(R)\subseteq \pre_R(\wnt(R))$. Thus, $\wnt(R)$ is
  a~recurrent set and hence $\wnt(R) \subseteq \wrs(R)$.
\qed}

The following lemma gives sufficient conditions under which $\wrs(R)$
can be computed as the limit $\bigcap_{n\geq1}
\pre^n_R(\zed^{\vec{x}})$ of the infinite descending Kleene sequence:
$$\pre_R(\zed^\vec{x}) \supseteq
\pre^2_R(\zed^\vec{x}) \supseteq 
\pre^3_R(\zed^\vec{x}) \ldots$$
\begin{lem}\label{wrs-kleene}
  Let $R \subseteq \zed^{\vec{x}} \times \zed^{\vec{x}}$ be a~relation
  such that at least one of the following holds: 
  \begin{enumerate}
  \item $\bigcap_{n\geq1} \pre^n_R(\zed^{\vec{x}})=\emptyset$, or
  \item $\pre^{n_2}_R(\zed^\vec{x})=\pre^{n_1}_R(\zed^{\vec{x}})$ for
    some $n_2>n_1\geq 1$, or
  \item $\pre_R$ is $\cap$-continuous.
  \end{enumerate}
  Then, we have $\wrs(R) = \bigcap_{n\geq1} \pre^n_R(\zed^{\vec{x}})$.
  Moreover, $\wrs(R)=\emptyset$ if (1) holds and 
  $\wrs(R)=\pre^{n_1}_R(\zed^\x)$ if (2) holds.
\end{lem}
\proof{By Lemma \ref{wrs-wnt-gfp}, $\wnt(R)=\wrs(R)=\gfp(\pre_R)$. Since $\gfp(\pre_R)$ is a fixpoint, it follows that $\gfp(\pre_R) = \pre_R^n(\gfp(\pre_R))$ for each $n\geq 1$. Since $\gfp(\pre_R) \subseteq \zed^\x$, it follows that $\pre_R^n(\gfp(\pre_R)) \subseteq \pre_R^n(\zed^\x)$ for each $n\geq 1$, by monotonicity of $\pre_R$ (Proposition \ref{prop:pre:properties}). Hence we obtain that $\gfp(\pre_R) \subseteq \pre_R^n(\zed^\x)$ for each $n\geq 1$ and consequently:
  \[ \wnt(R) = \wrs(R) = \gfp(\pre_R) \subseteq \bigcap_{n \geq1} \pre^n_R(\zed^{\vec{x}}) \]  
  We distinguish between the three cases from the hypothesis:
\noindent
\begin{enumerate}
  \item We have $\emptyset \subseteq \wrs(R) \subseteq \bigcap_{n \geq 1}
    \pre^n_R(\zed^{\vec{x}}) = \emptyset$. Hence, in this case we
    obtain $\wrs(R) = \bigcap_{n \geq 1} \pre^n_R(\zed^{\vec{x}}) =
    \emptyset$. 

  \item Since $\pre_R$ is a monotonic function, the sequence
    $\{\pre^n_R(\zed^{\vec{x}})\}_{n \geq 1}$ is descending:
    $$\pre^{n_1}_R(\zed^\vec{x}) \supseteq
    \pre^{n_1+1}_R(\zed^\vec{x}) \supseteq \ldots \supseteq
    \pre^{n_2}_R(\zed^\vec{x}) = \pre^{n_1}_R(\zed^{\vec{x}})$$ Hence,
    $\pre^{n_1}_R(\zed^\vec{x}) = \pre^{n}_R(\zed^\vec{x})$, for all
    $n \geq n_1$, i.e.\ $\pre^{n_1}_R(\zed^\vec{x})$ is a fixpoint of
    $\pre_R$, and thus we obtain: 
    $$\bigcap_{n \geq 0} \pre^n_R(\zed^{\vec{x}}) =
    \pre^{n_1}_R(\zed^\vec{x}) \subseteq \gfp(\pre_R)$$ Since
    $\gfp(\pre_R) \subseteq \bigcap_{n\geq1}
    \pre^n_R(\zed^{\vec{x}})$, we obtain: \[ \wrs(R) = \gfp(\pre_R) =
    \bigcap_{n \geq 1} \pre^n_R(\zed^{\vec{x}}) \]
    Since $\pre^{n_1}_R(\zed^\vec{x})$ is a fixpoint, then 
    \[ \wrs(R) = \bigcap_{n \geq 1} \pre^n_R(\zed^{\vec{x}}) = 
       \bigcap_{1\leq n \leq n_1} \pre^n_R(\zed^{\vec{x}}) = 
       \pre^{n_1}_R(\zed^\vec{x})
    \]

  \item If $\pre_R$ is $\cap$-continuous, then $\wrs(R) = \gfp(\pre_R)
    = \bigcap_{n \geq 1} \pre^n_R(\zed^{\vec{x}})$, by Kleene Fixpoint
    Theorem \cite{Kleene}.\qedhere
\end{enumerate}}\smallskip

\noindent In the next section, we show that Lemma \ref{wrs-kleene} is
applicable, for different reasons, to both octagonal (Definition
\ref{odbc}) and finite-monoid affine (Definition \ref{def:affine:rel})
relations: octagonal relations are either well founded (1), or their
descending Kleene sequences stabilize (2), and linear affine relations
are $\cap$-continuous (3). Thus one can compute the weakest
non-termination precondition for these classes as the limit of a
descending Kleene sequence. Next, we show that, for relations
satisfying one of the conditions of Lemma \ref{wrs-kleene}, one can
also define the weakest non-termination precondition in first order
arithmetic.

\begin{defi}\label{closed-form-def}
  Let $\{S_i\}_{i \geq 1}$ be an infinite sequence of valuation sets,
  $S_i \subseteq \zed^\x$, for all $i \geq 1$. The {\em closed form}
  of $\{S_i\}_{i \geq 1}$ is a formula $\widehat{S}(k,\x)$ such that,
  for all $n \geq 1$ and all $\nu \in \zed^\x$:
  $$\nu \in S_n \iff \nu \models \widehat{S}[n/k]$$
\end{defi}
In the rest of the paper, we shall define the weakest non-termination
precondition $\wnt(R)$ for relations $R$ that are octagonal or finite
monoid affine. Assuming that at least one of the hypotheses of Lemma
\ref{wrs-kleene} holds and that $\widehat{\pre_R}(k,\vec{x})$ is
a~closed form of the sequence $\{\pre^n_R(\zed^\x)\}_{n\geq1}$, the
weakest non-termination precondition of $R$ is equivalent to the
first-order arithmetic formula on the right hand side in the following
equivalence:
\begin{equation}\label{wrs}
(\wnt(R))(\vec{x}) \iff \forall k \geq 1 ~.~ \widehat{\pre_R}(k,\vec{x})
\end{equation}
In the upcoming developments, we will show that
$\widehat{\pre_R}(k,\vec{x})$ is Presburger definable, for octagonal
and finite monoid affine relations $R$. As a direct consequence of
(\ref{wrs}), the weakest non-termination precondition is definable in
Presburger arithmetic. Since satisfiability is decidable for
Presburger arithmetic \cite{presburger29}, the universal termination
problem for octagonal and finite-monoid affine relations is decidable
as well.

\begin{exa}
Consider the relation $R(x,x') \iff x \geq 0 \wedge x' = x - 1$. The
closed form of the sequence $\{\pre^n_R(\zed^\x)\}_{n\geq1}$ is
$\widehat{\pre_R}(k,x) \iff k\geq 1 \wedge x \geq k-1$. Then, by
(\ref{wrs}), we have:
\[ (\wnt(R))(\vec{x}) \iff \forall k \geq 1 ~.~ \widehat{\pre_R}(k,x) \iff 
\forall k \geq 1 ~.~ k\geq 1 \wedge x \geq k-1 \iff \false \] Hence
the relation $R$ is well founded. \qed
\end{exa}

\section{Octagonal Relations}
\label{sec:termination:octagons}

Octagonal constraints (also known as Unit Two Variables Per Inequality
or UTVPI, for short) appear in the context of abstract interpretation
where they have been extensively studied as an abstract domain
\cite{mine}. They are defined syntactically as conjunctions of
atomic propositions of the form $\pm x\pm y\leq c$, where $x$ and $y$
are variables and $c \in \zed$ is an integer constant. They are
a~generalization of the simpler notion of {\em difference bounds
  constraints}. Since most results concerning octagons rely on notions
related to difference bounds constraints, we introduce first the
latter, for reasons of self-containment.

\subsection{Difference Bounds Relations} 
\label{sec:background:dbrel}

Difference bounds constraints are also known as {\em zones} in the
context of timed automata verification \cite{alur-dill91} and abstract
interpretation \cite{mine,mine-thesis}. They are defined syntactically as
conjunctions of atomic propositions of the form $x-y\leq c$, where $x$
and $y$ are variables and $c \in \zed$ is an integer
constant. Difference bounds constraints can be represented as matrices
and graphs. These matrices (graphs) have a canonical form, which is
used for efficient inclusion checks, and can be computed by the
classical Floyd-Warshall shortest path algorithm \cite{cormen}.

\begin{defi}\label{dbc}
A formula $\phi(\vec{x})$ is a~\emph{difference bounds constraint} if
it is a~finite conjunction of atomic propositions of the form $x_i-x_j
\le a_{ij},~ 1 \le i,j \le N$, where $a_{ij}\in\zed$.
\end{defi}
\noindent
For example, the equality constraint $x - y = 5$ is equivalent to the
difference bounds constraint $x - y \leq 5 \wedge y - x \leq -5$. In
practice, difference bounds constraints are represented either as
matrices or as graphs:

\begin{defi}\label{dbc:dbm}
Let $\vec{x}=\{x_1, x_2, \ldots, x_N\}$ be a~set of variables ranging
over $\zed$ and $\phi(\vec{x})$ be a~difference bounds
constraint. Then the~{\em difference bounds matrix} (DBM) representing
$\phi$ is the matrix $M_\phi \in \zed_\infty^{N \times N}$ such that:
\[ 
(M_\phi)_{ij} = \begin{cases}
a_{ij} & \mbox{ if } (x_i-x_j\le a_{ij})\in Atom(\phi)\\
\infty & \mbox{ otherwise}
\end{cases}
\]
We denote by $\maxcoef{\phi} \stackrel{\scriptscriptstyle def}{=} \max\{ |c| ~|~
(x_i-x_j\leq c) \in Atom(\phi) \}$ the maximal absolute value over all
constants that appear in $\phi(\x)$.
\end{defi}

Weighted graphs are central to the upcoming developments. An
\emph{integer weighted digraph} is a tuple $G= \langle V, E \rangle$,
where $V$ is a~set of vertices, $E \subseteq V \times \zed \times V$
is a~set of integer-labeled edges. When $G$ is clear from the context,
we denote by $u \arrow{\alpha}{} v$ the fact that $(u, \alpha, v) \in
E$. A {\em path} in $G$ is a~sequence of the form $\pi: v_0
\arrow{\alpha_1}{} v_1 ~\cdots~ v_{p-1} \arrow{\alpha_p}{} v_p$ such
that $(v_{i-1}, \alpha_i, v_i) \in E$ for all $1 \leq i \leq
p$. A~path is \emph{elementary} if $v_i = v_j$ only if $i=1$ and
$j=p$. A \emph{cycle} is a path of length greater than zero, whose
source and destination vertices are the same. An \emph{elementary
  cycle} is a cycle who is elementary.

\begin{defi}\label{dbc:graph}
Let $\vec{x}=\{x_1, x_2, \ldots, x_N\}$ be a~set of variables ranging
over $\zed$ and $\phi(\vec{x})$ be a~difference bounds
constraint. Then $\phi$ can be represented as the~weighted graph
$\mathcal{G}_\phi = (\vec{x}, \rightarrow)$, where each vertex
corresponds to a~variable, and there is an edge $x_i \arrow{a_{ij}}{}
x_j$ in $\mathcal{G}_\phi$ if and only if there exists a~constraint
$x_i - x_j \leq a_{ij}$ in $\phi$, called the~{\em constraint graph}
of $\phi$. 
\end{defi}
Clearly, $M_\phi$ is the incidence matrix of $\mathcal{G}_\phi$.  If
$M \in \zed_\infty^{N \times N}$ is a~DBM, the corresponding
difference bounds constraint is defined as: 
\begin{equation}\label{dbm-formula}
\dbc{M} \equiv \bigwedge_{
  \begin{array}{c}
    \scriptstyle{1 \leq i,j \leq N} \\
    \scriptstyle{M_{ij} < \infty}
\end{array}} x_i - x_j \leq M_{ij}
\end{equation}
For two difference bounds matrices $M_1,M_2 \in \zed_\infty^{N\times
  N}$, let $\min(M_1,M_2) \in \zed_\infty^{N \times N}$ be the matrix
defined as $(\min(M_1,M_2))_{ij} = \min((M_1)_{ij}, (M_2)_{ij})$, for
all $1 \leq i,j \leq N$. We write $M_1=M_2$ if and only if
$(M_1)_{ij}=(M_2)_{ij}$ for all $1\leq i,j\leq N$ and $M_1\leq M_2$ if
and only if $(M_1)_{ij}\leq (M_2)_{ij}$ for all $1\leq i,j\leq N$. 
We write $M_1<M_2$ if and only if $M_1\leq M_2$ and $M_1\neq M_2$.
A~DBM $M$ is said to be \emph{consistent} if and only if its
corresponding constraint $\dbc{M}$ is consistent
(\ref{dbm-formula}). We denote in the following by $\botdb{N}$ any
inconsistent DBM of size $N \times N$. The next definition gives
a~canonical form for consistent DBMs.
\begin{defi}\label{dbm-closed}
A consistent DBM $M \in \zed_\infty^{N \times N}$ is said to be {\em
  closed} if and only if $M_{ii} = 0$ and $M_{ij} \le M_{ik} +
M_{kj}$, for all $1 \leq i, j, k \leq N$.
\end{defi}
Intuitively, the closure of a consistent DBM contains all information
induced by the triangle inequality $M_{ij} \le M_{ik} + M_{kj}$. It is
well known that, $M$ is consistent if and only if it does not contain
a negative weight circuit, i.e.\ there is no sequence of indices $1
\leq i_1, \ldots, i_p \leq N$ such that $M_{i_1i_2} + \ldots +
M_{i_{p-1}i_p} + M_{i_pi_1} < 0$. If $M$ is consistent, then its
closure is unique\footnote{See, e.g.\ \cite{mine}, Section 3.2}. Given
a~consistent DBM $M\in\zed_\infty^{N \times N}$, we denote by $M^*$
the (unique) closed DBM such that $\dbc{M} \iff \dbc{M^*}$. The
consistency of a DBM can be decided in PTIME by the classical
Floyd-Warshall shortest path algorithm (Algorithm \ref{alg:fw}), which
computes also the closure of consistent DBMs:

\begin{prop}\label{dbm-closure}
Let $M \in \zed_\infty^{N \times N}$ be a~DBM representing
a~difference bounds constraint $\phi$. If $M$ is consistent, the
output of Algorithm \ref{alg:fw} is its closure $M^*$. Otherwise, if
$M$ is inconsistent, Algorithm \ref{alg:fw} will report this fact. The
running time of the algorithm is of the order $\mathcal{O}(N^3
\cdot (N+ \log_2\maxcoef{\phi}))$.
\end{prop}
\proof{ 
The correctness proof of the Floyd-Warshall algorithm is standard, e.g.\ Theorem 3.3.5 in \cite{mine-thesis} proves that 
\begin{itemize}
\item eventually $M_{ii}<0$ for some $1\leq i\leq N$, if $M$ is inconsistent
\item the algorithm returns $M^*$, if $M$ is consistent
\end{itemize}
%
Note that inconsistency of $M$ is detected either on line \ref{alg:fw:test:1} or on line \ref{alg:fw:test:2}.

%
%

For each $1\leq i,j,k\leq N$, let $M^0_{ij}$ be the value of $M_{ij}$
after the loop on line \ref{alg:fw:loop:0} terminates and let
$M^k_{ij}$ be the value of $M_{ij}$ after the $k$-th iteration of the
outermost loop on line \ref{alg:fw:loop:k} terminates. For each $0\leq
k\leq N$, we define $\mu_k \stackrel{def}{=} \max\{ |M^k_{ij}| ~|~
M^k_{ij} < \infty \}$. For each $1\leq k\leq N$, we partition the set
$\{(i,j) ~|~ 1\leq i,j\leq N\}$ as follows:
\[\begin{array}{rcl}
  A_k & = & \{ (i,j) ~|~ i\neq k \wedge j\neq k \} \\
  B_k & = & \{ (i,j) ~|~ (i\neq k \wedge j=k) \vee (i=k \wedge j\neq k) \} \\
  C_k & = & \{ (k,k) \}
\end{array}
\]
We next analyze how the updated of matrix entries depend on one
another during the $k$-th iteration of the outermost loop and analyze
how the changes are propagated. Clearly, each $(i,j)\in A_k$ depends
on itself and on 2 entries $(i,k),(k,j)\in B_k$, each $(i,j)\in B_k$
depends on itself and on $(k,k)\in C_k$, and the entry $(k,k)\in C_k$
depends only on itself. It is easy to see, due to the test on line
\ref{alg:fw:test:2}, that before executing the update on line
\ref{alg:fw:update}, $M_{\ell\ell} = 0$ for each $1\leq \ell\leq
N$. Thus, the following holds for each $1\leq k\leq N$:
\[\begin{array}{lclcl}
  \forall (k,k)\in C_k ~.~ M^k_{kk} & = & \min(M^{k-1}_{kk},M^{k-1}_{kk}+M^{k-1}_{kk}) & = & 0 \\
  \forall (i,k)\in B_k ~.~ M^k_{ik} & = & \min(M^{k-1}_{ik},M^{k-1}_{ik}+M^{k-1}_{kk}) & = & M^{k-1}_{ik} \leq \mu_{k-1} \\
  \forall (k,j)\in B_k ~.~ M^k_{kj} & = & \min(M^{k-1}_{kj},M^{k-1}_{kk}+M^{k-1}_{kj}) & = & M^{k-1}_{kj} \leq \mu_{k-1} \\
  \forall (i,j)\in A_k ~.~ M^k_{ij} & = & \min(M^{k-1}_{ij},M^{k-1}_{ik}+M^{k-1}_{kj}) & \leq & 2\cdot \mu_{k-1}
\end{array}
\]
Hence, $\mu_k\leq 2\cdot \mu_{k-1}$ for each $1\leq k\leq N$ and
consequently, $\mu_N\leq 2^N \cdot \mu_0 = 2^N \cdot
\maxcoef{\phi}$. Thus, the $\min$ and sum operations at line
\ref{alg:fw:update} can be executed in time at most $\log_2 \mu_N$
which is of the order $\mathcal{O}(N + \log_2 \maxcoef{\phi})$. Since
line \ref{alg:fw:update} is iterated $N^3$ times, the complexity of
the nested loops at lines \ref{alg:fw:loop:k}--\ref{alg:fw:report:2}
is $\mathcal{O}(N^3 \cdot (N + \log_2 \maxcoef{\phi}))$. The loop at
lines \ref{alg:fw:loop:0}--\ref{alg:fw:init} does not add to this
factor.
%
\qed}

\begin{algorithm}
  \begin{algorithmic}[0]
    \State {\bf input} a difference bounds matrix $M \in
    \zed_\infty^{N \times N}$ 
    \State {\bf output} $M^*$ if $M$ is
    consistent, and report \textrm{''inconsistent''} otherwise
  \end{algorithmic}
  \begin{algorithmic}[1]
    \ForAll{$i=1,\ldots,N$} \label{alg:fw:loop:0}
    \If{$M_{ii} < 0$} \label{alg:fw:test:1}
    \textbf{report} \textrm{''inconsistent''} \label{alg:fw:report:1}
    \Else~ $M_{ii} \leftarrow 0$ \label{alg:fw:init}
    \EndIf
    \EndFor
    \ForAll{$k = 1,\ldots,N$} \label{alg:fw:loop:k}
    \ForAll{$i = 1,\ldots,N$} \label{alg:fw:loop:i}
    \ForAll{$j = 1,\ldots,N$} \label{alg:fw:loop:j}
    \State $M_{ij} \leftarrow \min(M_{ij}, M_{ik} + M_{kj})$ \label{alg:fw:update}
    \If{$i=j$ \textrm{ and } $M_{ii} < 0$} \label{alg:fw:test:2}
    \textbf{report} \textrm{''inconsistent''} \label{alg:fw:report:2}
    \EndIf
    \EndFor
    \EndFor
    \EndFor
  \end{algorithmic}
  \caption{The Floyd-Warshall shortest path algorithm}
  \label{alg:fw}
\end{algorithm}

The closure of DBMs is needed to check the equivalence and entailment
of two difference bounds constraints. Moreover, it is used for
quantifier elimination.

\begin{prop}\label{dbm-eq}
Let $\phi(\vec{x})$, $\phi_1$(\vec{x}) and $\phi_2(\vec{x})$, where
$\vec{x}=\{x_1,\dots,x_N\}$, be consistent difference bounds
constraints. Then the following hold:
\begin{enumerate}
  \item $\phi_1 \Leftrightarrow \phi_2$ if and only if $M_{\phi_1}^* = M_{\phi_2}^*$,
  \item $\phi_1 \Rightarrow \phi_2$ if and only if $M_{\phi_1}^* \leq M_{\phi_2}^*$.
  \item for any $1 \leq k \leq N$, there exists a difference bounds
    constraint $\psi(\vec{x} \setminus \{x_k\})$, such that $\psi \iff
    \exists x_k ~.~ \phi$, and $M^*_\psi \in \zed_\infty^{N-1 \times
      N-1}$ is obtained by eliminating the $k$-th line and column from
    $M^*_\phi$.
\end{enumerate}
\end{prop}
\proof{The points (1), (2) and (3), are equivalent to the Theorems
  3.4.1, 3.4.2 and 3.6.1 (second point) in \cite{mine-thesis},
  respectively. \qed}

{\em Difference bounds relations} are relations defined by difference
bounds constraints over primed and unprimed variables (e.g.\ $x-x'\leq
0$). Difference bounds relations have been studied by Comon and Jurski
who showed, in \cite{comon-jurski}, that their transitive closure is
Presburger definable. In the rest of this paper, for each difference
bounds relation $R \subseteq \zed^{\x} \times \zed^{\x}$, we denote by
$R(\x,\x')$ any difference bounds constraint that defines $R$. Each
DBM $M_{R(\x,\x')} \in \zed_\infty^{2N \times 2N}$ corresponding to
$R(\x,\x')$ is a matrix of dimension $2N \times 2N$, that can be split
into four matrices of dimension $N \times N$, corresponding to the
top-left, bottom-left, top-right and bottom-right corners, denoted as
$\topleft{M}_{R(\x,\x')},\botleft{M}_{R(\x,\x')},
\topright{M}_{R(\x,\x')},\botright{M}_{R(\x,\x')} \in \zed_\infty^{N
  \times N}$. Notice the equivalence $\dbc{\topleft{M^*}_{R(\x,\x')}}
\iff \exists \x' ~.~ R(\x,\x')$ for every consistent constraint
$R(\x,\x')$, by Proposition \ref{dbm-eq} (third point).  In the rest
of this section, we will often write $M_R$ instead of $M_{R(\x,\x')}$,
whenever the defining constraint $R(\x,\x')$ is clear from the
context. In the following, the projection operators are assumed to
have lower priority than closure operators, e.g.\ $\topleft{M^*_R}$
stands for $\topleft{(M^*_R)}$.

\begin{exa}\label{ex:dbm}
Figure~\ref{fig:rf:zigzag}(a) shows the constraint graph
$\mathcal{G}_R$ for the difference bounds relation defined as
$R(\x,\x') \equiv x_2\mi x'_1\leq -1 \wedge x_3\mi x'_2\leq 0 \wedge
x_1\mi x'_3\leq 0 \wedge x'_4\mi x_4\leq 0 \wedge x'_3\mi x_4\leq
0$. Figure~\ref{fig:rf:zigzag}(b) shows the closed DBM representation
of $R$.
\end{exa}

\newcommand{\symbolGone}[0]{
      \scalebox{0.85}{\begin{tikzpicture}
        \TermGridGenNoCapBoxed{0.0}{0.7}{2}{0.0}{0.4}{4}{0.4}{0.4}
        \foreach \ii in {1,...,1} {
          \pgfmathtruncatemacro\jj{\ii+1}
          \TermGridEdgeC{\ii}{1}{\jj}{3}{\tiny$0$}{above}
          \TermGridEdgeC{\jj}{4}{\ii}{4}{\tiny$0$}{above}
        }
      \end{tikzpicture}}
}
\newcommand{\symbolGtwo}[0]{
      \scalebox{0.85}{\begin{tikzpicture}
        \TermGridGenNoCapBoxed{0.0}{0.7}{2}{0.0}{0.4}{4}{0.4}{0.4}
        \foreach \ii in {1,...,1} {
          \pgfmathtruncatemacro\jj{\ii+1}
          \TermGridEdgeC{\ii}{3}{\jj}{2}{\tiny$0$}{above}
          \TermGridEdgeC{\jj}{4}{\ii}{4}{\tiny$0$}{above}
        }
      \end{tikzpicture}}
}
\newcommand{\symbolGthree}[0]{
      \scalebox{0.85}{\begin{tikzpicture}
        \TermGridGenNoCapBoxed{0.0}{0.7}{2}{0.0}{0.4}{4}{0.4}{0.4}
        \foreach \ii in {1,...,1} {
          \pgfmathtruncatemacro\jj{\ii+1}
          \TermGridEdgeC{\ii}{2}{\jj}{1}{\tiny$-\!1$}{above}
          \TermGridEdgeC{\jj}{4}{\ii}{4}{\tiny$0$}{above}
        }
      \end{tikzpicture}}
}
\newcommand{\symbolGfour}[0]{
      \scalebox{0.85}{\begin{tikzpicture}
        \TermGridGenNoCapBoxed{0.0}{0.7}{2}{0.0}{0.4}{4}{0.4}{0.4}
        \foreach \ii in {1,...,1} {
          \pgfmathtruncatemacro\jj{\ii+1}
          \TermGridEdgeC{\ii}{1}{\jj}{3}{\tiny$0$}{above}
          \TermGridEdgeC{\jj}{3}{\ii}{4}{\tiny$0$}{above}
        }
      \end{tikzpicture}}
}
\newcommand{\symbolGeps}[0]{
      \scalebox{0.85}{\begin{tikzpicture}
        \TermGridGenNoCapBoxed{0.0}{0.7}{2}{0.0}{0.4}{4}{0.4}{0.4}
        \foreach \ii in {1,...,1} {
          \pgfmathtruncatemacro\jj{\ii+1}
        }
      \end{tikzpicture}}
}
\newcommand{\symbolGfive}[0]{
      \scalebox{0.85}{\begin{tikzpicture}
        \TermGridGenNoCapBoxed{0.0}{0.7}{2}{0.0}{0.4}{4}{0.4}{0.4}
        \foreach \ii in {1,...,1} {
          \pgfmathtruncatemacro\jj{\ii+1}
          \TermGridEdgeC{\ii}{3}{\jj}{2}{\tiny$0$}{above}
          \TermGridEdgeC{\jj}{3}{\ii}{4}{\tiny$0$}{above}
        }
      \end{tikzpicture}}
}
\newcommand{\symbolGsix}[0]{
      \scalebox{0.85}{\begin{tikzpicture}
        \TermGridGenNoCapBoxed{0.0}{0.7}{2}{0.0}{0.4}{4}{0.4}{0.4}
        \foreach \ii in {1,...,1} {
          \pgfmathtruncatemacro\jj{\ii+1}
          \TermGridEdgeC{\ii}{2}{\jj}{1}{\tiny$-\!1$}{above}
          \TermGridEdgeC{\jj}{3}{\ii}{4}{\tiny$0$}{above}
        }
      \end{tikzpicture}}
}

\begin{figure}[h!]
\centering{
\begin{tabular}{c}
 \begin{tabular}{cc}
  \mbox{\begin{minipage}{2.5cm}
      \scalebox{0.85}{\begin{tikzpicture}
        \TermGridGenDifferentCap{0.0}{1.0}{2}{0.0}{0.75}{4}{0.7}{0.5}{0}
        \foreach \ii in {1,...,1} {
          \pgfmathtruncatemacro\jj{\ii+1}
          \TermGridEdgeC{\ii}{2}{\jj}{1}{\tiny$-\!1$}{right}
          \TermGridEdgeC{\ii}{3}{\jj}{2}{\tiny$0$}{left}
          \TermGridEdgeC{\ii}{1}{\jj}{3}{\tiny$0$}{right}
          \TermGridEdgeC{\jj}{4}{\ii}{4}{\tiny$0$}{below}
          \TermGridEdgeC{\jj}{3}{\ii}{4}{\tiny$0$}{right}
        }
      \end{tikzpicture}}
  \end{minipage}} & 
  \mbox{\begin{minipage}{8cm}
      \scalebox{0.85}{\begin{tikzpicture}
        \TermGridGen{0.0}{1.0}{9}{0.0}{0.75}{4}{0.7}{0.5}{0}
        \foreach \ii in {1,...,8} {
          \pgfmathtruncatemacro\jj{\ii+1}
          \TermGridEdgeC{\ii}{2}{\jj}{1}{\tiny$-\!1$}{right}
          \TermGridEdgeC{\ii}{3}{\jj}{2}{\tiny$0$}{left}
          \TermGridEdgeC{\ii}{1}{\jj}{3}{\tiny$0$}{right}
          \TermGridEdgeC{\jj}{4}{\ii}{4}{\tiny$0$}{below}
          \TermGridEdgeC{\jj}{3}{\ii}{4}{\tiny$0$}{right}
        }
      \end{tikzpicture}}
  \end{minipage}}
  \\
  (a) $\mathcal{G}_R$ -- the constraint graph of $R$ & 
  (c) $\mathcal{G}_R^8$ -- the 8-times unfolding of $\mathcal{G}_R$
 \end{tabular}
\\
\begin{tabular}{cc}
 \scalebox{0.85}{\mbox{\begin{minipage}{7cm}
 $\bordermatrix{
    ~    & x_1 & x_2 & x_3 & x_4 & x_1' & x_2' & x_3' & x_4' \cr
    x_1  & 0 & \infty & \infty & 0 & \infty & \infty & 0 & \infty \cr
    x_2  & \infty & 0 & \infty & \infty & -1 & \infty & \infty & \infty \cr
    x_3  & \infty & \infty & 0 & \infty & \infty & 0 & \infty & \infty \cr
    x_4  & \infty & \infty & \infty & 0 & \infty & \infty & \infty & \infty \cr
    x_1' & \infty & \infty & \infty & \infty & 0 & \infty & \infty & \infty \cr
    x_2' & \infty & \infty & \infty & \infty & \infty & 0 & \infty & \infty \cr
    x_3' & \infty & \infty & \infty & 0 & \infty & \infty & 0 & \infty \cr
    x_4' & \infty & \infty & \infty & 0 & \infty & \infty & \infty & 0 \cr
 }$
 \end{minipage}}}
 &
 \mbox{\begin{minipage}{6.2cm}\mbox{\begin{tikzpicture}

\tikzset{
  sState/.style={draw=black,rectangle,rounded corners=4pt,inner sep=1.5pt,semithick},
  sInitial/.style={initial,initial text=}
}

\node[sState,sInitial] (n1) at (10mm,25mm) {\shortstack{$\bot$\\$r$\\$\bot$\\$l$}};
\node[sState] (n2) at (25mm,3mm){\shortstack{$r$\\$\bot$\\$\bot$\\$l$}};
\node[sState] (n3) at (-5mm,3mm){\shortstack{$\bot$\\$\bot$\\$r$\\$l$}};
\node[sState,accepting] (n4) at (45mm,3mm){\shortstack{$\bot$\\$\bot$\\$rl$\\$\bot$}};
\node[sState,accepting] (n5) at (33mm,25mm){\shortstack{$\bot$\\$\bot$\\$\bot$\\$\bot$}};

\node at (10mm,14mm){$q_2$};
\node at (25mm,-8mm){$q_0$};
\node at (-5mm,-8mm){$q_1$};
\node at (45mm,-8mm){$q_3$};
\node at (33mm,14mm){$q_4$};

\path[->,bend angle=5] 
    (n1) edge [bend left] (n2)
    (n2) edge [bend left] (n3) edge [bend right] (n4)
    (n3) edge [bend left] (n1)
    (n4) edge [bend right] (n5)
    (n5) edge [loop right,out=-40,in=40,looseness=7] (n5);

\node at (22mm,21mm) {\scalebox{0.85}{\symbolGthree}};
\node at (22mm,29mm) {\scalebox{0.75}{$G_3$}};
\node at (10mm,-5mm) {\scalebox{0.85}{\symbolGone}};
\node at (17mm,-7mm) {\scalebox{0.75}{$G_1$}};
\node at (-4mm,19mm) {\scalebox{0.85}{\symbolGtwo}};
\node at (-4mm,27mm) {\scalebox{0.75}{$G_2$}};
\node at (33mm,-5mm) {\scalebox{0.85}{\symbolGfour}};
\node at (40mm,-7mm) {\scalebox{0.75}{$G_4$}};
\node at (45mm,31mm) {\scalebox{0.75}{\symbolGeps}};
\node at (51mm,29mm) {\scalebox{0.75}{$G_5$}};
\node at (45mm,19mm) {\scalebox{0.75}{\symbolGeps}};
\node at (51mm,17mm) {\scalebox{0.75}{$G_5$}};

\end{tikzpicture}}\end{minipage}}
 \\
 (b) $M^*_R$ -- the difference bounds matrix of $R$ & (d) Zigzag automaton $\mathcal{A}_{2,4}$
\end{tabular}
\\
\begin{tabular}{cccccccccccccccccccccc}
  \symbolGone & \symbolGtwo & \symbolGthree & \symbolGfour & \symbolGeps & \symbolGfive & \symbolGsix
  \\
  $G_1$ & $G_2$ & $G_3$ & $G_4$ & $G_5$ & $G_6$ & $G_7$
\end{tabular}
\\
(e) The zigzag alphabet $\Sigma_R=\{G_1,\dots,G_7\}$
\\
\scalebox{1.0}{\begin{tikzpicture}
  \scriptsize
        \TermGridGen{0.0}{0.8}{9}{0.0}{0.4}{4}{0.7}{0.5}{0}
        \TermGridEdgeC{1}{2}{2}{1}{$-\!1$}{above}
        \TermGridEdgeC{2}{1}{3}{3}{$0$}{above}
        \TermGridEdgeC{3}{3}{4}{2}{$0$}{above}
        \TermGridEdgeC{4}{2}{5}{1}{$-\!1$}{above}
        \TermGridEdgeC{5}{1}{6}{3}{$0$}{above}
        \TermGridEdgeC{6}{3}{7}{2}{$0$}{above}
        \TermGridEdgeC{7}{2}{8}{1}{$-\!1$}{above}
        \TermGridEdgeC{8}{1}{9}{3}{$0$}{above}
        \TermGridEdgeC{9}{3}{8}{4}{$0$}{above}
        \TermGridEdgeC{8}{4}{7}{4}{$0$}{above}
        \TermGridEdgeC{7}{4}{6}{4}{$0$}{above}
        \TermGridEdgeC{6}{4}{5}{4}{$0$}{above}
        \TermGridEdgeC{5}{4}{4}{4}{$0$}{above}
        \TermGridEdgeC{4}{4}{3}{4}{$0$}{above}
        \TermGridEdgeC{3}{4}{2}{4}{$0$}{above}
        \TermGridEdgeC{2}{4}{1}{4}{$0$}{above}
      \end{tikzpicture}}
\\
(f) A~path from $\xxki{0}{2}$ to $\xxki{0}{4}$ in $\mathcal{G}_R^8$ (Fig.\ \ref{fig:rf:zigzag}\,(b))
\\
\mbox{\begin{minipage}{12cm}
\hspace{0cm}\mbox{\scalebox{1.0}{\begin{tikzpicture}
  \TermZrBase{1}{4}{4}{0.7}{0.65}{0.5}{1}{3}{1}
  \TermZrBase{5}{8}{4}{0.7}{0.65}{0.5}{4}{3}{1}


  \newarray\aCaption
  \readarray{aCaption}{$q_2$ & $q_0$ & $q_1$ & $q_2$ & $q_0$ & $q_1$ & $q_2$ & $q_0$ & $q_3$}
  \TermZrCapState{0}{8}{4}{0.7}{0.65}{0.5}
  \readarray{aCaption}{$G_3$ & $G_1$ & $G_2$ & $G_3$}
  \TermZrCapGraph{1}{4}{4}{0.7}{0.65}{0.5}
  \readarray{aCaption}{$G_1$ & $G_2$ & $G_3$ & $G_4$}
  \TermZrCapGraph{5}{8}{4}{0.7}{0.65}{0.5}

  \TermZrStateElem{0}{1}{\bot}
  \TermZrStateElem{0}{2}{r}
  \TermZrStateElem{0}{3}{\bot}
  \TermZrStateElem{0}{4}{l}

  \TermZrStateElem{1}{1}{r}
  \TermZrStateElem{1}{2}{\bot}
  \TermZrStateElem{1}{3}{\bot}
  \TermZrStateElem{1}{4}{l}

  \TermZrStateElem{2}{1}{\bot}
  \TermZrStateElem{2}{2}{\bot}
  \TermZrStateElem{2}{3}{r}
  \TermZrStateElem{2}{4}{l}

  \TermZrStateElem{3}{1}{\bot}
  \TermZrStateElem{3}{2}{r}
  \TermZrStateElem{3}{3}{\bot}
  \TermZrStateElem{3}{4}{l}

  \TermZrStateElem{4}{1}{r}
  \TermZrStateElem{4}{2}{\bot}
  \TermZrStateElem{4}{3}{\bot}
  \TermZrStateElem{4}{4}{l}

  \TermZrStateElem{5}{1}{\bot}
  \TermZrStateElem{5}{2}{\bot}
  \TermZrStateElem{5}{3}{r}
  \TermZrStateElem{5}{4}{l}

  \TermZrStateElem{6}{1}{\bot}
  \TermZrStateElem{6}{2}{r}
  \TermZrStateElem{6}{3}{\bot}
  \TermZrStateElem{6}{4}{l}

  \TermZrStateElem{7}{1}{r}
  \TermZrStateElem{7}{2}{\bot}
  \TermZrStateElem{7}{3}{\bot}
  \TermZrStateElem{7}{4}{l}

  \TermZrStateElem{8}{1}{\bot}
  \TermZrStateElem{8}{2}{\bot}
  \TermZrStateElem{8}{3}{rl}
  \TermZrStateElem{8}{4}{\bot}

 {\scriptsize
  \TermZrEdgeFWLab{1}{2}{1}{-1}
  \TermZrEdgeBWLab{1}{4}{4}{0}

  \TermZrEdgeFWLab{2}{1}{3}{0}
  \TermZrEdgeBWLab{2}{4}{4}{0}

  \TermZrEdgeFWLab{3}{3}{2}{0}
  \TermZrEdgeBWLab{3}{4}{4}{0}

  \TermZrEdgeFWLab{4}{2}{1}{-1}
  \TermZrEdgeBWLab{4}{4}{4}{0}

  \TermZrEdgeFWLab{5}{1}{3}{0}
  \TermZrEdgeBWLab{5}{4}{4}{0}

  \TermZrEdgeFWLab{6}{3}{2}{0}
  \TermZrEdgeBWLab{6}{4}{4}{0}

  \TermZrEdgeFWLab{7}{2}{1}{-1}
  \TermZrEdgeBWLab{7}{4}{4}{0}

  \TermZrEdgeFWLab{8}{1}{3}{0}
  \TermZrEdgeBWLab{8}{3}{4}{0}
 }

\end{tikzpicture}}}
\end{minipage}}
\\
(g) A~run of $\mathcal{A}_{2,4}$ (Fig.\ \ref{fig:rf:zigzag}\,(d)) accepting the word $G_3.(G_1.G_2.G_3)^2.G_4\in\Sigma_R^+$ (Fig.\ \ref{fig:rf:zigzag}\,(e))
\\
which encodes the path from Fig.\ \ref{fig:rf:zigzag}\,(f)
\end{tabular}
}
\caption{ Illustration of various notions for a~difference bounds
  relation $R \iff x_2\mi x'_1\leq -1 \wedge x_3\mi x'_2\leq 0 \wedge
  x_1\mi x'_3\leq 0 \wedge x'_4\mi x_4\leq 0 \wedge x'_3\mi x_4\leq
  0$. }
\label{fig:rf:zigzag}
\end{figure}
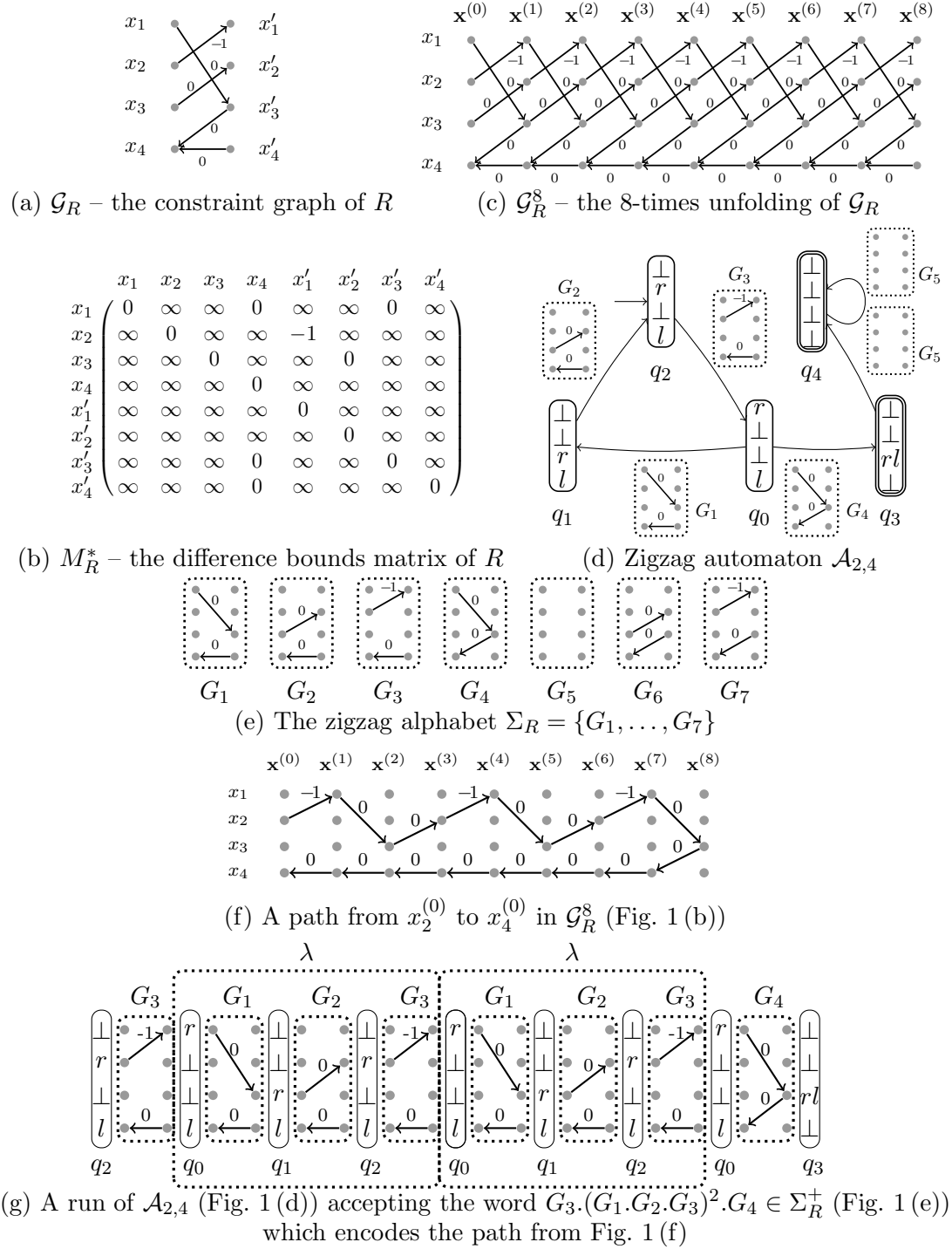

We show next that the composition of two difference bounds relations
encoded as DBMs can be computed in PTIME using Algorithm
\ref{alg:fw}. Let $R_1,R_2 \subseteq \zed^{\x} \times \zed^{\x}$ be
two difference bounds relations. We write $M_1$ and $M_2$ for
$M_{R_1(\x,\x')}$ and $M_{R_2(\x,\x')}$, i.e.\ the DBMs corresponding
to the difference bounds constraints $R_1(\x,\x')$ and $R_2(\x,\x')$,
respectively. Let $\mathcal{M}_{12} \in \zed^{3N \times 3N}$ be the
following matrix:
\begin{equation}\label{m-one-two}
\mathcal{M}_{12} = \left(\begin{array}{ccc}
  \scriptstyle{\topleft{M_1}} & \scriptstyle{\topright{M_1}} & \infty
  \cr \scriptstyle{\botleft{M_1}} & \scriptstyle{\min(\botright{M_1},
    \topleft{M_2})} & \scriptstyle{\topright{M_2}} \cr \infty &
  \scriptstyle{\botleft{M_2}} & \scriptstyle{\botright{M_2}}
\end{array}\right)
\end{equation}
and let $M_1 \odot M_2 \in \zed^{2N \times 2N}$ be the matrix obtained
by erasing the lines and columns $N+1,\ldots,2N$ from the closure
$\mathcal{M}^*_{12}$, if $\mathcal{M}_{12}$ is consistent, and
$\botdb{2N}$, otherwise. 

\begin{prop}\label{dbrel-comp}
  Let $R_1,R_2 \subseteq \zed^{\vec{x}} \times \zed^{\vec{x}}$ be two
  relations defined by the difference bounds constraints $R_1(\x,\x')$
  and $R_2(\x,\x')$, respectively. Then $\dbc{M_{R_1(\x,\x')} \odot
    M_{R_2(\x,\x')}}$ defines the composition $R_1 \circ R_2 \subseteq
  \zed^{\vec{x}} \times \zed^{\vec{x}}$. Moreover, $M_{R_1(\x,\x')}
  \odot M_{R_2(\x,\x')}$ can be computed in time $\mathcal{O}(N^3
  \cdot (N + \log_2(\max(\maxcoef{R_1},\maxcoef{R_2}))))$.
\end{prop}
\proof{The composition $R_1 \circ R_2$ is defined by the formula
  $\exists \vec{y} ~.~ R_1(\vec{x},\vec{y}) \wedge
  R_2(\vec{y},\vec{x'})$. It is easy to see that $\mathcal{M}_{12}$ is
  the DBM corresponding to the conjunction $R_1(\vec{x},\vec{y})
  \wedge R_2(\vec{y},\vec{x'})$, after the elimination of the
  redundant constraints on $\vec{y}$, i.e.\ the replacement of any
  conjunction of the form $x_i - x_j \leq c \wedge x_i - x_j \leq d$
  by $x_i - x_j \leq \min(c,d)$. The existential quantifiers are
  eliminated by checking the consistency of $\mathcal{M}_{12}$,
  computing its closure, and erasing the lines and columns
  $N+1,\ldots,2N$ (by Proposition \ref{dbm-eq}, third point). The time
  complexity upper bound is a direct consequence of the complexity of
  Algorithm \ref{alg:fw} (Proposition \ref{dbm-closure}) used to
  compute $\mathcal{M}^*_{12}$. \qed}

In general, for a DBM $M \in \zed_\infty^{2N \times 2N}$, we define
$M^{\odot^1} = M$ and $M^{\odot^n} = M^{\odot^{n-1}} \odot M$, for any
$n > 1$. An inductive argument shows that the difference bounds
constraint $\dbc{M_{R(\x,\x')}^{\odot^n}}$ defines $R^n$, for any
difference bounds relation $R \subseteq \zed^{\x} \times \zed^{\x}$
and $n > 0$. In the following, we write $R^n(\x,\x')$ for
$\dbc{M_{R(\x,\x')}^{\odot^n}}$.

\subsection{Zigzag Automata} \label{sec:zigzag}

In this section we introduce an automata-theoretic model for reasoning
about the powers of a difference bounds relation. Since a difference
bounds relation $R \subseteq \zed^{\x} \times \zed^{\x}$ is
represented by a difference constraint formula $R(\x,\x')$, which, in
turn, can be seen as a constraint graph $\mathcal{G}_R$ (Definition
\ref{dbc:graph}), the $m$-th power of $R$ can be seen as a constraint
graph consisting of $m$ copies of $\mathcal{G}_R$:

\begin{defi} \label{dbc:graph:unfolding}
  Let $R\subseteq\zed^\x \times \zed^\x$ be a~difference bounds
  relation, where $\vec{x}=\{x_1,\dots,x_N\}$, and $\mathcal{G}_R$ be
  the constraint graph of a difference bounds constraint $R(\x,\x')$
  defining $R$. The {\em $n$-times unfolding} of $\mathcal{G}_R$ is
  defined for every $n>0$ as: \[\mathcal{G}^n_R = (\bigcup_{k=0}^{n}
  \xk{k}, \rightarrow)\] where $\arrow{}{}~ \subseteq
  (\bigcup_{k=0}^{n}\xk{k}) \times \zed \times
  (\bigcup_{k=0}^{n}\xk{k})$, $\xk{k}=\{\xxki{k}{i} ~|~ 1 \leq i \leq
  N\}$ and for all $0\leq k < n$, there is an edge:
  \begin{itemize}
    \item $\xxki{k}{i} \arrow{c}{} \xxki{k}{j}$ if and only if
      $(x_i-x_j\leq c) \in Atom(R(\x,\x'))$
    \item $\xxki{k}{i} \arrow{c}{} \xxki{k+1}{j}$ if and only if
      $(x_i-x'_j\leq c) \in Atom(R(\x,\x'))$
    \item $\xxki{k+1}{i} \arrow{c}{} \xxki{k}{j}$ if and only if
      $(x'_i-x_j\leq c) \in Atom(R(\x,\x'))$
    \item $\xxki{k+1}{i} \arrow{c}{} \xxki{k+1}{j}$ if and only if
      $(x'_i-x'_j\leq c) \in Atom(R(\x,\x'))$
  \end{itemize}
  where $\xxki{k}{i} \arrow{c}{} \xxki{\ell}{j}$ stands for
  $(\xxki{k}{i}, c, \xxki{\ell}{j}) \in ~\arrow{}{}$.
\end{defi}
Each constraint in $R^n(\x,\x')$ corresponds to a~path between
extremal\footnote{A vertex $v$ is said to be {\em extremal} in
  $\mathcal{G}_R^n$ if $v\in(\xk{0}\cup\xk{n})$.} vertices in
$\mathcal{G}_R^n$. Notice that, since difference bounds relations are
closed under composition (Proposition \ref{dbrel-comp}), then $R^n$ is
a difference bounds relation, for any $n>0$. For any given integer
$n>0$, assuming that $R^n$ is consistent, $R^n$ is defined by the
following difference constraint:
\begin{equation}\label{dbm-min-paths}
\begin{array}{rcl}
\bigwedge_{\scriptscriptstyle{1 \leq i, j \leq N}} & x_i - x_j \leq
\min_{[\mathcal{G}_R^n]}\{\xxki{0}{i} \arrow{}{} \xxki{0}{j}\} \wedge
x'_i - x'_j \leq \min_{[\mathcal{G}_R^n]}\{\xxki{n}{i} \arrow{}{} \xxki{n}{j}\} ~\wedge \\ 
& x_i - x_j' \leq \min_{[\mathcal{G}_R^n]}\{\xxki{0}{i} \arrow{}{} \xxki{n}{j}\} \wedge 
x'_i - x_j \leq \min_{[\mathcal{G}_R^n]}\{\xxki{n}{i} \arrow{}{} \xxki{0}{j}\}
\end{array}
\end{equation}
\noindent
where $\min_{[\mathcal{G}_R^n]}\{\xxki{p}{i} \arrow{}{} \xxki{q}{j}\}$
stands for the minimal weight between all paths among the extremal
vertices $\xxki{p}{i}$ and $\xxki{q}{j}$ in $\mathcal{G}_R^n$, for
$p,q \in \{0,n\}$.

\begin{exa}\label{ex:unfold-dbm}
Figure~\ref{fig:rf:zigzag}(c) depicts the $8$-times unfolding of
$\mathcal{G}_R$ for the relation $R(\x,\x') \equiv x_2\mi x'_1\leq -1
\wedge x_3\mi x'_2\leq 0 \wedge x_1\mi x'_3\leq 0 \wedge x'_4\mi
x_4\leq 0 \wedge x'_3\mi x_4\leq 0$ from Example \ref{ex:dbm}.
\end{exa}

The set of paths between any two extremal vertices in the unfolding
graph $\mathcal{G}_R^n$ of a difference bounds relation $R$, for some
$n > 0$, can be seen as words over the finite alphabet of subgraphs of
$\mathcal{G}_R$ that are accepted by a~finite weighted automaton
called \emph{zigzag automaton} \cite{fundamenta}. Intuitively, a
zigzag automaton reads, at step $i$ in the computation, all edges
between $\xk{i}$ and $\xk{i+1}$ simultaneously. The weight of a
transition fired by the zigzag automaton at step $i$ is the sum of the
weights of these edges. A run of a zigzag automaton of length $n> 0$
will thus encode a~path between the extremal vertices in
$\mathcal{G}_R^n$. Since we are interested in the minimal weight paths
(\ref{dbm-min-paths}), we aim at computing the minimal weight among
all runs of length $n$, as a function of $n$. One of the results of
\cite{fundamenta} is that the minimal weight functions are definable
in Presburger arithmetic, hence the transitive closures of difference
bounds relations are Presburger definable as well. Moreover, one of
the results of \cite{cav10} is that these functions generate {\em
  periodic} sequences. In this paper we use zigzag automata to define
the closed form of the sequence
$\{\pre^n_R(\zed^{\vec{x}})\}_{n\geq1}$ of sets (preconditions) from
which larger and larger executions, of length $n=1,2,\ldots$ are
possible. This section is concerned with the formal definition of
zigzag automata.

\subsubsection{The Zigzag Alphabet}

Without losing generality, we work with a~simplified, yet equivalent,
form of difference bounds relations. Let $R \subseteq \zed^{\vec{x}}
\times \zed^{\vec{x}}$ be a difference bounds relation, and
$R(\x,\x')$ be a difference bounds constraint defining $R$. We can
replace all atomic propositions of the form $x - y \leq c$ in
$R(\x,\x')$ by conjunctions $x - z' \leq c ~\wedge~ z' - y \leq 0$,
and all atomic propositions of the form $x' - y' \leq c$ by
conjunctions $x' - z \leq c ~\wedge~ z - y' \leq 0$, for some
variables $z \in \vec{x} \setminus FV(R(\x,\x'))$, one for each
replaced atomic proposition, not occurring initially in
$R(\x,\x')$. We assume further on that any given difference bounds
constraint $R(\x,\x')$ does not contain atomic propositions of the form
$x - y \leq c$ or $x' - y' \leq c$, and that its constraint graph
$\mathcal{G}_R$ is \emph{bipartite}, i.e.\ it does only contain edges
from $\vec{x}$ to $\vec{x'}$ or vice versa.

 We define the zigzag automaton that is used to define the closed form
 of precondition sequences $\{\pre^n_R(\zed^{\vec{x}})\}_{n \geq 0}$,
 where $\pre^n_R(\zed^{\vec{x}}) \subseteq \zed^{\vec{x}}$ are sets
 defined only by constraints between unprimed variables. Since
 $\pre^n_R(\zed^{\vec{x}}) = \pre_{R^n}(\zed^{\vec{x}})$, and taking
 into account the definition of the $n$-th powers of $R$
 (\ref{dbm-min-paths}), these constraints correspond to minimal weight
 paths of the form $\xxki{0}{i} \arrow{}{} \xxki{0}{j}$ in
 $\mathcal{G}^n_R$. These paths are represented by words $w=w_1\ldots
 w_n$, as follows: the symbol $w_i$ represents \emph{simultaneously}
 all edges of $\pi$ that involve only nodes from $\vec{x}^{(i)} \cup
 \vec{x}^{(i+1)}$, for all $0 \leq i < n$. With these considerations,
 the alphabet $\Sigma_R$ is the set of graphs $G$ satisfying the
 following conditions:
\begin{enumerate}
\item the set of nodes of $G$ is $\vec{x} \cup \vec{x'}$

\item for any $x, y \in \vec{x} \cup \vec{x'}$, there is an edge
  labeled with $c \in \zed$ from $x$ to $y$ only if $(x - y \leq c)
  \in Atom(\phi)$

\item the in-degree and out-degree of each node are at most one

\item the number of edges from $\vec{x}$ to $\vec{x'}$ equals the
  number of edges from $\vec{x'}$ to $\vec{x}$
\end{enumerate}
We denote by $\Sigma^+_R$ the set of all non-empty words using symbols
from $\Sigma_R$. The weight of any symbol $G \in \Sigma_R$, denoted
$\omega(G)$, is the sum of the weights that occur on its edges.
For a~word $w = w_1 w_2 \ldots w_n \in \Sigma_R^+$, we define its 
weight as $\omega(w) = \sum_{i=1}^n \omega(w_i)$.

\begin{exa}\label{example:alphabet}
Figure~\ref{fig:rf:zigzag}(e) shows the zigzag alphabet $\Sigma_R$
for the difference bounds relation $R \iff x_2\mi x'_1\leq -1 \wedge
x_3\mi x'_2\leq 0 \wedge x_1\mi x'_3\leq 0 \wedge x'_4\mi x_4\leq 0
\wedge x'_3\mi x_4\leq 0$ from Example \ref{ex:dbm}.
\end{exa}

\subsubsection{The Transition Table of Zigzag Automata}


For each pair of variables $x_i, x_j \in \vec{x} = \{x_1, \ldots,
x_N\}$, we define an automaton $\mathcal{A}_{ij}$ that encodes all
paths from $\mathcal{G}^n_R$, starting in $\xxki{0}{i}$ and ending in
$\xxki{0}{j}$, for some $n > 0$. These automata share the same
alphabet and transition table, and differ only by the choice of the
sets of initial and final states. The common transition table is
defined as $T_R = \langle Q, \delta \rangle$, where the set of states
$Q$ is the set of $N$-tuples $\vec{q} = \langle \vec{q}_1, \ldots,
\vec{q}_N \rangle$ of symbols $\vec{q}_i \in \{\ell, r, \ell r, r\ell,
\bot\}$ capturing the direction of the incoming and outgoing edges of
the alphabet symbols: $\ell$ for a~path traversing from right to left,
$r$ for a~path traversing from left to right, $\ell r$ for a~right
incoming and right outgoing path, $r\ell$ for a~left incoming and left
outgoing path, and $\bot$ when there are no incoming nor outgoing
edges from that node (see Figure~\ref{fig:rf:zigzag}(g) for an example
of the use of states in a zigzag automaton). The set of transitions
$\delta$ is the set of transitions of the form $\vec{q} \arrow{G}{}
\vec{q'}$ such that for every $1 \leq i \leq N$:
\begin{itemize}
\item $\vec{q}_i = \ell$ iff $G$ has one edge whose destination is $x_i$,
  and no other edge involving $x_i$,
\item $\vec{q'}_i = \ell$ iff $G$ has one edge whose source is $x_i'$,
  and no other edge involving $x_i'$,
\item $\vec{q}_i = r$ iff $G$ has one edge whose source is $x_i$, and no
  other edge involving $x_i$,
\item $\vec{q'}_i = r$ iff $G$ has one edge whose destination is
  $x_i'$, and no other edge involving $x_i'$,
\item $\vec{q}_i = \ell r$ iff $G$ has exactly two edges involving $x_i$, one
  having $x_i$ as source, and another as destination,
\item $\vec{q'}_i = r\ell$ iff $G$ has exactly two edges involving
  $x_i'$, one having $x_i'$ as source, and another as destination,
\item $\vec{q'}_i \in \{\ell r, \bot\}$ iff $G$ has no edge involving
  $x_i'$,
\item $\vec{q}_i \in \{r\ell, \bot\}$ iff $G$ has no edge involving
  $x_i$.
\end{itemize}
The weight of each transition $\vec{q} \arrow{G}{} \vec{q'}$ from
$\delta$ is the weight of its symbol $\omega(G)$.  The weight of a run
$\pi: q_1\arrow{a_1}{}q_2\arrow{a_2}{}\dots \arrow{a_n}{}q_{n+1}$,
$n\geq 1$, is defined as $\omega(\pi) \stackrel{def}{=} \sum_{i=1}^{n}
\omega(a_i)$.

The zigzag automaton recognizing paths from $\xxki{0}{i}$ to
$\xxki{0}{j}$, for two distinct indices $1 \leq i,j \leq N$, $i \neq
j$, is defined as $\mathcal{A}_{ij} = \langle T_R, I_{ij}, F \rangle$,
where $I_{ij}, F \subseteq \{\ell, r, \ell r, r\ell, \bot\}^N$ are the
sets of initial and final states, respectively:
\[\begin{array}{ccl}
I_{ij} & = & \{\vec{q} ~|~ \vec{q}_i = r,~ \vec{q}_j = \ell,~ 
\vec{q}_h \in \{\ell r, \bot\},~ \forall h \in \{1,\ldots,N\} \setminus \{i,j\}\} \\
F & = & \{r\ell, \bot\}^N 
\end{array}\]
The zigzag automaton recognizing elementary cycles that traverse $\xxki{0}{i}$ for some $1\leq i\leq N$, is defined as $\mathcal{A}_{ii} = \langle T_R,I_{ii},F \rangle$ where $T_R$ and $F$ are as defined previously and 
\[ I_{ii} = \{\vec{q} ~|~ \vec{q}_i = \ell r,~
\vec{q}_h \in \{\ell r, \bot\},~ \forall h \in \{1,\ldots,N\} \setminus \{i\}\} \]

Since the set of states of a zigzag automaton is the set of tuples
$\{\ell, r, \ell r, r\ell, \bot\}^N$, then the number of states
reachable from an initial state, and co-reachable from a final state
is bounded by $5^N$. In the following, we denote runs of the form
$q_1\arrow{a_1}{}q_2\arrow{a_2}{}\dots \arrow{a_n}{}q_{n+1}$ in the
zigzag automata by $q_1\arrow{a_1\dots a_n}{}q_{n+1}$. Given words
$w_1,w_2 \in \Sigma_R^*$ and runs $\pi_1=q_1\arrow{w_1}{}q_2$ and
$\pi_2=q_2\arrow{w_2}{}q_3$ of some zigzag automaton
$\mathcal{A}_{ij}$, we write $\pi=\pi_1.\pi_2$ to denote their
concatenation $q_1 \arrow{w_1}{}q_2 \arrow{w_2}{} q_3$.

\begin{exa}
Figure~\ref{fig:rf:zigzag}(d) shows the zigzag automaton
$\mathcal{A}_{24}$ of the difference bounds relation $R \iff x_2\mi
x'_1\leq -1 ~\wedge~ x_3\mi x'_2\leq 0 ~\wedge~ x_1\mi x'_3\leq 0
~\wedge~ x'_4\mi x_4\leq 0 ~\wedge~ x'_3\mi x_4\leq 0$ from Example
\ref{ex:dbm} and Example \ref{example:alphabet}. Note that
useless\footnote{A control state is {\em useless} if it is not
  reachable from an initial state or no final state is reachable from
  it.} control states are not shown and hence the alphabet symbols
$G_6$ and $G_7$ are not used. Figure~\ref{fig:rf:zigzag}(f) shows a
path $\xxki{0}{2} \arrow{}{} \ldots \arrow{}{} \xxki{0}{4}$ from
$\mathcal{G}^8_R$ which is encoded by the word $\gamma =
G_3.(G_1.G_2.G_3)^2.G_4$. Figure~\ref{fig:rf:zigzag}(g) shows a~run of
$\mathcal{A}_{24}$ that accepts $\gamma$. The weights of the symbols
in the word are $\omega(G_1)\!=\!\omega(G_2)\!=\!\omega(G_4)\!=\!0$, $\omega(G_3)\!=\!-1$,
hence $\omega(\gamma)=-3$.
\end{exa} 

\subsubsection{Language and Periodicity of Zigzag Automata}

We recall that $\mathcal{G}_R^n$ denotes the constraint graph obtained
by concatenating the constraint graph of $R$ to itself $n>0$ times.  A
run of the zigzag automaton $\mathcal{A}_{ij} = \langle T_R, I_{ij}, F
\rangle$, for some $1 \leq i,j \leq N$ is said to be {\em accepting}
if it starts with a state from $I_{ij}$ and it ends with a state from
$F$. The following lemma relates certain paths in
$\mathcal{G}^n_R$ to runs in zigzag automata.

\begin{lem}[\cite{fundamenta}]\label{zigzag-automata}
Let $R(\x,\x')$ be a~difference bounds constraint defining a~relation
and let $\mathcal{G}_R$ be its constraint graph. Then for any $n\geq
1$ such that $R^n(\x,\x')$ is consistent and any $1 \leq i,j \leq N$,
$i \neq j$, $\mathcal{A}_{ij}$ has an accepting run of length $n$ if
and only if there exists a~path in $\mathcal{G}_R^n$, from
$\xxki{0}{i}$ to $\xxki{0}{j}$. Moreover,
\[
\begin{array}{lcl}
  (M^*_{R^n})_{ij}
  & = & \min\{ \omega(\pi) ~|~ \textrm{$\pi$ is an accepting run in $\mathcal{A}_{ij}$ of length $n$} \}
\end{array}
\]
Furthermore, for any $n\geq 1$, $R^n(\x,\x')$ is inconsistent if and
only if $A_{ii}$ has an accepting run $\pi$ such that $|\pi|=n$ and
$\omega(\pi)<0$ for some $1\leq i\leq N$.
\end{lem}
\proof{See \cite{fundamenta}, Lemma 4.3. \qed}

The formula (\ref{dbm-min-paths}) defining the powers of a difference
bounds relation $R$ says that, if $R^n$ is consistent, for a given $n
> 0$, then $R^n$ is definable by a closed DBM\footnote{Since the
  coefficients of the DBM are minimal weight paths, the triangle
  inequality holds.}  $M_{R^n} \in \zed^{2N \times 2N}$. It follows
that the set $\pre^n_R(\zed^\x)$ is defined by $\topleft{M_n}$, for
any $n > 0$. Moreover, by (\ref{dbm-min-paths}),
$(\!\topleft{M_n})_{ij}$ is the minimum weight among all accepting
runs of length $n$ of $\mathcal{A}_{ij}$. In the following, we show
that the sequence of matrices $\{\!\topleft{M_{R^n}}\}_{n\geq1}$ is
{\em periodic} in the following sense:

\begin{defi}\label{up-matrix}
  An infinite sequence of integers $\{m_k\}_{k=1}^\infty \in \zed$ is
  said to be {\em periodic} if and only if:
  $$\exists b \geq 1~ \exists c \geq 1~ \exists
  \lambda_0,\lambda_1,\ldots,\lambda_{c-1} \in \zed ~.~ m_{b+(k+1)c+i}
  = \lambda_i + m_{b+kc+i}$$ for all $k \geq 1$ and $i =
  0,1,\ldots,c-1$. An infinite sequence of matrices
  $\{M_k\}_{k=1}^\infty \in \zed_\infty^{N \times N}$ is said to be
  {\em periodic} if and only if:
  $$\exists b \geq 1~ \exists c \geq 1~ \exists
  \Lambda_0,\Lambda_1,\ldots,\Lambda_{c-1} \in \zed_\infty^{N \times
    N} ~.~ M_{b+(k+1)c+i} = \Lambda_i + M_{b+kc+i}$$ for all $k \geq
  1$ and $i = 0,1,\ldots,c-1$. The smallest $b,c$ for which the above
  holds are called the {\em prefix} and {\em period} of the periodic
  sequence, respectively. $\Lambda_0,\Lambda_1,\ldots,\Lambda_{c-1}$
  are called the {\em rates} of the periodic sequence.
\end{defi}
Intuitively, the elements situated at equal distances ($c \geq 1$) beyond
a certain threshold ($b \geq 1$) in a periodic sequence, differ by
equal quantities. The following proposition establishes the
equivalence between periodic sequences of integers and matrices:

\begin{prop}\label{up-int-matrix}
   An infinite sequence of matrices $\{M_k\}_{k=1}^\infty \in
   \zed_\infty^{N \times N}$ is periodic if and only if the sequences
   $\{(M_k)_{ij}\}_{k=1}^\infty \in \zed_\infty$ are periodic, for all
   $1 \leq i,j \leq N$. Moreover, the prefix, period and rates of the
   $\{M_k\}_{k=1}^\infty$ sequence are effectively computable given
   the prefix, period and rates of the $\{(M_k)_{ij}\}_{k=1}^\infty$
   sequences, respectively.
\end{prop}
\proof{See Lemma 1 in \cite{cav10}. \qed}

Periodicity of integer sequences is preserved by several arithmetic
operations, as shown by the following lemma:

\begin{lem}\label{up-sum-min-half}
  Let $\{s_k\}_{k=1}^\infty \in \zed_\infty$ and
  $\{t_k\}_{k=1}^\infty$ be two periodic sequences of integers, of
  given prefix, period and rates. Then the sequences
  $\{\min(s_k,t_k)\}_{k=1}^\infty$, $\{s_k+t_k\}_{k=1}^\infty$ and
  $\{\lfloor\frac{s_k}{2}\rfloor\}_{k=1}^\infty$ are periodic, and
  moreover, their prefix, period and rates are effectively
  computable, respectively.
\end{lem}
\proof{See Lemma 6 in \cite{cav10}. \qed}

Formally, a \emph{weighted digraph} is a tuple $G= \langle V, E,
\omega \rangle$, where $V$ is a~set of vertices, $E \subseteq V \times
V$ is a~set of edges, and $\omega: E \rightarrow \zed$ is a~weight
function. The following theorem shows that the matrices giving the
weights of the minimal weight paths of a given length in a weighted
graph form a periodic sequence of matrices.

\begin{thm}\label{periodic-tropical-matrix}
  Let $G = \langle V, E, \omega \rangle$ be a weighted graph, $V =
  \{v_1, \ldots, v_N\}$ be its set of vertices, and let $\{A_n\}_{n
    \geq 1}$ be the sequence of matrices $A_n \in \zed_\infty^{N
    \times N}$, where for all $1 \leq i,j \leq N$, $(A_n)_{ij}$ is the
  minimal weight among all paths of length $n$ from $v_i$ to $v_j$ in
  $G$. Then $\{A_n\}_{n \geq 1}$ is a periodic sequence, and its
  prefix, period and rates are effectively computable.
\end{thm}
\proof{See, e.g.\ Theorem 3.3 in \cite{de-schutter}. \qed}

An important consequence of Theorem \ref{periodic-tropical-matrix} is
that, for a $*$-consistent difference bounds relation $R$, the
sequence of sets $\{\pre^n_R(\zed^\vec{x})\}_{n \geq 1}$ is definable
by a periodic sequence of difference bounds matrices.

\begin{cor}\label{periodic-pre-dbm}
  Let $R \subseteq \zed^\x \times \zed^\x$, where $\x = \{x_1, \ldots,
  x_N\}$, be a $*$-consistent difference bounds relation. Then, for all
  $n \geq 1$, the difference bounds constraint
  $\dbc{\!\topleft{M^*_{R^n}}}$ defines $\pre^n_R(\zed^\x)$. Moreover,
  the sequence $\{\!\topleft{M^*_{R^n}}\}_{n\geq1}$ is periodic, and
  its prefix, period and rates are all effectively computable.
\end{cor}
\proof{Since $R$ is $*$-consistent, $\mathcal{G}^n_R$ does not have
  negative cycles, for any $n > 0$, hence the minimum
  $\min_{[\mathcal{G}^n_R]}\{\xxki{0}{i} \arrow{}{} \xxki{0}{j}\}$ is
  well defined, for all $1 \leq i,j \leq N$, $i \neq j$. Since $R^n$
  is defined by the difference bounds constraint
  (\ref{dbm-min-paths}), and since the triangle inequality: 
  $$\min_{[\mathcal{G}^n_R]}\{\xxki{0}{i} \arrow{}{} \xxki{0}{j}\}
  \leq \min_{[\mathcal{G}^n_R]}\{\xxki{0}{i} \arrow{}{} \xxki{0}{k}\}
  + \min_{[\mathcal{G}^n_R]}\{\xxki{0}{k} \arrow{}{} \xxki{0}{j}\}$$
  holds for all pairwise distinct indices $1 \leq i,j,k \leq N$, then
  we have: 
  $$(\!\topleft{M^*_{R^n}})_{ij} = \min_{[\mathcal{G}^n_R]}\{\xxki{0}{i}
  \arrow{}{} \xxki{0}{j}\}$$ for all $1 \leq i, j \leq N$, where $i
  \neq j$, by the uniqueness of the closure for DBMs. Clearly, 
  $$(\!\topleft{M^*_{R^n}})_{ii} = 0$$ for all $1 \leq i \leq N$, by
  Definition \ref{dbm-closed}. Then $\pre^n_R(\zed^\x) =
  \pre_{R^n}(\zed^\x)$ is defined by the constraint $\exists \x' ~.~
  R^n(\x,\x') \iff \dbc{\!\topleft{M^*_{R^n}}}$.

  To prove that the sequence of matrices
  $\{\!\topleft{M^*_{R^n}}\}_{n\geq1}$ is periodic, it is enough to
  show that, for all $1 \leq i,j \leq N$, the sequence of integers
  $\{\!(\topleft{M^*_{R^n}})_{ij}\}_{n\geq1}$ is periodic (by
  Proposition \ref{up-int-matrix}). Clearly
  $\{\!(\topleft{M^*_{R^n}})_{ii}\}_{n\geq1}$ is periodic, because
  $(\topleft{M^*_{R^n}})_{ii} = 0$, for all $1 \leq i \leq N$ and all
  $n \geq 1$ (Definition \ref{dbm-closed}). 

  Let $T_R = \langle Q, \delta, \omega \rangle$, $Q = \{q_1, \ldots,
  q_{5^N}\}$, be the common transition table of all zigzag automata
  $\mathcal{A}_{ij} = \langle T_R, I_{ij}, F \rangle$ for $R$. Then,
  by Theorem \ref{periodic-tropical-matrix}, the sequence
  $\{\mathcal{T}_m\}_{m \geq 0}$ is periodic, where $\mathcal{T}_m \in
  \zed^{5^N \times 5^N}$ is the matrix defined as:
  $(\mathcal{T}_m)_{k\ell}$ is the minimum weight among all paths of
  length $m$ between $q_k$ and $q_\ell$ in $T_R$, $1 \leq k,\ell \leq
  5^N$. By Lemma \ref{zigzag-automata}, we have:
  \[\begin{array}{rcl} 
  (\!\topleft{M^*_{R^n}})_{ij} & = & \min\{\omega(\rho) ~|~ \rho
  ~\mbox{is an accepting run of length $n$ in $\mathcal{A}_{ij}$}\} \\
  & = & \min\{(\mathcal{T}_m)_{k\ell} ~|~ q_k \in I_{ij}, q_\ell \in F\}
  \end{array}\]
  By Lemma \ref{up-sum-min-half},
  we obtain that the sequence
  $\{\!\topleft{(M^*_{R^n})}_{ij}\}_{n\geq1}$ is periodic. The
  effective computability of the prefix, period, and rates of the
  $\{\!\topleft{M^*_{R^n}}\}_{n\geq1}$ sequence follows from the
  constructive arguments of Theorem \ref{periodic-tropical-matrix},
  Proposition \ref{up-int-matrix} and Lemma \ref{up-sum-min-half},
  respectively. \qed}

\begin{figure}[h]
\centering{
{\tiny
\[\begin{array}{cccccccccccccc}
\begin{array}{ccccccccccccccccc}
  \scalebox{1.0}{\mbox{\begin{minipage}{3cm}\bordermatrix{
    \topleft{M_{R^1}^*} & x_1 & x_2 & x_3 & x_4 \cr
    x_1 & 0 & \infty & \infty & 0 \cr
    x_2 & \infty & 0 & \infty & \infty \cr
    x_3 & \infty & \infty & 0 & \infty \cr
    x_4 & \infty & \infty & \infty & 0 \cr
  }\end{minipage}}}
  &
  \scalebox{1.0}{\mbox{\begin{minipage}{3cm}\bordermatrix{
    \topleft{M_{R^2}^*} & x_1 & x_2 & x_3 & x_4 \cr
    x_1 & 0 & \infty & \infty & 0 \cr
    x_2 & \infty & 0 & \infty & -1 \cr
    x_3 & \infty & \infty & 0 & \infty \cr
    x_4 & \infty & \infty & \infty & 0 \cr
  }\end{minipage}}}
\end{array}
\\
\begin{array}{ccccccccccccccccc}
  \scalebox{1.0}{\mbox{\begin{minipage}{3cm}\bordermatrix{
    \topleft{M_{R^3}^*} & x_1 & x_2 & x_3 & x_4 \cr
    x_1 & 0 & \infty & \infty & -1 \cr
    x_2 & \infty & 0 & \infty & -1 \cr
    x_3 & \infty & \infty & 0 & -1 \cr
    x_4 & \infty & \infty & \infty & 0 \cr
  }\end{minipage}}}
  &
  \scalebox{1.0}{\mbox{\begin{minipage}{3cm}\bordermatrix{
    \topleft{M_{R^4}^*} & x_1 & x_2 & x_3 & x_4 \cr
    x_1 & 0 & \infty & \infty & -1 \cr
    x_2 & \infty & 0 & \infty & -2 \cr
    x_3 & \infty & \infty & 0 & -1 \cr
    x_4 & \infty & \infty & \infty & 0 \cr
  }\end{minipage}}}
  &
  \scalebox{1.0}{\mbox{\begin{minipage}{3cm}\bordermatrix{
    \topleft{M_{R^5}^*} & x_1 & x_2 & x_3 & x_4 \cr
    x_1 & 0 & \infty & \infty & -1 \cr
    x_2 & \infty & 0 & \infty & -2 \cr
    x_3 & \infty & \infty & 0 & -2 \cr
    x_4 & \infty & \infty & \infty & 0 \cr
  }\end{minipage}}}
  \\
  \scalebox{1.0}{\mbox{\begin{minipage}{3cm}\bordermatrix{
    \topleft{M_{R^6}^*} & x_1 & x_2 & x_3 & x_4 \cr
    x_1 & 0 & \infty & \infty & -2 \cr
    x_2 & \infty & 0 & \infty & -2 \cr
    x_3 & \infty & \infty & 0 & -2 \cr
    x_4 & \infty & \infty & \infty & 0 \cr
  }\end{minipage}}}
  &
  \scalebox{1.0}{\mbox{\begin{minipage}{3cm}\bordermatrix{
    \topleft{M_{R^7}^*} & x_1 & x_2 & x_3 & x_4 \cr
    x_1 & 0 & \infty & \infty & -2 \cr
    x_2 & \infty & 0 & \infty & -3 \cr
    x_3 & \infty & \infty & 0 & -2 \cr
    x_4 & \infty & \infty & \infty & 0 \cr
  }\end{minipage}}}
  &
  \scalebox{1.0}{\mbox{\begin{minipage}{3cm}\bordermatrix{
    \topleft{M_{R^8}^*} & x_1 & x_2 & x_3 & x_4 \cr
    x_1 & 0 & \infty & \infty & -2 \cr
    x_2 & \infty & 0 & \infty & -3 \cr
    x_3 & \infty & \infty & 0 & -3 \cr
    x_4 & \infty & \infty & \infty & 0 \cr
  }\end{minipage}}}
  \\
  \scalebox{1.0}{\mbox{\begin{minipage}{3cm}\bordermatrix{
    \topleft{M_{R^9}^*} & x_1 & x_2 & x_3 & x_4 \cr
    x_1 & 0 & \infty & \infty & -3 \cr
    x_2 & \infty & 0 & \infty & -3 \cr
    x_3 & \infty & \infty & 0 & -3 \cr
    x_4 & \infty & \infty & \infty & 0 \cr
  }\end{minipage}}}
  &
  \scalebox{1.0}{\mbox{\begin{minipage}{3cm}\bordermatrix{
    \topleft{M_{R^{10}}^*} & x_1 & x_2 & x_3 & x_4 \cr
    x_1 & 0 & \infty & \infty & -3 \cr
    x_2 & \infty & 0 & \infty & -4 \cr
    x_3 & \infty & \infty & 0 & -3 \cr
    x_4 & \infty & \infty & \infty & 0 \cr
  }\end{minipage}}}
  &
  \scalebox{1.0}{\mbox{\begin{minipage}{3cm}\bordermatrix{
    \topleft{M_{R^{11}}^*} & x_1 & x_2 & x_3 & x_4 \cr
    x_1 & 0 & \infty & \infty & -3 \cr
    x_2 & \infty & 0 & \infty & -4 \cr
    x_3 & \infty & \infty & 0 & -4 \cr
    x_4 & \infty & \infty & \infty & 0 \cr
  }\end{minipage}}}
\end{array}
\\\\
\begin{array}{ccccccccccccccccc}
  b=3,c=3,
  \Lambda_0=\Lambda_1=\Lambda_2=
  \left(\begin{array}{cccccc}
    0 & 0 & 0 & -1 \\
    0 & 0 & 0 & -1 \\
    0 & 0 & 0 & -1 \\
    0 & 0 & 0 & 0 
  \end{array}\right)
\end{array}
\end{array}\]
}
}
\caption{Periodic behavior of the infinite sequence $\{\topleft{M^*_{R^n}}\}_{n\geq 1}$ where \\$R \iff x_2\mi x'_1\leq -1 \wedge x_3\mi x'_2\leq 0 \wedge x_1\mi x'_3\leq 0 \wedge x'_4\mi x_4\leq 0 \wedge x'_3\mi x_4\leq 0$. }
\label{fig:ex:periodicity}
\end{figure}

\begin{exa} \label{ex:periodicity}
Consider the difference bounds constraint $R(\x,\x') \equiv x_2\mi
x'_1\leq -1 \wedge x_3\mi x'_2\leq 0 \wedge x_1\mi x'_3\leq 0 \wedge
x'_4\mi x_4\leq 0 \wedge x'_3\mi x_4\leq 0$ from Example
\ref{ex:dbm}. We compute the sequence $\{ \topleft{M_{R^n}^*}
\}_{n\geq0}$. Since $R$ is $*$-consistent, the DBM
$\topleft{M_{R^n}^*}$ can be defined for each $n\geq 1$ as
\[ 
  (\topleft{M_{R^n}^*})_{ij}= \left\{\begin{array}{ll} 0 & \textrm{ if
} i=j \\ \min\{ \omega(\rho) ~|~ \rho \textrm{ is a~path from }
\xxki{0}{i} \textrm{ to } \xxki{0}{j} \textrm{ in } \mathcal{G}_R^n
\}\cup\{\infty\} & \textrm{ if } i\neq j
    \end{array}\right.
\]
by (\ref{dbm-min-paths}) (see Fig.\ \ref{fig:rf:zigzag} for $\mathcal{G}_R^8$). The first 11 elements of the sequence are depicted in Figure~\ref{fig:ex:periodicity}. The periodic behavior can be observed for prefix $b=3$, period $c=3$, and rates $\Lambda_0,\Lambda_1,\Lambda_2$ defined in Figure~\ref{fig:ex:periodicity}. For example, $\topleft{M_{R^{6}}^*} = \topleft{M_{R^{3}}^*} + \Lambda_0$, $\topleft{M_{R^{9}}^*} = \topleft{M_{R^{6}}^*} + \Lambda_0$, etc.
\qed\end{exa}

\subsection{Octagonal Constraints}
\label{sec:background:oct:constr}

Octagonal constraints are a~generalization of difference bounds
constraints to conjunctions of atomic propositions of the form $\pm x
\pm y \leq c$, $c \in \zed$. An octagonal constraint
$\phi(x_1,\dots,x_N)$ is usually represented by a difference bounds
constraints $\phi(y_1,\dots,y_{2N})$ where $y_{2i-1}$ stands for
$+x_i$ and $y_{2i}$ stands for $-x_i$, with the implicit requirement
that $y_{2i-1}=-y_{2i}$, for each $1\leq i\leq N$. It is important to
notice that this implicit condition cannot be directly represented as
a difference constraint. The class of integer octagonal constraints is
formally defined as follows:

\begin{defi}\label{odbc}
  A~formula $\phi(\vec{x})$ is an \emph{octagonal constraint} if it is
  a~finite conjunction of terms of the form $x_i - x_j \le a_{ij}$,
  $x_i + x_j \le b_{ij}$ or $-x_i - x_j \le c_{ij}$ where $a_{ij},
  b_{ij}, c_{ij} \in \zed$, for all $1 \le i,j\le N$.
\end{defi}

We represent octagons as difference bounds constraints over the dual
set of variables $\vec{y} = \{y_1,y_2,\ldots,y_{2N}\}$, with the
convention that $y_{2i-1}$ stands for $x_i$ and $y_{2i}$ for $-x_i$,
respectively. For example, the octagonal constraint $x_1+x_2=3$ is
represented as $y_1 - y_4 \leq 3 \wedge y_2 - y_3 \leq -3$.
In order to handle the $\vec{y}$ variables in the following, we define
$\bar{\imath} = i-1$, if $i$ is even, and $\bar{\imath}=i+1$ if $i$ is
odd. Obviously, we have $\bar{\bar{\imath}} = i$, for all $i \in
\zed,~ i \geq 1$. We denote by $\overline{\phi}(\y)$ the difference
bounds constraint over $\y$ that represents $\phi(\x)$ and which is
defined as follows:

\begin{defi} \label{def:oct:to:dbc}
  Given an octagonal constraint $\phi(\x)$, $\x=\{x_1,\dots,x_N\}$,
  its difference bounds representation $\overline{\phi}(\y)$, where
  $\y=\{y_1,\dots,y_{2N}\}$ is a~conjunction of the following
  difference bounds constraints where $1\leq i,j \leq N$, $c\in\zed$.
  \[ \begin{array}{lcl}
    (x_i-x_j\leq c)\in Atom(\phi) & \Leftrightarrow & (y_{2i-1}-y_{2j-1}\leq c), (y_{2j}-y_{2i}\leq c)\in Atom(\overline{\phi}) \\
    (-x_i+x_j\leq c)\in Atom(\phi) & \Leftrightarrow & (y_{2j-1}-y_{2i-1}\leq c), (y_{2i}-y_{2j}\leq c)\in Atom(\overline{\phi}) \\
    (-x_i-x_j\leq c)\in Atom(\phi) & \Leftrightarrow & (y_{2i}-y_{2j-1}\leq c), (y_{2j}-y_{2i-1}\leq c)\in Atom(\overline{\phi}) \\
    (x_i+x_j\leq c)\in Atom(\phi) & \Leftrightarrow & (y_{2i-1}-y_{2j}\leq c), (y_{2j-1}-y_{2i}\leq c)\in Atom(\overline{\phi}) 
  \end{array} \]
\end{defi}

\noindent An octagonal constraint $\phi$ is equivalently represented by the DBM
$M_{\overline{\phi}} \in \zed_\infty^{2N \times 2N}$, corresponding to
$\overline{\phi}$. We sometimes write $M_\phi$ instead of
$M_{\overline{\phi}}$. We say that a~DBM $M \in \zed_\infty^{2N \times
  2N}$ is \emph{coherent} iff $M_{ij} = M_{\bar{\jmath}\bar{\imath}}$
for all $1 \leq i,j \leq 2N$. This property is needed since, for
example, an atomic proposition $x_i - x_j \leq a_{ij}$, $1 \leq i,j
\leq N$, can be represented as both $y_{2i-1} - y_{2j-1} \leq a_{ij}$
and $y_{2j} - y_{2i} \leq a_{ij}$. Dually, a~coherent DBM $M \in
\zed_\infty^{2N \times 2N}$ corresponds to the following octagonal
constraint:
\begin{equation}\label{dbm-octagon}
  \begin{array}{rcl}
    \oct{M} & \equiv &  
    \bigwedge_{
      \begin{array}{l}
        \scriptscriptstyle{1 \leq i,j \leq N} \\
        \scriptscriptstyle{M_{2i-1,2j-1} < \infty}
      \end{array}
    } {x_i - x_j \leq M_{2i-1,2j-1}} ~\wedge~ \\
    && \bigwedge_{
      \begin{array}{l}
        \scriptscriptstyle{1 \leq i,j \leq N} \\
        \scriptscriptstyle{M_{2i-1,2j} < \infty}
      \end{array}
    } {x_i + x_j \leq M_{2i-1,2j}} ~\wedge~ \\
    && \bigwedge_{
      \begin{array}{l}
        \scriptscriptstyle{1 \leq i,j \leq N} \\
        \scriptscriptstyle{M_{2i,2j-1} < \infty}
      \end{array}
    } {-x_i-x_j \leq M_{2i,2j-1}}
    \end{array}
\end{equation}
Given an octagonal constraint $\phi(\x)$, we have the following equivalences:
\begin{equation}\label{dbc-to-oct}
 \begin{array}{rcl}
  \phi(\vec{x}) & \iff & (\exists y_2, y_4, \ldots, y_{2N} ~.~
  \overline{\phi}(\y) \wedge \bigwedge_{i=1}^N y_{2i-1} = -y_{2i})
           [x_i/y_{2i-1}]_{i=1}^N \\ 
           & \iff & \overline{\phi}(\y)[x_i/y_{2i-1},-x_i/y_{2i}]_{i=1}^{N} \\ 
           & \iff & \oct{M_{\overline{\phi}}}
 \end{array}
\end{equation} 
A coherent DBM $M$ is said to be \emph{octagonal-consistent} if and
only if $\oct{M}$ is consistent.

For each octagonal constraint $\phi(\x)$, we define $\maxcoef{\phi}$
to be the maximal absolute value over all constants that appear in
$\phi(\x)$, formally: $\maxcoef{\phi} \stackrel{def}{=} \max\{ |c| ~|~
(\pm x_i \pm x_j \leq c)\in Atom(\phi) \}$.

\begin{defi}\label{tclose}
An octagonal-consistent coherent DBM $M \in \zed_\infty^{2N \times
  2N}$ is said to be \emph{tightly closed} if and only if it is closed
and $M_{ij} \le \lfloor \frac{M_{i\bar{\imath}}}{2} \rfloor + \lfloor
\frac{M_{\bar{\jmath}j}}{2} \rfloor$, for all $1 \leq i,j \leq N$.
\end{defi}
The last condition from Definition \ref{tclose} ensures that the
knowledge induced by the implicit conditions $y_i + y_{\bar\imath} =
0$, which cannot be represented as difference constraints, has been
propagated through the DBM. Since $2y_i = y_i - y_{\bar\imath} \leq
M_{i\bar\imath}$ and $-2y_j = y_{\bar\jmath} - y_j \leq M_{\bar\jmath
  j}$, we have $y_i \leq \lfloor\frac{M_{i\bar\imath}}{2}\rfloor$ and
$-y_j \leq \lfloor\frac{M_{\bar\jmath j}}{2}\rfloor$, which implies
$y_i - y_j \leq \lfloor\frac{M_{i\bar\imath}}{2}\rfloor +
\lfloor\frac{M_{\bar\jmath j}}{2}\rfloor$, thus $M_{ij} \leq
\lfloor\frac{M_{i\bar\imath}}{2}\rfloor + \lfloor\frac{M_{\bar\jmath
    j}}{2}\rfloor$ must hold, if $M$ is supposed to be the most
precise DBM representation of an octagonal constraint. Moreover, by
taking $j = \bar\imath$ in the previous, we have $M_{i\bar\imath} \leq
2 \lfloor \frac{M_{i\bar\imath}}{2} \rfloor$, implying that
$M_{i\bar\imath}$ is necessarily even, if $M$ is tightly closed.

The following theorem from \cite{bagnara-hill-zaffanella08} provides
an effective way of testing octagonal-consistency and computing the
tight closure of a~coherent DBM. Moreover, it shows that the tight
closure of a~given DBM is unique and can also be computed with the
same worst-case time complexity as the DBM closure.

\begin{thm} (\cite{bagnara-hill-zaffanella08}) \label{bhz}
Let $M \in \zed_\infty^{2N \times 2N}$ be a~coherent DBM. Then $M$ is
octagonal-consistent if and only if $M$ is consistent and $\lfloor
\frac{M^*_{i\bar{\imath}}}{2} \rfloor + \lfloor
\frac{M^*_{\bar{\imath}i}}{2} \rfloor \geq 0$, for all $1 \leq i \leq
2N$. Moreover, if $M$ is octagonal-consistent,
the tight closure of $M$ is the DBM $M^t \in
\zed_\infty^{2N \times 2N}$ defined as: $$M^t_{ij}=\min\left\{M^*_{ij},
\left\lfloor \frac{M^*_{i\bar{\imath}}}{2} \right\rfloor +
\left\lfloor \frac{M^*_{\bar{\jmath}j}}{2} \right\rfloor\right\}$$ for
all $1 \leq i,j \leq 2N$ where $M^* \in \zed_\infty^{2N \times 2N}$
is the closure of $M$.
\end{thm}

\begin{cor}\label{oct:consistency:test}
Let $\phi(\x)$ be an octagonal constraint for some
$\x=\{x_1,\dots,x_N\}$ and $N\geq 1$. Then, consistency of $\phi$ can
be decided in at most $\mathcal{O}(N^3\cdot(N+\maxcoef{\phi}))$
time. Moreover, it $\phi$ is consistent, $M^t_{\overline{\phi}}$ can
be computed in at most $\mathcal{O}(N^3\cdot(N+\maxcoef{\phi}))$ time
as well.
\end{cor}
\proof{
An immediate consequence of Theorem \ref{bhz} and Proposition \ref{dbm-closure}.
\qed}

Given an octagonal-consistent coherent DBM $M\in\zed_\infty^{2N\times 2N}$, we
denote by $M^t$ the (unique) tightly closed DBM such that $\oct{M}
\iff \oct{M^t}$. The tight closure of DBMs is needed for checking
equivalence and entailment between octagonal constraints. 

\begin{prop}\label{oct-eq}
  Let $\phi_1(\x)$ and $\phi_2(\x)$ be two consistent octagonal constraints. Then,
  \begin{enumerate}
  \item $\phi_1 \iff \phi_2$ if and only if $M^t_{\overline{\phi_1}} = M^t_{\overline{\phi_2}}$,
  \item $\phi_1 \Rightarrow \phi_2$ if and only if $M^t_{\overline{\phi_1}} \leq M^t_{\overline{\phi_2}}$.
  \end{enumerate}
\end{prop}
\proof{Points (1) and (2) are Theorem 4.4.1 (points 4 and 5,
  respectively) in \cite{mine-thesis}.  \qed}
Moreover, the following proposition shows that octagonal
constraints are closed under existential quantification.

\begin{prop}\label{odbc:qelim}
Let $\phi(\x)$, where $\x=\{x_1,\dots,x_N\}$, be a consistent
octagonal constraint. Further, let $1 \leq k \leq N$ and $M'$ be the
DBM obtained from $M^t_{\overline{\phi}}$ by eliminating the lines and
columns $2k-1$ and $2k$. Then, $M'$ is tightly closed, and
\begin{itemize}
 \item $\oct{M'} \Leftrightarrow \exists x_k.\phi(\x)$
 \item $\exists x_k ~.~ \phi(\x) \iff \big(\exists y_{2k-1},y_{2k} ~.~ \overline{\phi}(\y)\big)[x_i/y_{2i-1},-x_i/y_{2i}]_{i\in\{1,\dots,N\}\setminus\{k\}}$
\end{itemize}
\end{prop}
\proof{
For the first point, see Theorem 2 in \cite{tacas09}. For the second point, let us define the substitution $\sigma \stackrel{def}{=} [x_i/y_{2i-1},-x_i/y_{2i}]_{i\in\{1,\dots,N\}}$. We first prove that $\dbc{P^*}[\sigma] \iff \dbc{P^t}[\sigma]$ for every octagonal-consistent coherent DBM $P\in \zed_\infty^{2N\times 2N}$. By Theorem~\ref{bhz}, it is sufficient to prove that for every $1\leq i,j\leq 2N$ such that $P^*_{i\bar{\imath}} < \infty$ and $P^*_{\bar{\jmath}j} < \infty$, the following holds: 
\begin{equation}\label{eq:tcl:1}
\dbc{M^*}[\sigma] \Rightarrow \big(y_i-y_j \leq \lfloor\frac{P^*_{i\bar{\imath}}}{2}\rfloor + \lfloor\frac{P^*_{\bar{\jmath}j}}{2}\rfloor\big)[\sigma]
\end{equation}
Clearly, there exists $1\leq k,\ell\leq N$ such that either of the following holds:
\[\begin{array}{llllll}
 (1) & i=2k-1, & j=2\ell-1 & 
 (3) & i=2k, & j=2\ell-1 \\
 (2) & i=2k-1, & j=2\ell &
 (4) & i=2k, & j=2\ell
\end{array}\]
We give the proof for the first case (the other being symmetric). Then, \eqref{eq:tcl:1} is equivalent to $\dbc{M^*}[\sigma] \Rightarrow x_k-k_\ell \leq \lfloor\frac{P^*_{i\bar{\imath}}}{2}\rfloor + \lfloor\frac{P^*_{\bar{\jmath}j}}{2}\rfloor$. Clearly, $\dbc{P^*} \Rightarrow (y_{i}-y_{\bar{\imath}}\leq P^*_{i\bar{\imath}}) ~\wedge~ (y_{\bar{\jmath}}-y_{j}\leq P^*_{\bar{\jmath}j})$ and consequently,
\[\begin{array}{lcl}
\dbc{P^*}[\sigma] & \Rightarrow &
 (x_k+x_k \leq P^*_{i\bar{\imath}}) ~\wedge~ (-x_\ell-x_\ell \leq P^*_{\bar{\jmath}j})
\\ & \Rightarrow & x_k \leq \lfloor \frac{P^*_{i\bar{\imath}}}{2} \rfloor ~\wedge~ -x_\ell \leq \lfloor \frac{P^*_{\bar{\jmath}j}}{2} \rfloor
\\ & \Rightarrow & x_k-x_\ell \leq \lfloor\frac{P^*_{i\bar{\imath}}}{2}\rfloor + \lfloor\frac{P^*_{\bar{\jmath}j}}{2}\rfloor
\end{array}\]
Hence, \eqref{eq:tcl:1} holds.

Let $M^*_p$ ($M^t_p$, respectively) be the restriction of $M^*_{\overline{\phi}}$ (of $M^t_{\overline{\phi}}$, respectively) to $\y\setminus\{y_{2k-1},y_{2k}\}$ and let $\sigma_p \stackrel{def}{=} [x_i/y_{2i-1},-x_i/y_{2i}]_{i\in\{1,\dots,N\}\setminus\{k\}}$. By Theorem \ref{bhz}, it is easy to see that $M^t_p$ is the tight closure of $M^*_p$ and thus $\dbc{M^t_p}[\sigma_p]=\dbc{M^*_p}[\sigma_p]$, by the previous observation. By the first point of this proposition, $\exists x_k ~.~ \phi(\x) \iff \oct{M^t_p}$. By Proposition \ref{dbm-eq} (third point), $\dbc{M^*_p} \iff \exists y_{2k-1},y_{2k} ~.~ \overline{\phi}(\y)$. Next, we observe that $\oct{P} \iff \dbc{P}[\sigma]$ for every coherent DBM $P\in\zed_\infty^{2N\times 2N}$ and hence $\oct{M^t_p} \iff \dbc{M^t_p}[\sigma_p]$. Finally, we combine the equivalences:
\[
\exists x_k ~.~ \phi(\x) 
\iff \oct{M^t_p}
\iff \dbc{M^t_p}[\sigma_p]
\iff \dbc{M^*_p}[\sigma_p]
\iff \big(\exists y_{2k-1},y_{2k} ~.~
\overline{\phi}(\y)\big)[\sigma_p]
\eqno{\qEd}
\]
}

\noindent A relation $R \subseteq \zed^{\vec{x}} \times \zed^{\vec{x}}$ over a
set of variables is an {\em octagonal relation} if it can be defined
by an octagonal constraint. The problem of computing the closed forms
of octagonal relations has been studied first in \cite{tacas09}, where
it was shown that the transitive closures of octagonal relations are
Presburger definable. In \cite{cav10} we show that the sequence of
tightly closed DBM encodings of the powers of an octagonal relations
is periodic, in the sense of Definition \ref{up-int-matrix}. Moreover,
the prefix, period and rates of this sequence of matrices are
effectively computable. This result is crucial in showing that the
weakest non-termination preconditions $\wrs(R)$ are Presburger
definable and effectively computable, and moreover, that the
well-foundedness problem for octagonal relations is decidable.

\begin{figure}
\begin{tabular}{c@{\extracolsep{0mm}}c}
\mbox{\begin{minipage}{6.0cm}
  \begin{tikzpicture}
    \tiny
    \tikzset{
      sState/.style={draw=black,circle,inner sep=1.5pt,semithick}
    }

    \node[sState] (xpm) at (0mm,0mm) {$y_2'$};
    \node[sState] (xm) at (15mm,0mm) {$y_2$};
    \node[sState] (ym) at (30mm,0mm) {$y_4$};
    \node[sState] (ypm) at (45mm,0mm) {$y_4'$};

    \node[sState] (xpp) at (0mm,15mm) {$y_1'$};
    \node[sState] (xp) at (15mm,15mm) {$y_1$};
    \node[sState] (yp) at (30mm,15mm) {$y_3$};
    \node[sState] (ypp) at (45mm,15mm) {$y_3'$};

    \path[->,bend angle=5] 
       (xpp) edge [bend left] node [below] {$-2$} (xp)
       (xm) edge [bend left] node [above] {$-2$} (xpm)
       (ypp) edge [bend right] node [below] {$-3$} (yp)
       (ym) edge [bend right] node [above] {$-3$} (ypm)
       (xp) edge [bend right] node {} (ym)
       (yp) edge [bend left] node {} (xm)
       (ypp) edge [bend angle=25,bend right] node [below] {$1$} (xpp)
       (xpm) edge [bend angle=25,bend right] node [above] {$1$} (ypm)
       (ypp) edge [bend left,dashed] node [left] {\shortstack{\st{$1$} \\ $0$ }} (ypm);

    \node at (20mm,11mm) {$5$};
    \node at (25mm,11mm) {$5$};

  \end{tikzpicture}
\end{minipage}}
&
\mbox{\begin{minipage}{6.5cm}
\scalebox{0.85}{\bordermatrix{
   ~ & y_1 & y_2 & y_3 & y_4 & y_1' & y_2' & y_3' & y_4' \cr
   y_1 & 0 & \infty & \infty & 5 & \infty & \infty & \infty & 2 \cr
   y_2 & \infty & 0 & \infty & \infty & \infty & -2 & \infty & -1 \cr
   y_3 & \infty & 5 & 0 & \infty & \infty & 3 & \infty & 4 \cr
   y_4 & \infty & \infty & \infty & 0 & \infty & \infty & \infty & -3 \cr
   y_1' & -2 & \infty & \infty & 3 & 0 & \infty & \infty & 0 \cr
   y_2' & \infty & \infty & \infty & \infty & \infty & 0 & \infty & 1 \cr
   y_3' & -1 & 2 & -3 & 4 & 1 & 0 & 0 & 0 \cr
   y_4' & \infty & \infty & \infty & \infty & \infty & \infty & \infty & 0 
}}
\end{minipage}}
\\[2mm]
$\mathcal{G}_{\overline{R}}$ &
$M_{\overline{R}}^t$
\end{tabular}
\caption{Graph and matrix representation of the difference bounds
  representation $\overline{R}(\y,\y')$ of an octagonal relation
  $R(\x,\x') \equiv x_1+x_2 \leq 5 ~\wedge~ x_1'-x_1 \leq -2 ~\wedge~
  x_2'-x_2 \leq -3 ~\wedge~ x_2'-x_1' \leq 1$.}
\label{oct:ex}
\end{figure}
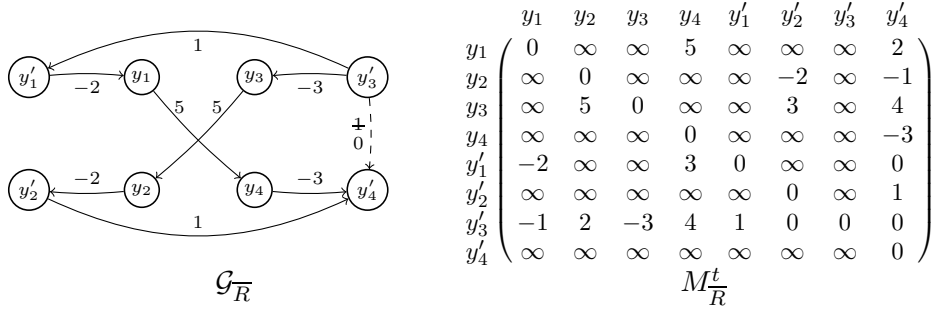

\begin{exa}
Consider the octagonal relation $R(x_1,x_2,x_1',x_2') \equiv x_1+x_2
\leq 5 \wedge x_1'-x_1 \leq -2 \wedge x_2'-x_2 \leq -3 \wedge
x_2'-x_1' \leq 1$.  Its difference bounds representation is
$\overline{R}(\y,\y') \iff y_1-y_4 \leq 5 \wedge y_3-y_2 \leq 5 \wedge
y_1'-y_1 \leq -2 \wedge y_2-y_2' \leq -2 \wedge y_3'-y_3 \leq -3
\wedge y_4-y_4' \leq -3 \wedge y_3'-y_1' \leq 1 \wedge y_2'-y_4' \leq
1$, where $\y=\{y_1,\dots,y_4\}$. Figure \ref{oct:ex}(a) shows the graph
representation $\mathcal{G}_R$. Note that the implicit constraint
$y_3'-y_4' \leq 1$ (represented by a~dashed edge in Figure
\ref{oct:ex}(a) is not tight. The tightening step replaces the bound
$1$ (crossed in Figure \ref{oct:ex}(a)) with $0$. Figure \ref{oct:ex}(b)
shows the tightly closed DBM representation of $R$, denoted $M^t_R$.
\end{exa}

\begin{prop}\label{oct:to:dbm:consistency}
Let $R(\x,\x')$, where $\x=\{x_1,\dots,x_N\}$, be an octagonal constraint and $\overline{R}(\y,\y')$, where $\y=\{y_1,\dots,y_{2N}\}$, be its difference bounds representation. Then, for each $n\geq 1$, consistency of $R^n(\x,\x')$ implies consistency of $\overline{R}^{\,n}(\y,\y')$. Consequently, $*$-consistency of $R(\x,\x')$ implies $*$-consistency of $\overline{R}(\y,\y')$.
\end{prop}
\proof{
It follows from the definition of consistency of octagonal and difference bounds constraints that:
\[\begin{array}{lllll}
  R^n(\x,\x') \textrm{ is consistent} & \iff & 
    R(\vec{x}_0,\vec{x}_1) \wedge\dots\wedge R(\vec{x}_{n-1},\vec{x}_n) \textrm{ is consistent} \\
    & \iff & M_{\overline{R(\vec{x}_0,\vec{x}_1) \wedge\dots\wedge R(\vec{x}_{n-1},\vec{x}_n)}} \textrm{ is octagonal-consistent} \\
    & \Rightarrow & M_{\overline{R(\vec{x}_0,\vec{x}_1) \wedge\dots\wedge R(\vec{x}_{n-1},\vec{x}_n)}} \textrm{ is consistent} \\
    & \iff & \overline{R(\vec{x}_0,\vec{x}_1) \wedge\dots\wedge R(\vec{x}_{n-1},\vec{x}_n)} \textrm{ is consistent} \\
    & \iff & \overline{R(\vec{x}_0,\vec{x}_1)} \wedge\dots\wedge \overline{R(\vec{x}_{n-1},\vec{x}_n)} \textrm{ is consistent} \\
    & \iff & \overline{R(\x,\x')}^{\,n} \textrm{ is consistent} \\
\end{array}\]
Thus, for each $n\geq 1$, consistency of $R^n(\x,\x')$ implies consistency of $\overline{R}^{\,n}(\y,\y')$. Thus, if $R(\x,\x')$ is $*$-consistent, then $\overline{R}(\y,\y')$ is $*$-consistent too.
\qed}

The next proposition shows that the composition of two octagonal
relations is octagonal, and moreover, can be computed in PTIME using
the tight closure method of Theorem \ref{bhz}. If $R_1, R_2 \subseteq
\zed^{\x} \times \zed^{\x}$ are two octagonal relations, defined by
two octagonal constraints $R_1(\x,\x')$ and $R_2(\x,\x')$, then let
$M_1, M_2 \in \zed^{4N \times 4N}$ be the DBM encodings of
$\overline{R_1}(\y,\y')$ and $\overline{R_2}(\y,\y')$,
respectively. Then $\mathcal{M}_{12} \in \zed^{6N \times 6N}$ is the
matrix defined by Equation (\ref{m-one-two}), and let $M_1 \odot_t M_2
\in \zed^{4N \times 4N}$ be the matrix obtained by erasing lines and
columns $2N + 1, \ldots, 4N$ from $\mathcal{M}^t_{12}$, if
$\mathcal{M}_{12}$ is octagonal-consistent, and $\botdb{4N}$,
otherwise.

\begin{prop}\label{oct:rel:comp}
  Let $R_1(\vec{x},\vec{x'})$ and $R_2(\vec{x},\vec{x'})$ be two
  octagonal constraints defining two relations $R_1,R_2 \subseteq
  \zed^{\vec{x}} \times \zed^{\vec{x}}$, respectively. Then the
  octagonal constraint $\oct{M_{\overline{R_1}(\y,\y')} \odot_t
  M_{\overline{R_2}(\y,\y')}}$ defines the composition $R_1 \circ
  R_2$. Moreover, $M_{\overline{R_1}} \odot_t M_{\overline{R_2}}$ can
  be computed in time $\mathcal{O}(N^3 \cdot (N + \log_2(\max(\maxcoef{R_1},\maxcoef{R_2}))))$.
\end{prop}
\proof{Among the lines of the proof of Proposition
  \ref{dbrel-comp}. An easy check shows that, if $M_1$ and $M_2$ are
  coherent, then $\mathcal{M}_{12}$ is coherent as well. The
  consistency of $\mathcal{M}_{12}$ can be checked in time
  $\mathcal{O}(N^3 \cdot (N + \log_2 (\max(\maxcoef{R_1},\maxcoef{R_2})) ))$ by Algorithm \ref{alg:fw},
  and its closure $\mathcal{M}^*_{12}$ can be computed during this
  check. The octagonal consistency of $\mathcal{M}_{12}$ is checked
  applying Theorem \ref{bhz}, and the same can be done to compute the
  tight closure $\mathcal{M}^t_{12}$. Clearly, these steps do not add
  to the previous complexity upper bound. Finally, the existential
  quantifier from $\exists \vec{x''} ~.~ R_1(\x,\vec{x''}) \wedge
  R_2(\vec{x''}, \x')$ can be eliminated using Proposition
  \ref{odbc:qelim}. \qed}

In general, for a DBM $M \in \zed_\infty^{4N \times 4N}$ encoding an
octagonal constraint $R(\x,\x')$, where $\x=\{1,\ldots,N\}$, we define
$M^{\odot_t^1} = M$ and $M^{\odot_t^n} = M^{\odot_t^{n-1}} \odot_t M$, 
for $n > 1$. A simple inductive argument based on Proposition
\ref{oct:rel:comp} shows that the $n$-th power $R^n$ of the relation
$R \subseteq \zed^\x \times \zed^\x$ is defined by the octagonal
constraint $\oct{M^{\odot_t^n}}$, for all $n > 0$. In the following,
we denote the formula $\oct{M^{\odot_t^n}}$ by $R^n(\x,\x')$. As
usual, let $\overline{R}(\y,\y')$ be the difference bounds constraint
encoding $R(\x,\x')$, and $\overline{R}^n(\y,\y')$ be the difference
bounds constraint defining the $n$-th power of the relation defined by
$\overline{R}(\y,\y')$. The following lemma establishes an essential
connection between the DBMs
$M^t_{\overline{R^n}},M^t_{\overline{R}^n},M^*_{\overline{R}^n} \in
\zed^{4N \times 4N}$, leading to a method for the computation of the
transitive closures for octagonal relations \cite{tacas09}.

\begin{lem}\label{theorem-tacas09}
  Let $\x=\{x_1,\ldots,x_N\}$ be a set of variables and $R \subseteq
  \zed^{\x} \times \zed^{\x}$ be a $*$-consistent octagonal
  relation. Then the following hold, for all integers $n > 0$:
  \begin{enumerate}
  \item $M^t_{\overline{R^n}} = M^t_{\overline{R}^n}$, and
  \item $(M^t_{\overline{R}^n})_{ij} = \min
    \left\{(M^*_{\overline{R}^n})_{ij},
    \left\lfloor\frac{(M^*_{\overline{R}^n})_{i\bar\imath}}{2}\right\rfloor
    + \left\lfloor\frac{(M^*_{\overline{R}^n})_{\bar\jmath
        j}}{2}\right\rfloor\right\}$, for all $1 \leq i,j \leq 4N$.
  \end{enumerate}
\end{lem}
\proof{We prove the first point by induction on $n > 0$. The base case 
  $n=1$ is immediate. For the induction step $n > 1$, we have $R^{n+1}(\x,\x') =
  \oct{M_{\overline{R}}^{\odot_t^{n+1}}}$, hence:
\[\begin{array}{rcll}
M^t_{\overline{R^{n+1}}} & = & M_{\overline{R}}^{\odot_t^{n+1}} \\
& = & M_{\overline{R}}^{\odot_t^{n}} \odot_t M_{\overline{R}} \\
& = & M^t_{\overline{R^n}} \odot_t M_{\overline{R}} \\
& = & M^t_{\overline{R}^n} \odot_t M_{\overline{R}} & \mbox{by the induction hypothesis} \\
& = & M^t_{\overline{R}^{n+1}} & \mbox{by Proposition \ref{oct:rel:comp}}
\end{array}\]
Since $R$ is $*$-consistent, then $M^t_{\overline{R}^n}$ is an 
octagonal-consistent DBM and we can directly apply Theorem \ref{bhz}
to prove the second point.
\qed}

The following result shows that the sequence $\{\pre^n_R(\zed^\x)\}_{n \geq
  1}$, of a $*$-consistent octagonal relation $R$ is defined by a
periodic sequence of matrices. 

\begin{lem}\label{periodic-pre-oct}
  Let $R \subseteq \zed^\x \times \zed^\x$ be a $*$-consistent
  octagonal relation. Then, for all $n \geq 1$, the octagonal
  constraint $\oct{\!\topleft{M^t_{\overline{R^n}}}}$ defines the set
  $\pre^n_R(\zed^\x)$. Moreover, the sequence
  $\{\!\topleft{M^t_{\overline{R^n}}}\}_{n\geq1}$ is periodic, and its
  prefix, period and rates are all effectively computable. 
\end{lem}
\proof{By Lemma \ref{theorem-tacas09}, for all $1 \leq i,j \leq 2N$ we
  have:
  $$(M^t_{\overline{R^n}})_{ij} = \min
  \left\{(M^*_{\overline{R}^n})_{ij},
  \left\lfloor\frac{(M^*_{\overline{R}^n})_{i\bar\imath}}{2}\right\rfloor
  + \left\lfloor\frac{(M^*_{\overline{R}^n})_{\bar\jmath
      j}}{2}\right\rfloor\right\}$$ By Corollary
  \ref{periodic-pre-dbm}, the sequence of matrices
  $\{\!\topleft{M^*_{\overline{R}^n}}\}_{n\geq1}$ is periodic, hence
  the sequence of integers
  $\{(\!\topleft{M^*_{\overline{R}^n}})_{ij}\}_{n\geq1}$ is periodic,
  for all $1 \leq i,j \leq 2N$. By Lemma \ref{up-sum-min-half},
  the sequence of integers $(M^t_{\overline{R^n}})_{ij}$ is also
  periodic, hence the sequence of matrices
  $\{\!\topleft{M^t_{\overline{R^n}}}\}_{n\geq1}$ is periodic,
  by Proposition \ref{up-int-matrix}.
  The effective computability of the prefix, period, and rates of
  the sequence follows from the constructive arguments of Lemma
  \ref{up-sum-min-half} and Proposition \ref{up-int-matrix}. \qed}

\subsection{Computing Weakest non-termination preconditions in Polynomial Time}
\label{sec:term:polynomial:alg}

In the rest of this section, let $R(\x,\x')$, where
$\x=\{x_1,\dots,x_N\}$ for some $N\geq 1$, be an octagonal relation
and $\overline{R}(\y,\y')$, where $\y=\{y_1,\dots,y_{2N}\}$, be its
difference bounds representation. Recall that $\maxcoef{R}
\stackrel{def}{=} \max\{ |c| ~|~ (\pm x_i \pm x_j \leq c) \in Atom(R)
\}$.

The main result of this section is an algorithm (Algorithm
\ref{alg:recurset}) that computes the weakest recurrent set of an
octagonal relation $R$ in at most $\mathcal{O}(N^4\cdot(N+\log_2
(\maxcoef{R})))$ time. The main insight of the algorithm is that the
Kleene sequence $\{ \pre_R^n(\zed^\x) \}_{n\geq 1}$ either
(1) never stabilizes, in which case \[\pre_R^1(\zed^\x)\supsetneq
\pre_R^2(\zed^\x) \supsetneq \pre_R^3(\zed^\x) \supsetneq \dots\] and
$\wrs(R)=\emptyset$, or (2) stabilizes after at most $5^{2N}$ steps,
in which
case \[\wrs(R)=\pre_R^{5^{2N}}(\zed^\x)=\pre_R^{5^{2N}+1}(\zed^\x)=\pre_R^{5^{2N}+2}(\zed^\x)=\dots\]
Then, the stability of the sequence can be checked by checking
equality between its $5^{2N}$-th element with the $(5^{2N}+1)$-th
element. These elements can be computed by fast exponentiation by
applying at most $\mathcal{O}(\lceil \log_2 5^{2N}+1
\rceil)=\mathcal{O}(N)$ relational compositions. We then show that the
absolute values of the coefficients of the octagonal constraint
defining the set $\pre_R^n(\zed^\x)$ is of the order
$\mathcal{O}(\maxcoef{R}\cdot N\cdot n)$. Consequently, each of the
octagonal compositions performed during fast exponentiation takes at
most $\mathcal{O}(N^3\cdot(\log_2 (\maxcoef{R} \cdot N\cdot 5^{2N})))
= \mathcal{O}(N^4\cdot(N+\log_2 (\maxcoef{R})))$ time, by Proposition
\ref{oct:rel:comp}. As a~direct consequence of the correctness of this
algorithm, one obtains a~decision procedure for the termination
problem with the same worst-case complexity, simply by testing the
computed $\wrs(R)$, itself an octagonal constraint, for
consistency. 


The correctness argument of Algorithm \ref{alg:recurset} for
$*$-consistent octagonal relations depends on Lemmas \ref{lem:dp1},
\ref{lem:core:results:db}, and \ref{lem:core:results}. First, Lemma
\ref{lem:dp1} proves that the weakest recurrent set of
an~$*$-consistent octagonal relation $R(\x,\x')$ is the limit of the
Kleene sequence $\{ \pre_R^n(\zed^\x) \}_{n\geq 1}$ and moreover, that
the limit is either empty or stabilizes after a~finite number of steps.
Next, Lemma \ref{lem:core:results:db} gives two equivalent conditions
for checking well-foundedness of an arbitrary $*$-consistent
difference bounds relation $R(\x,\x')$. Its main insight is that the
instability of the sequence $\{ \pre_R^n(\zed^\x) \}_{n\geq 1}$ (and
thus well-foundedness of $R$) is equivalent to existence of
a~negative-weight cycle in zigzag automata. Moreover, it proves that
the instability manifests already after $5^{N}$ steps ($5^N$ is an
upper bound on the size of elementary cycles in zigzag
automata). Then, Lemma \ref{lem:core:results} proves that an octagonal
relation $R(\x,\x')$, where $\x=\{x_1,\dots,x_N\}$, is well founded if
and only if its difference bounds representation
$\overline{R}(\y,\y')$, where $\y=\{y_1,\dots,y_{2N}\}$, is well
founded. Hence the stability stability bound of $5^{2N}$ applies for
octagonal relations, as a consequence of Lemma
\ref{lem:core:results:db}.

The following proposition gives an alternative characterization of
periodic sequences of matrices.
\begin{prop}\label{periodic-characterization}
  A sequence of matrices $\{M_k \in \zed_\infty^{N \times
    N}\}_{k=1}^{\infty}$ is periodic if and only if there exist
  integers $b\geq 1$, $c\geq 1$, and matrices
  $\Lambda_0,\dots,\Lambda_{c-1} \in \zed_\infty^{m \times m}$ such
  that
  \[ M_{nc+b+i} = n \cdot \Lambda_i + M_{b+i} \]
  for all $n \geq 0$ and for all $0 \leq i < c$.
\end{prop}
\proof{
  By induction on $n\geq 0$, we prove that $M_{nc+b+i} = n \cdot \Lambda_i + M_{b+i}$, for all $n \geq 0$ and for all $0 \leq i < c$. The base case trivially holds. For the induction step, observe that
  \[ M_{b+i+(n+1)c} = \Lambda_i + M_{b+i+nc} = 
     \Lambda_i + n \cdot \Lambda_i + M_{b+i} = 
     (n+1) \cdot \Lambda_i + M_{b+i} \textrm{.} 
  \]
  The first equality is by Definition \ref{up-matrix}, the second is by the induction hypothesis.
\qed}

Given a~$*$-consistent octagonal relation $R(\x,\x')$ and integers $b\geq 1,c\geq 1$, we denote by $\widehat{\pre_{R,b,c}}(k,\vec{x})$ the closed form of the sequence $\{ \pre_R^{b+nc}(\zed^\x)\}_{n\geq 0}$. 
Given a~$*$-consistent octagonal relation $R$ and integers $b,c$ such that $b$ is the prefix and $c$ is the period of the sequence $\{{\topleft{M^t_{\overline{R^n}}}}\}_{n\geq 1}$, the following lemma proves that the closed form $\widehat{\pre_{R,b,c}}(k,\vec{x})$ can be computed and moreover, one can perform a simple syntactical check on $\widehat{\pre_{R,b,c}}(k,\vec{x})$ to compute the weakest recurrent set, which is either $\emptyset$ or $\pre_R^b(\zed^\x)$. For a~set $\vec{v}$ of variables, let $OctTerm(\vec{v}) = \{\pm v_1 \pm v_2 ~|~ v_1, v_2 \in \vec{v}\}$ denote the set of octagonal terms over $\vec{v}$. 
\begin{lem} \label{lem:dp1}
Let $R(\x,\x')$ be an octagonal constraint defining a~$*$-consistent relation $R\subseteq\zed^\x\times\zed^\x$, let $b$ be the prefix and $c$ the period of $\{{\topleft{M^t_{\overline{R^n}}}}\}_{n\geq 1}$. Then, there exists a~set of octagonal terms $U\subseteq OctTerm(\x)$ such that 
\begin{equation} \label{cf:octagons}
 \widehat{\pre_{R,b,c}}(k,\vec{x}) \iff \bigwedge_{u\in U} u\leq a_u + d_u\cdot k
\end{equation}
for some $a_u\in\zed$, $d_u\leq 0$. Moreover, the set $U$ and the coefficients $a_u,d_u$, $u\in U$, are effectively computable. Furthermore, 
\[
  \wrs(R) = \bigcap_{n\geq 1} \pre_R^{n}(\zed^{\vec{x}}) = \left\{\begin{array}{ll}
    \emptyset & \textrm{ if $d_u<0$ for some $u\in U$} \\
    \pre_R^{b}(\zed^{\vec{x}}) & \textrm{ otherwise }
  \end{array}\right.
\]
\end{lem}
\proof{
The sequence $\{{\topleft{M^t_{\overline{R^n}}}}\}_{n\geq 1}$ is periodic, by Lemma \ref{periodic-pre-oct}. Let $\Lambda_0,\dots,\Lambda_{c-1}$ be its rates. 
For each $u\in OctTerm(\x)$, we define indices $i_u,j_u$ as:
\[\begin{array}{lll}
  i_u = 2k-1, & j_u=2\ell-1 & \textrm{ if } u=x_k-x_\ell \textrm{ for some $1\leq k,\ell\leq N$ } \\
  i_u = 2k-1, & j_u=2\ell & \textrm{ if } u=x_k+x_\ell \textrm{ for some $1\leq k,\ell\leq N$ } \\
  i_u = 2k, & j_u=2\ell-1 & \textrm{ if } u=-x_k-x_\ell \textrm{ for some $1\leq k,\ell\leq N$ }
\end{array}\]
Then, the set of octagonal terms which are bounded in $pre_R^{b}(\zed^\x)$ is:
\[ U \stackrel{def}{=} \{ u\in OctTerm(\x) ~|~ (\topleft{M^t_{R^b}})_{i_uj_u} < \infty \} \]
Since $R^n$ is consistent and $\topleft{M^t_{\overline{R^n}}}$ is coherent for all $n\geq 1$, we have:
\begin{equation} \label{eq:oct:cf:bc}
\begin{array}{lcll}
\pre_R^{b+nc}(\zed^\x) 
  & \iff & \oct{\topleft{M^t_{\overline{R^{b+nc}}}}} & \textrm{(by Proposition \ref{odbc:qelim})}
\\ & \iff & \oct{\topleft{M^t_{\overline{R^{b}}}} + n \cdot \Lambda_0} & \textrm{(by Proposition \ref{periodic-characterization})}
\\ & \iff & \bigwedge_{u\in U} u \leq (\topleft{M^t_{\overline{R^{b}}}})_{i_uj_u} + n \cdot (\Lambda_0)_{i_uj_u} & \textrm{(by Equation \eqref{dbm-octagon})}
\end{array}
\end{equation}
for every $n\geq 0$. Clearly, $(\topleft{M^t_{\overline{R^{b}}}})_{i_uj_u} < \infty$ for each $u\in U$, by definition of $U$. We prove that $(\Lambda_0)_{i_uj_u} \leq 0$. By contradiction, if $(\Lambda_0)_{i_uj_u} > 0$, then 
\[ (\topleft{M^t_{\overline{R^{b+c}}}})_{i_uj_u} = (\topleft{M^t_{\overline{R^{b}}}})_{i_uj_u} + (\Lambda_0)_{i_uj_u} > (\topleft{M^t_{\overline{R^b}}})_{i_uj_u} \]
by Proposition \ref{periodic-characterization}. By Proposition \ref{prop:pre:properties}, $\pre^{b+c}_R(\zed^\x) \subseteq \pre^{b}_R(\zed^\x)$. By Proposition \ref{oct-eq}, we infer that $\topleft{M^t_{\overline{R^{b+c}}}} \leq \topleft{M^t_{\overline{R^{b}}}}$. Contradiction with $(\topleft{M^t_{\overline{R^{b+c}}}})_{i_uj_u} > (\topleft{M^t_{\overline{R^{b+c}}}})_{i_uj_u}$. Hence, we can define the coefficients $a_u\in\zed,d_u\leq 0$ for each $u\in U$ as
\[
  a_u \stackrel{def}{=} (\topleft{M^t_{\overline{R^{b}}}})_{i_uj_u}
  \hspace{10mm}
  d_u \stackrel{def}{=} (\Lambda_0)_{i_uj_u}
\]
By Lemma \ref{periodic-pre-oct}, the prefix $b$, the period $c$, and the rate $\Lambda_0$ are effectively computable. Consequently, the set $U$ and coefficients $a_u,d_u$, $u\in U$, defined above are effectively computable too. It follows from \eqref{eq:oct:cf:bc} that the closed form of $\{ \pre_R^{b+nc}(\zed^\x) \}_{n\geq 0}$ can now be defined as
\[
\widehat{\pre_{R,b,c}}(k,\vec{x}) \stackrel{def}{=}
\bigwedge_{u\in U} u \leq a_u + d_u \cdot k
\]

By Proposition \ref{prop:pre:properties}, $pre_R^{n_1}(\zed^\vec{x}) \supseteq \pre_R^{n_2}(\zed^\vec{x})$ for all $n_1 \leq n_2$.  Consequently, we have that $\bigcap_{n \geq 1} \pre_R^{n}(\zed^{\vec{x}}) = \bigcap_{n\geq 0} \pre_R^{b+cn}(\zed^{\vec{x}})$. The latter set can now be defined as
$\forall k \geq 0 ~.~ \widehat{\pre_{R,b,c}}(k,\vec{x})$
which is equivalent to 
\[ \bigwedge_{u \in U} u \leq \inf~\{a_u+d_un ~|~ n\geq 0 \} \]
We have  
\[
  \inf~ \{a_u+d_un ~|~ n\geq 0 \} = 
  \left\{\begin{array}{cl}
     -\infty & \mbox{if $d_u < 0$,} \\ 
     a_u & \mbox{otherwise.} 
  \end{array}\right.
\]
Hence $\bigcap_{n \geq 1} \pre_R^{n}(\zed^{\vec{x}})$ is the empty set, if $d_u < 0$ for some $u \in U$. In this case, condition $3$ of Lemma \ref{wrs-kleene} holds. Otherwise, we obtain $\bigcap_{n\geq 1} \pre_R^{n}(\zed^{\vec{x}}) \equiv \bigwedge_{u \in U} u \leq a_u$. However, this is exactly the set $\pre_R^{b}(\zed^{\vec{x}})$, since $\bigwedge_{u \in U} (u \leq a_u) \iff \widehat{\pre_{R,b,c}}(k,\vec{x})[0/k]$. In this case, condition $2$ of Lemma \ref{wrs-kleene} holds. Thus, we can apply Lemma \ref{wrs-kleene} in both cases and conclude that $\wrs(R) = \bigcap_{n \geq 1} \pre_R^{n}(\zed^{\vec{x}})$. To summarize, $\wrs(R) = \emptyset$ if $d_u < 0$ for some $u \in U$. Otherwise, $\wrs(R) = \pre_R^{b}(\zed^{\vec{x}})$.
\qed}

The following proposition proves that the Kleene sequence $\{ \pre_R^n(\zed^\x)\}_{n\geq 1}$ is strictly descending for arbitrary relation that is both $*$-consistent and well founded.
\begin{prop}\label{prop:starcons:wf}
Let $R\subseteq \zed^\x\times\zed^\x$ be a~$*$-consistent and well-founded relation. Then,
$\pre_R^{n_1}(\zed^\x) \supsetneq \pre_R^{n_2}(\zed^\x)$ for all $1\leq n_1<n_2$. Consequently, the sequence $\{ \pre_R^n(\zed^\x)\}_{n\geq 1}$ is strictly descending.
\end{prop}
\proof{
By Proposition \ref{prop:pre:properties}, $\pre_R^{n_1}(\zed^\x) \supseteq \pre_R^{n_2}(\zed^\x)$ for all $1\leq n_1<n_2$. For a~proof by contraposition, suppose that $\pre_R^{n_1}(\zedToX) = \pre_R^{n_2}(\zedToX)$ some $n_2>n_1\geq 1$. Then $\wrs(R) = \pre_R^{n_1}(\zed^\x)$, by Lemma \ref{wrs-kleene}. Since $R$ is $*$-consistent, then clearly $\wrs(R)= \pre_R^{n_1}(\zedToX)\neq\emptyset$ and $R$ is not well founded.
\qed}

The following two lemmas give several equivalent conditions for checking that a~difference bounds (Lemma \ref{lem:core:results:db}) or an octagonal relation (Lemma \ref{lem:core:results}) is well founded. These conditions will later be used to design an efficient polynomial time algorithm that computes the weakest recurrent set of an octagonal relation. These conditions also provide the basis for the proof of existence of a~linear ranking functions for well-founded octagonal relations, which we give in the next section.

\begin{lem}\label{lem:core:results:db}
Let $R(\x,\x')$, where $\x=\{x_1,\dots,x_N\}$, be a~difference bounds constraint defining a $*$-consistent relation $R\subseteq\zed^\x\times\zed^\x$ and let $T_R=\langle Q,\delta,\omega \rangle$ be the transition table of zigzag automata. Then, the following statements are equivalent:
  \begin{enumerate}
    \item $R$ is well founded,
    \item $\pre_R^{n_2}(\zedToX) \subsetneq \pre_R^{n_1}(\zedToX)$ for some $n_2>n_1\geq 5^N$,
    \item there exists a~zigzag automaton $\mathcal{A}_{ij}=\langle T_R,I_{ij},F \rangle$ for some $1 \leq i, j\leq N, i\neq j$ with an accepting run $\mu.\lambda.\mu'$ where $\lambda$ is a cycle such that $|\lambda|>0$ and $\omega(\lambda)<0$.
  \end{enumerate}
\end{lem}
\proof{
  \medskip\noindent\textbf{($1\Rightarrow 2$)} Follows immediately from Proposition \ref{prop:starcons:wf}.

  \medskip\noindent\textbf{($2\Rightarrow 3$)}
Let $n_2>n_1\geq 5^N$ be integers such that $\pre_R^{n_2}(\zedToX) \subsetneq \pre_R^{n_1}(\zedToX)$. 
Then, $\topleft{M^*_{R^{n_1}}} > \topleft{M^*_{R^{n_2}}}$ by Proposition \ref{dbm-eq}. Since $R$ is $*$-consistent, $(\topleft{M^*_{R^{n_1}}})_{ii} = (\topleft{M^*_{R^{n_2}}})_{ii} = 0$ for each $1\leq i\leq N$ and hence $(\topleft{M^*_{R^{n_1}}})_{ij} > (\topleft{M^*_{R^{n_2}}})_{ij}$ for some $1\leq i,j\leq N, i\neq j$. By Lemma \ref{zigzag-automata}, $\mathcal{A}_{ij}$ has an accepting run $\pi$ of length $|\pi|=n_2$ and weight $\omega(\pi)=(\topleft{M^*_{R^{n_2}}})_{ij}$. 

Let $\pi_0 \stackrel{def}{=} \pi$. We next define, iteratively for $i=1,2,\dots$, an accepting run $\pi_i$ by erasing an arbitrary cycle $\lambda_i$ from $\pi_{i-1}$. Note that if $|\pi_{i-1}|\geq 5^N$, then $\pi_{i-1}$ must contain at least one cycle $\lambda_i$, by pigeonhole principle (since $5^N$ is the cardinality of the set of control states in $\mathcal{A}_{ij}$). Clearly $|\pi_p|<5^N$ for some $p\geq 1$. Let $n\stackrel{def}{=}|\pi_p|$.
%
%
We next prove that 
\[ \big(\sum_{i=1}^p \omega(\lambda_i)\big) < 0 \] 
For a proof by contradiction, suppose that $(\sum_{i=1}^p
\omega(\lambda_i)) \geq 0$. Then $\omega(\pi_p)\leq\omega(\pi)$, since
$\omega(\pi_p)=\omega(\pi)-(\sum_{i=1}^p \omega(\lambda_i))$. Observe
that (the first inequality is by Lemma \ref{zigzag-automata}):
  \[ (\topleft{M^*_{R^n}})_{ij} \leq \omega(\pi_p) \leq \omega(\pi) = (\topleft{M^*_{R^{n_2}}})_{ij} \]
Since $n<5^{N}\leq n_1<n_2$, then $\pre_R^{n}(\zed^\x) \supseteq
\pre_R^{n_1}(\zed^\x) \supseteq \pre_R^{n_2}(\zed^\x)$, by Proposition
\ref {prop:pre:properties}. Consequently, by Proposition \ref{dbm-eq}:
  \[ (\topleft{M^*_{R^{n}}})_{ij} \geq (\topleft{M^*_{R^{n_1}}})_{ij} \geq (\topleft{M^*_{R^{n_2}}})_{ij} \]
Combining the above inequalities, we obtain that
$(\topleft{M^*_{R^{n}}})_{ij} = (\topleft{M^*_{R^{n_1}}})_{ij} =
(\topleft{M^*_{R^{n_2}}})_{ij}$. Contradiction with
$(\topleft{M^*_{R^{n_1}}})_{ij} > (\topleft{M^*_{R^{n_2}}})_{ij}$.

Thus, $(\sum_{i=1}^p \omega(\lambda_i)) < 0$ and consequently, there
exists $1\leq k\leq p$ such that $\omega(\lambda_k)<0$. By definition
of $\pi_k$, there exists $\mu,\mu'$ such that
$\pi_k=\mu.\lambda_k.\mu'$. Since $\omega(\lambda_k)<0$, the run
$\mu.\lambda_k.\mu'$ satisfied the requirements of the lemma.

  \medskip\noindent\textbf{($3\Rightarrow 1$)} Let us denote
  $d=|\mu.\mu'|$ and $e=|\lambda|$. Since $\omega(\lambda)<0$, the
  infinite sequence $\{\omega(\mu.\lambda^n.\mu')\}_{n\geq 0}$ is
  strictly descending and thus
  $\inf\{\omega(\mu.\lambda^n.\mu')\}_{n\geq 0}=-\infty$. By Lemma
  \ref{zigzag-automata}, $(\topleft{M^*_{R^{d+ne}}})_{ij} \leq
  \omega(\mu.\lambda^n.\mu')$ for all $n\geq 0$ and hence,
  $\inf\{(\topleft{M^*_{R^{d+ne}}})_{ij}\}_{n\geq 0}=-\infty$. By
  Lemma \ref{lem:dp1}, $\wrs(R) = \bigcap_{n\geq 1}
  \pre_{R}^n(\zed^\x)$. Next, observe that since $R$ is
  $*$-consistent, $\dbc{\topleft{M^*_{R^n}}}$ defines
  $\pre_{R}^n(\zed^\x)$ for each $n\geq 1$. Hence, any formula that
  defines $\wrs(R)$ must imply $x_i-x_j \leq
  \inf\{(\topleft{M^*_{R^{d+ne}}})_{ij}\}_{n\geq 0}=-\infty$. Since
  this formula is inconsistent, it follows that $\wrs(R)=\emptyset$
  and $R$ is well founded.
\qed}

\begin{lem}\label{lem:core:results}
  Let $R(\x,\x')$, where $\x=\{x_1,\dots,x_N\}$, be an octagonal constraint defining a~$*$-consistent relation $R\subseteq\zed^\x\times\zed^\x$, and let $\overline{R}(\y,\y')$, where $\y=\{y_1,\dots,y_{2N}\}$, be the difference bounds encoding of $R(\x,\x')$. Then, the following statements are equivalent.
  \begin{enumerate}
    \item $R$ is well founded
    \item $\overline{R}$ is well founded
    \item $\pre_R^{n_1}(\zedToX) \supsetneq \pre_R^{n_2}(\zedToX)$ for some integers $n_1,n_2$ such that $5^{2N}\leq n_1<n_2$
  \end{enumerate}
\end{lem}
\proof{
Observe that since $R$ is $*$-consistent, $\overline{R}$ is $*$-consistent too, by Proposition \ref{oct:to:dbm:consistency}.

  \medskip\noindent\textbf{($1\Rightarrow 3$)} Follows immediately from Proposition \ref{prop:starcons:wf}.

  \medskip\noindent\textbf{($3\Rightarrow 2$)} 
We first prove that $\pre_R^{n_1}(\zedToX) \supsetneq \pre_R^{n_2}(\zedToX)$ implies that $\pre_{\overline{R}}^{n_1}(\zedToY) \supsetneq \pre_{\overline{R}}^{n_2}(\zedToY)$. For a proof by contraposition, suppose that $\pre_{\overline{R}}^{n_1}(\zedToY) \subseteq \pre_{\overline{R}}^{n_2}(\zedToY)$. By Proposition \ref{prop:pre:properties}, $\pre_{\overline{R}}^{n_1}(\zedToY) \supseteq \pre_{\overline{R}}^{n_2}(\zedToY)$ and consequently, $\pre_{\overline{R}}^{n_1}(\zedToY) = \pre_{\overline{R}}^{n_2}(\zedToY)$. Then, $\topleft{M^*_{\overline{R}^{n_1}}} = \topleft{M^*_{\overline{R}^{n_2}}}$, by Proposition \ref{dbm-eq}.  This implies that $\topleft{M^t_{\overline{R}^{n_1}}} = \topleft{M^t_{\overline{R}^{n_2}}}$, by Lemma \ref{theorem-tacas09}.  Consequently, $\pre_R^{n_1}(\zedToX) = \pre_R^{n_2}(\zedToX)$, by Proposition \ref{oct-eq}.

Since $5^{2N}\leq n_1<n_2$ and $\pre_{\overline{R}}^{n_1}(\zedToY) \supsetneq \pre_{\overline{R}}^{n_2}(\zedToY)$, then $\overline{R}$ is well founded, by Lemma \ref{lem:core:results:db}.

  \medskip\noindent\textbf{($2\Rightarrow 1$)}
The sequence $\{ \pre^n_{\overline{R}}(\zed^\y) \}_{n\geq 1}$ is strictly descending, by Proposition \ref{prop:starcons:wf}. Hence $\pre^1_{\overline{R}}(\zed^\y) \supsetneq \pre^2_{\overline{R}}(\zed^\y) \supsetneq \pre^3_{\overline{R}}(\zed^\y) \supsetneq \dots$ and it follows from Proposition \ref{dbm-eq} that
\[ \topleft{M^*_{\overline{R}^1}} > \topleft{M^*_{\overline{R}^2}} > \topleft{M^*_{\overline{R}^3}} > \dots \]
For each $n\geq 1$, let $1\leq i_n,j_n\leq 2N$ be arbitrary integers such that $(\topleft{M^*_{\overline{R}^n}})_{i_nj_n} > (\topleft{M^*_{\overline{R}^{n+1}}})_{i_nj_n}$. Clearly, there exist integers $1\leq i,j\leq 2N$ such that $i=i_n$ and $j=j_n$ for infinitely many $n\geq 1$. Consequently, for each $n\geq 1$ there exists $m>n$ such that $(\topleft{M^*_{\overline{R}^n}})_{ij} > (\topleft{M^*_{\overline{R}^m}})_{ij}$ and hence
\[ \inf\{ (\topleft{M^*_{\overline{R}^n}})_{ij} \}_{n\geq 1} = -\infty \]
By Lemma \ref{theorem-tacas09}, the following holds for each $n\geq 1$
\[ (\topleft{M^t_{\overline{R^n}}})_{ij} = \min\left\{ (\topleft{M^*_{\overline{R}^n}})_{ij} , \big\lfloor\frac{(\topleft{M^*_{\overline{R}^n}})_{i\bar{\imath}}}{2}\big\rfloor + \big\lfloor\frac{(M^*_{\overline{R}^n})_{\bar{\jmath}j}}{2}\big\rfloor \right\} \]
Thus clearly, since $\inf\{ (\topleft{M^*_{\overline{R}^n}})_{ij} \}_{n\geq 1} = -\infty$, then $\inf\{ (\topleft{M^t_{\overline{R}^n}})_{ij} \}_{n\geq 1} = -\infty$ too.
%
By Equation \eqref{dbm-octagon} and coherency of tight encoding, there exist integers $1\leq k,\ell\leq N$ such that for each $n\geq 1$, $\oct{\topleft{M^t_{\overline{R^n}}}}$ implies:
\[\begin{array}{llll}
(1) & x_k - x_\ell \leq (\topleft{M^t_{\overline{R^n}}})_{2k-1,2\ell-1} = (\topleft{M^t_{\overline{R^n}}})_{ij}
   & \textrm{if } i=2k-1,j=2\ell-1 
\\ 
(2) & x_k + x_\ell \leq (\topleft{M^t_{\overline{R^n}}})_{2k-1,2\ell} = (\topleft{M^t_{\overline{R^n}}})_{ij}
   & \textrm{if } i=2k-1,j=2\ell
\\
(3) & -x_k - x_\ell \leq (\topleft{M^t_{\overline{R^n}}})_{2k,2\ell-1} = (\topleft{M^t_{\overline{R^n}}})_{ij}
   & \textrm{if } i=2k,j=2\ell-1  
\\
(4) & x_\ell - x_k \leq (\topleft{M^t_{\overline{R^n}}})_{2\ell-1,2k-1} = (\topleft{M^t_{\overline{R^n}}})_{2k,2\ell} = (\topleft{M^t_{\overline{R^n}}})_{ij}
   & \textrm{if } i=2k,j=2\ell 
\end{array}\]
Let $u\in OctTerm(\x)$ be the octagonal term from above (i.e.\ of the form $\pm x_k \pm x_\ell$). By Lemma \ref{lem:dp1}, $\wrs(R) = \bigcap_{n\geq 1} \pre_R^n(\zed^\x)$. Since $R$ is $*$-consistent, $\pre_{R}^n(\zed^\x)$ is defined by $\oct{\topleft{M^t_{\overline{R^n}}}}$ for each $n\geq 1$. Thus, any formula that defines $\wrs(R)$ must imply $u \leq \inf\{(\topleft{M^t_{\overline{R^n}}})_{ij}\}_{n\geq 1}$. This formula is inconsistent, since $\inf\{(\topleft{M^t_{\overline{R^n}}})_{ij}\}_{n\geq 1} = -\infty$. Consequently, $\wrs(R)=\emptyset$ and $R$ is thus well founded.
\qed}

The main result of this section is Algorithm \ref{alg:recurset} which computes the weakest non-termination precondition of an octagonal relation, in time polynomial in the number of variables and logarithmic in the maximal absolute value among all coefficients of the relation. As an auxiliary procedure, it uses Algorithm \ref{alg:fast:power} to compute exponentially large powers in polynomial time.

\begin{algorithm}
\begin{algorithmic}[0]
\State {\bf input} An octagonal constraint $R(\x,\x')$ and an integer $n\geq 1$
\State {\bf output} An octagonal constraint representing $R^n(\x,\x')$
\end{algorithmic}
\begin{algorithmic}[1]
  \Function{FastPower}{$R,n$} 
  \If{$R \iff \textrm{false}$} \label{alg:fastpow:consistent:0}
  \State \textbf{return} \textrm{false} \label{alg:fastpow:return:0}
  \EndIf
  \State $P \leftarrow M^t_{\overline{R^0}}$ \label{alg:fastpow:R0}
  \State $Q \leftarrow M^t_{\overline{R^1}}$ \label{alg:fastpow:R1}
  \For {$i = 1,\ldots,\lceil\log_2 n\rceil$} \label{alg:fastpow:loop}
  \If{$\oct{Q} \iff \textrm{false}$} \label{alg:fastpow:consistent}
  \State \textbf{return} \textrm{false} \label{alg:fastpow:return:1}
  \EndIf
  \If {the $i$-th least significant bit of $n$ is $1$} \label{alg:fastpow:test:bit}
  \State $P \leftarrow P \odot_t Q$ \label{alg:fastpow:comp:1}
  \EndIf 
  \State $Q \leftarrow Q \odot_t Q$ \label{alg:fastpow:comp:2} \hfill [at this point $\oct{Q} \iff R^{2^i}(\x,\x')$]
  \EndFor 
  \State\textbf{return} $\oct{P}$
  \EndFunction
\end{algorithmic}
\caption{Fast Exponentiation Algorithm \label{alg:fast:power}}
\end{algorithm}

\begin{lem}\label{lem:fast:power}
Given an octagonal constraint $R(\x,\x')$, where
$\x=\{x_1,\dotsm,\x_N\}$ for some $N\geq 1$, and an integer $n\geq 1$,
Algorithm \ref{alg:fast:power} computes $R^n(\x,\x')$ in at most
$\mathcal{O}(\lceil\log_2 n\rceil \cdot N^3\cdot(N+\log_2 \maxcoef{R}
+ \lceil\log_2 n\rceil))$ time. Moreover, $\maxcoef{R^n}$ is of the
order $\mathcal{O}(\maxcoef{R} \cdot N \cdot n)$.
\end{lem}
\proof{ Let $\mu_{P,i}$ (respectively $\mu_{Q,i}$) be the maximal
  absolute value over all integer entries of $P$ (respectively $Q$)
  before executing line \ref{alg:fastpow:test:bit} during the $i$-th
  iteration for $i = 1,\ldots,\lceil\log_2 n\rceil$. Further, let
  $n_i\geq 0$ be an integer such that $\oct{P} \iff R^{n_i}(\x,\x')$
  at line \ref{alg:fastpow:consistent} during the $i$-th iteration.
Notice that before executing line \ref{alg:fastpow:consistent},
$\oct{Q} \iff R^{2^{i-1}}(\x,\x')$ and $\oct{P} \iff R^{n_i}(\x,\x')$
where $n_i\leq 2^{i-1}$. It is easy to see that $\oct{Q}$ is
consistent before executing line \ref{alg:fastpow:test:bit}. Since
$n_i\leq 2^{i-1}$, it then follows that $\oct{P}$ is consistent before
executing line \ref{alg:fastpow:comp:1} too. Thus, compositions on
lines \ref{alg:fastpow:comp:1} and \ref{alg:fastpow:comp:2} are always
applied to two consistent relations.

If the test on line \ref{alg:fastpow:consistent} passes, then $\oct{Q}
\iff R^{2^{i-1}} \iff \textbf{false}$ and consequently, since
$2^{i-1}<n$, $R^n \iff \textbf{false}$ too. Thus, the algorithm
returns the correct result on line \ref{alg:fastpow:return:1}. The
correctness of the rest of the algorithm is easy to see.

Lines \ref{alg:fastpow:consistent:0}--\ref{alg:fastpow:R1} take at
most $\mathcal{O}(N^3 \cdot (N+ \log_2\maxcoef{R}))$ time, by
Corollary \ref{oct:consistency:test}.
Since the graph unfolding $\mathcal{G}^{2^{i}}_{\overline{R}}$,
corresponding to ${\overline{R}}^{2^{i}}$ for each $i\geq 1$, has $2N
\cdot 2^{i}$ nodes, each elementary path in this graph is of length at
most $2N \cdot 2^{i}$. Thus, $(M^*_{\overline{R}^{2^i}})_{k\ell} \leq
\maxcoef{R}\cdot 2N\cdot 2^i$ for all $1\leq k,\ell\leq 4N$ whenever
$R^{2^{i}}$ is consistent. Tightening clearly does not change this
bound. Since $Q \iff R^{2^{i-1}} \not\iff \textbf{false}$ on line
\ref{alg:fastpow:test:bit}, then $\mu_{Q,i} \leq \maxcoef{R} \cdot 2N
\cdot 2^{i-1}$. By Proposition \ref{oct:rel:comp}, composition on line
\ref{alg:fastpow:comp:2} can be computed in time
$\mathcal{O}(N^3\cdot(N+\log_2 (\mu_{Q,i})))$. Since $i\leq
\lceil\log_2 n\rceil$, this simplifies to
$\mathcal{O}(N^3\cdot(N+\log_2 \maxcoef{R} + \lceil\log_2
n\rceil))$. Since $n_i\leq 2^{i-1}$, then $\mu_{P,i}\leq \mu_{Q,i}$
and the same bound applies for the composition on line
\ref{alg:fastpow:comp:1}. By the definition of the composition
operator $\odot_t$ and the tight closure operator, the
octagonal-consistency check on line \ref{alg:fastpow:consistent} can
be taken care of during the preceding assignment to $Q$, i.e.\ on line
\ref{alg:fastpow:comp:2} (composition) or on line \ref{alg:fastpow:R1}
(tight closure). Thus, the overall running time of the algorithm is in
the order of $\mathcal{O}(\lceil\log_2 n\rceil \cdot N^3\cdot(N+\log_2
\maxcoef{R} + \lceil\log_2 n\rceil))$. Finally, $\maxcoef{R^n}$ is
asymptotically bounded by $\mathcal{O}(\maxcoef{R} \cdot N \cdot n)$.
\qed}

\begin{algorithm}
\begin{algorithmic}[0]
\State {\bf input} An octagonal constraint $R(\x,\x')$ where $\x = \{x_1,\ldots,x_N\}$ 
\State {\bf output} An octagonal constraint representing $\wnt(R)$
\end{algorithmic}
\begin{algorithmic}[1]
  \Function{WNT}{$R$}
  \State $V(\x,\x') \leftarrow \textsc{FastPower($R(\x,\x')$,$5^{2N}$)}$ \label{alg:wrs:asgn:1}
  \State $W(\x,\x') \leftarrow \textsc{FastPower($R(\x,\x')$,$5^{2N}+1$)}$ \label{alg:wrs:asgn:2}
  \If{$W \iff \textrm{false}$ \textbf{or} $\topleft{M^t_V} > \topleft{M^t_W}$ } \label{alg:wrs:test}
  \State \textbf{return} $\textrm{false}$
  \Else 
  \State \textbf{return} $\oct{\topleft{M^t_V}}$
  \EndIf
  \EndFunction
\end{algorithmic}
\caption{Weakest non-termination precondition for Octagonal Relations}
\label{alg:recurset}
\end{algorithm}

\begin{thm} \label{th:oct:wrs}
Let $R(\x,\x')$, where $\x=\{x_1,\dots,x_N\}$ for some $N\geq 1$, be
an octagonal constraint defining a relation $R \subseteq
\zed^{\vec{x}} \times \zed^{\vec{x}}$. Then, Algorithm
\ref{alg:recurset} returns an octagonal constraint $\phi(\x)$ that
defines $\wrs(R)$ in at most $\mathcal{O}(N^4\cdot(N+\log_2
\maxcoef{R}))$ time. Also,
$\maxcoef{\phi}=\mathcal{O}(\maxcoef{R}\cdot N\cdot 2^N)$.
\end{thm}
\proof{ By Lemma \ref{lem:fast:power}, lines \ref{alg:wrs:asgn:1} and
  \ref{alg:wrs:asgn:2} of the algorithm compute $V \iff R^{5^{2N}}$
  and $W \iff R^{5^{2N}+1}$ in at most $\mathcal{O}(N^4\cdot(N+\log_2
  \maxcoef{R}))$ time and moreover, $\maxcoef{V}$ and $\maxcoef{W}$
  are of the order $\mathcal{O}(\maxcoef{R} \cdot N \cdot 2^{N})$.

By Corollary \ref{oct:consistency:test}, the test $W \iff
\textrm{false}$ can be performed in at most $\mathcal{O}(N^3 \cdot (N+
\log_2\maxcoef{W}))$ time. If the test fails, the algorithm returns
\textbf{false}. Otherwise, $W$ is consistent and moreover, since
$5^{2N}<5^{2N}+1$, $V$ is consistent too. Then, $\topleft{M^t_V}$ and
$\topleft{M^t_W}$ can be computed and the test $\topleft{M^t_V} >
\topleft{M^t_W}$ can be performed in at most $\mathcal{O}(N^3 \cdot
(N+ \log_2\maxcoef{W}))$ time, by Proposition \ref{oct-eq} and
Corollary \ref{oct:consistency:test}. Also,
$\maxcoef{\oct{\topleft{M^t_V}}}$ inherits the upper bound of
$\maxcoef{V}$, by Proposition \ref{odbc:qelim}.

Consider first the case when $R$ is $*$-consistent. Then clearly $W
\not\iff \textrm{false}$. Notice that the test $\topleft{M^t_V} >
\topleft{M^t_W}$ is equivalent to $\pre_R^{5^{2N}}(\zed^\x) \supsetneq
\pre_R^{5^{2N}+1}(\zed^\x)$. If this test passes, $R$ is well founded,
by Lemma \ref{lem:core:results}, and the algorithm correctly returns
\textbf{false}. Otherwise, if this test fails, then
$\pre_R^{5^{2N}}(\zed^\x) = \pre_R^{5^{2N}+1}(\zed^\x)$ and
consequently, $\wrs(R) = \pre_R^{5^{2N}}(\zed^\x)$ by Lemma
\ref{wrs-kleene} and the algorithm correctly returns
$\oct{\topleft{M^t_V}}$.

Second, consider the case when $R$ is not $*$-consistent. Then clearly
$\wrs(R)=\emptyset$. Hence, if the test on line \ref{alg:wrs:test}
passes, the algorithm returns the correct result. To see that the test
on line \ref{alg:wrs:test} cannot fail, let us assume, by
contradiction, that
$\pre_R^{5^{2N}}(\zed^\x)=\pre_R^{5^{2N}+1}(\zed^\x)$ and
$\pre_R^{5^{2N}+1}(\zed^\x)\neq\emptyset$. Then,
$\wrs(R)=\pre_R^{5^{2N}+1}(\zed^\x)$, by Proposition
\ref{wrs-kleene}. Since $\pre_R^{5^{2N}+1}(\zed^\x) \neq \emptyset$,
then $\wrs(R)\neq\emptyset$. Contradiction with $\wrs(R)=\emptyset$.
\qed}

An immediate consequence of Theorem \ref{th:oct:wrs} is that the
termination problem is decidable.
\begin{thm}\label{theorem:term:oct}
Let $R(\x,\x')$, where $\x=\{x_1,\dots,x_N\}$ for some $N\geq 1$, be
an octagonal constraint defining a relation $R \subseteq
\zed^{\vec{x}} \times \zed^{\vec{x}}$. The well-foundedness of
$R(\x,\x')$ can be decided in at most
$\mathcal{O}(N^4\cdot(N+\log_2\maxcoef{R}))$ time.
\end{thm}
\proof{ By Theorem \ref{th:oct:wrs}, Algorithm \ref{alg:recurset}
  computes an octagonal constraint $\phi(\x)$ that defines $\wrs(R)$
  in $\mathcal{O}(N^4\cdot(N+\log_2\maxcoef{R}))$ time and moreover
  $\maxcoef{\phi}$ is in the order of $\mathcal{O}(\maxcoef{R}\cdot
  N\cdot 2^N)$. Well-foundedness of $R$ can be decided by checking
  whether $\phi(\x)$ is consistent. This check can be performed in
  time $\mathcal{O}(N^3 \cdot (N+ \log_2\maxcoef{\phi}))$, by
  Corollary \ref{oct:consistency:test}, which simplifies to
  $\mathcal{O}(N^3 \cdot (N+ \log(\maxcoef{R}\cdot N\cdot 2^N))) =
  \mathcal{O}(N^3 \cdot (N + \log_2\maxcoef{R}))$.  \qed}

\subsection{On the Existence of Linear Ranking Functions}
\label{sec:ranking-function}

We first define the notion of a~{\em linear ranking function}, using
the following notation: if $f(\x)$ is a~linear term over $\x$ of the
form $f(\x)=a_0+\sum_{i=1}^N a_ix_i$ where $a_0,\dots,a_N\in\zed$,
then $f(\x')$ denotes the corresponding term over $\x'$ defined as
$f(\x')\stackrel{def}{=}a_0+\sum_{i=1}^N a_ix'_i$.
\begin{defi}\label{lrf}
Given a~relation defined by $R(\x,\x')$, a~{\em linear ranking
  function} $f: \x \rightarrow \zed$ for $R(\x,\x')$ is a~linear term
$f(\x)$ such that the following holds:
\[\exists h \forall \x \forall \x' ~.~ R(\x,\x') ~\Rightarrow~ f(\x) > f(\x') ~\wedge~ f(\x) \geq h \]
\end{defi}
\noindent Intuitively, $R(\x,\x') \Rightarrow f(\x) > f(\x')$ requires
that $f$ is {\em decreasing} and $R(\x,\x') \Rightarrow f(\x) \geq h$ requires that $f$ is {\em bounded}.

A ranking function for a~given relation $R$ constitutes a~proof of the
fact that $R$ is well founded. In this section, we show that for any
well-founded octagonal relation $R(\x,\x')$, where
$\x=\{x_1,\dots,x_N\}$, the (strengthened) relation $V$ defined as
$V(\x,\x') \equiv \relRFd$ has a~linear ranking function if and only
if $R$ is well founded. Note that if $R$ is well founded, then $V$ is
guaranteed to have a~linear ranking function even when $R$ alone does
not have one. Moreover, we show that such a~linear ranking function can
be computed in polynomial time. The proof is organized as
follows. First, we show in Lemma \ref{strenghten} that for each $m\geq
1$, strengthening $R(\x,\x')$ with $\exists \x' . R^m(\x,\x')$
preserves the (conditional) termination problem, formally:
$\wrs(R)=\wrs(R_m)$ where $R_m$ is defined by $\relRFm$. As a
consequence, $\wrs(R)=\wrs(V)$.

In Section \ref{sec:rf:db}, we study the case when $R(\x,\x')$ is
a~well-founded difference bounds constraint. Here, we first generalize
Lemma \ref{lem:core:results:db} and show that the zigzag automaton of
$R$ is guaranteed to have a negative-weight cycle, whenever the
$5^N$-th power of $R$ is consistent. Lemma \ref{decreasing} and Lemma
\ref{bounded} use the structure of this cycle, representing several of
the constraints in $R$, to show the existence of the linear ranking
function for the witness relation $\relRFc$.

Section \ref{sec:rf:oct} then studies octagonal relations. Given an
octagonal constraint $R(\x,\x')$, where $\x=\{x_1,\dots,x_N\}$, with
its difference bounds representation $\overline{R}(\y,\y')$, where
$\y=\{y_1,\dots,y_{2N}\}$, such that $R$ is well founded and the
$5^{2N}$-th power of $R$ is consistent, we first apply the above
result and immediately infer that $\relRFcDbOct$ has a~linear ranking
function $\overline{f}(\y)$. Then, we prove in Proposition
\ref{prop-rf-db-to-oct-2} that the function defined as
$f\stackrel{def}{=}\overline{f}(\y)[x_i/y_{2i-1},-x_i/y_{2i}]_{i=1}^{N}$
is a~linear ranking function for $\relRFcOct$. For the case when the
$5^{2N}$-th power is not consistent, it follows easily that
$\relRFcOct$ is not consistent either and hence, trivially, has a
linear ranking function. Then, since the sequence $\{
\pre_R^n(\zed^\x) \}_{n\geq 1}$ is descending, it follows that
$\exists \x'.R^{5^{2N}}(\x,\x') \Rightarrow \exists
\x'.R^{4N^2}(\x,\x')$ and one can thus show that $f$ is also a ranking
function for $V$. Finally, we summarize this reasoning in Theorem
\ref{theorem:rf} and prove that such a~	linear ranking function can be
found in polynomial time.


\begin{lem}\label{strenghten}
Let $R\subseteq\zed^\x\times\zed^\x$ be a~relation defined by a~formula $R(\x,\x')$, and $m\geq 1$ be an integer. Then
$\wrs(R) = \wrs(R_m)$, where $R_m$ is the relation defined by
$\relRFm$.
\end{lem}
\proof{ ``$\subseteq$'' By Proposition \ref{prop:pre:properties},
  $\pre_{R'}(S) \subseteq \pre_R(S)$ for any set $S$ and relations
  $R,R'$ such that $R'\subseteq R$. Since $R_m\subseteq R$, then
  $\pre_{R_m}(\zed^\x)\subseteq \pre_R(\zed^\x)$. Applying this
  argument $n$-times, we infer that $\pre^n_{R_m}(\zed^\x) \subseteq
  \pre^n_R(\zed^\x)$. Thus, we have:
\[\begin{array}{lclll}
\wrs(R_m) & = & \bigcap_{n\geq 1} \pre^n_{R_m}(\zed^\x) & \textrm{by Lemma \ref{lem:dp1}}
\\ & \subseteq & \bigcap_{n\geq 1} \pre^n_{R}(\zed^\x)
\\ & = & \wrs(R) & \textrm{by Lemma \ref{lem:dp1}}
\end{array}\]
``$\supseteq$'' We prove the dual. Assume that $\wrs(R) \neq
\emptyset$, i.e.\ there exists an infinite sequence of valuations
$\sigma = \{ \nu_i \in \zed^x \}_{i\geq 0}$ such that
$(\nu_i,\nu_{i+1}) \in R$, for all $i \geq 0$. Then each $\nu_i$
belong to the set defined by $\exists \x' ~.~ R^m(\x,\x')$, hence
$\sigma$ is an infinite sequence for the relation defined by $\relRFm$
as well.  \qed}

\subsubsection{Linear Ranking Function for Difference Bounds Relation}
\label{sec:rf:db}

In the rest of this section, let us fix the set of variables $\x=\{x_1,\dots,x_N\}$ for some constant $N\geq 1$. We first prove the existence of a~negative-weight cycle in a~zigzag automaton whenever the $5^N$-th power of a well-founded difference bounds relation $R(\x,\x')$ is consistent.
\begin{lem}\label{incons:neg:cycle}
Let $R(\x,\x')$ be a~well-founded difference bounds relation such that $R^{5^N}(\x,\x')$ is consistent. Then, there exists a~zigzag automaton $\mathcal{A}_{ij}=\langle T_R,I_{ij},F \rangle$ for some $1 \leq i, j\leq N$ with an accepting run $\mu.\lambda.\mu'$ where $\lambda$ is a cycle such that $|\lambda|>0$ and $\omega(\lambda)<0$.
\end{lem}
\proof{
If $R(\x,\x')$ is $*$-consistent, then the result follows immediately from Lemma \ref{lem:core:results:db}. In the rest of the proof, let $R(\x,\x')$ be a~$*$-inconsistent relation such that $R^{5^N}(\x,\x')$ is consistent. We first define $n \stackrel{def}{=} \min\{ i\geq 1 ~|~ R^i(\x,\x') \textrm{ is inconsistent} \}$. Clearly, $n>5^N$. By Lemma \ref{zigzag-automata}, there exists $1\leq i\leq N$ such that $\mathcal{A}_{ii}$ has an accepting run $\pi$ such that $|\pi|=n$ and $\omega(\pi)<0$. Since $n>5^N\geq |Q|$, there must be at least one cycle $\lambda$ in $\pi$, formally: $\pi=\mu.\lambda.\mu'$ for some paths $\mu,\mu'$ and a~cycle $\lambda$. Let us denote $m=|\mu.\mu'|$. Clearly $m<n$. We prove that $\omega(\lambda)<0$. By contradiction, suppose that $\omega(\lambda)\geq 0$. Then $\omega(\mu.\mu') = \omega(\pi)-\omega(\lambda) < 0$ and hence, by Lemma \ref{zigzag-automata}, $R^m(\x,\x')$ is not consistent. Since $m<n$, this contradicts the definition of $n$ as the minimal inconsistent power. Thus, $\omega(\lambda)<0$ and the run $\mu.\lambda.\mu'$ of $\mathcal{A}_{ii}$ has the property required by the lemma.
\qed}

We next prove the existence of a~linear decreasing function, based on the existence of a~negative-weight cycle in the zigzag automaton.

\begin{lem} \label{decreasing}
Let $R(\x,\x')$, where $\x=\{x_1,\dots,x_N\}$, be a~difference bounds constraint defining a~well-founded relation $R\subseteq\zed^\x\times\zed^\x$ such that $R^{5^N}(\x,\x')$ is consistent. Then, there exists a~linear function $f(\vec{x})$ such that $\forall \x,\x' ~.~ R(\x,\x') \Rightarrow f(\x) > f(\x')$ is valid.
\end{lem}
\proof{
By Lemma \ref{incons:neg:cycle}, there exist integers $1\leq i,j\leq N$ such that the zigzag automaton $\mathcal{A}_{ij}$ has an accepting run $\mu.\lambda.\mu'$ where $\lambda$ is a cycle such that $|\lambda|>0$ and $\omega(\lambda)<0$. Let us write $\lambda$ as $\lambda = q_0\arrow{G_0}{}q_1\arrow{G_1}{}q_2\dots q_{p-1}\arrow{G_{p-1}}{} q_0$ where $p=|\lambda|$ and $G_j=(\x\cup \x',E_j)$ for some set of edges $E_j$, $0\leq j<p$. Recall that $G_j$ is a~bipartite graph for each $0\leq j<p$ and therefore contains edges of the form $x_i\arrow{}{}x'_j$ or $x'_i\arrow{}{}x_j$. Consider the following sum of all constraints represented by edges appearing in $\lambda$ (note that the sum of weights of these edges equals $\omega(\lambda)$): 
\begin{equation}\label{rf:eq1}
\sum\limits_{\substack{0\leq j<p \\ 1\leq i,k\leq N \\ (x_k\arrow{}{}x'_i)\in E_j}} (x_k - x'_i) ~+ \sum\limits_{\substack{0\leq j<p \\ 1\leq i,k\leq N \\ (x'_k\arrow{}{}x_i)\in E_j}} (x'_k - x_i) \leq \sum\limits_{\substack{0\leq j<p \\ e\in E_j}} \omega(e)  = \omega(\lambda)
\end{equation}
Notice that for each $0\leq j<p$, there exists an accepting run of the form
\[ q \arrow{w}{} q_{(j-1)\mmod p} \arrow{G_{(j-1)\mmod p}}{} q_j \arrow{G_j}{} q_{(j+1)\mmod p} \arrow{G_{(j+1)\mmod p}}{} q_{(j+2)\mmod p} \arrow{w'}{} q' \]
for some $q,q'\in Q$ and $w,w'\in \Sigma_R^*$.
It follows from the definition of zigzag automata that for each edge $e_1: x'_k\arrow{}{}x_i \in E_j$, there exists a~unique ``successor`` $e_2$ which is of either of the following forms:
\begin{equation}\label{rf:eq3}\begin{array}{llll}
 \textrm{either} & e_2: x_i\arrow{}{}x'_m \in E_{j} & \textrm{if $(q_j)_i=\ell r$,} \\
 \textrm{or} & e_2: x_i'\arrow{}{}x_m \in E_{(j-1)\mmod p} & \textrm{if $(q_j)_i=\ell$.}
\end{array}\end{equation}
Dually, $e_1$ is said to be the unique ``predecessor`` of $e_2$. Similarly, 
for each edge $e_1: x_k\arrow{}{}x'_i \in E_j$, there exists a~unique successor $e_2$ which is of either of the following forms:
\begin{equation}\label{rf:eq2}\begin{array}{llll}
 \textrm{either} & e_2: x'_i\arrow{}{}x_m \in E_{j} & \textrm{if $(q_{(j+1)\mmod p})_i=r \ell$,} \\
 \textrm{or} & e_2: x_i\arrow{}{}x'_m \in E_{(j+1)\mmod p} & \textrm{if $(q_{(j+1)\mmod p})_i=r$.}
\end{array}\end{equation}
Consider the following sum:
\begin{equation}\label{rf:eq4}
\sum\limits_{\substack{0\leq j<p \\ 1\leq i,k,m \leq N \\ (x_k\arrow{}{}x'_i)\in E_j \\ (x_i\arrow{}{}x'_m)\in E_{(j+1)\mmod p}}} \hspace{-8mm}(-x'_i+x_i) ~+
\sum\limits_{\substack{0\leq j<p \\ 1\leq i,k,m \leq N \\ (x_k\arrow{}{}x'_i)\in E_j \\ (x'_i\arrow{}{}x_m)\in E_{j}}} \hspace{-5mm}(-x'_i+x'_i) ~+
\sum\limits_{\substack{0\leq j<p \\ 1\leq i,k,m \leq N \\ (x'_k\arrow{}{}x_i)\in E_j \\ (x'_i\arrow{}{}x_m)\in E_{(j-1)\mmod p}}} \hspace{-8mm}(-x_i+x'_i) ~+
\sum\limits_{\substack{0\leq j<p \\ 1\leq i,k,m \leq N \\ (x'_k\arrow{}{}x_i)\in E_j \\ (x_i\arrow{}{}x'_m)\in E_{j}}} \hspace{-5mm}(-x_i+x_i)
\end{equation}
and note that every edge $e=(x_k\arrow{}{}x'_i)\in E_j$, where $1\leq i,j\leq N, 0\leq j<p$, is considered exactly twice in \eqref{rf:eq4}, since
\begin{itemize} 
 \item $e$ has a unique successor and therefore contributes with the $-x'_i$ term in \eqref{rf:eq4}
 \item $e$ has a unique predecessor and therefore contributes with the $+x_k$ term in \eqref{rf:eq4}
\end{itemize}
Similarly, every edge $(x'_k\arrow{}{}x_i)\in E_j$ is considered twice and contributes with terms $-x_i$ and $+x'_k$. Hence, the sum \eqref{rf:eq4} is equivalent to the left-hand side of \eqref{rf:eq1}.
Clearly, the second and the fourth sum in \eqref{rf:eq4} evaluate to zero. It follows from Equations \eqref{rf:eq2} and \eqref{rf:eq3} that the remaining two sums can be written equivalently as 
\begin{equation}\label{rf:eq5}
\sum\limits_{\substack{0\leq j<p \\ (q_j)_i=r}} (-x'_i+x_i) ~+
\sum\limits_{\substack{0\leq j<p \\ (q_j)_i=\ell}} (-x_i+x'_i)
\end{equation}
Thus, \eqref{rf:eq1} can be written equivalently as 
\begin{equation}\label{rf:eq6}
\sum\limits_{\substack{0\leq j<p \\ (q_j)_i=r}} (-x'_i+x_i) ~+
\sum\limits_{\substack{0\leq j<p \\ (q_j)_i=\ell}} (-x_i+x'_i)
\leq \omega(\lambda)
\end{equation}
Let $f(\x)$ denote the negated sum of all unprimed terms in \eqref{rf:eq5} and $g(\x')$ denote the sum of all primed terms in \eqref{rf:eq5}. Clearly, $f(\x) = g(\x')[\x/\x']$ (i.e.\ $g(\x')$ is the primed counterpart of $f(\x)$) and \eqref{rf:eq6} can be written as $g(\x')-f(\x) \leq \omega(\lambda)$. Recall that $f(\x')\stackrel{def}{=}a_0+\sum_{i=1}^N a_ix'_i$ and hence $f(\x')=g(\x')$. We thus obtain:
\begin{equation}\label{rf:sum4} f(\x')-f(\x) \leq \omega(\lambda) < 0 \end{equation} 
Hence, $f(\x)$ is strictly decreasing, formally: $R(\x,\x') \Rightarrow f(\x) > f(\x')$.
\qed}

\begin{figure}
\begin{center}
\begin{tabular}{cc}
\mbox{\begin{minipage}{7.5cm}
\mbox{\begin{tikzpicture}
  \TermZrBase{1}{5}{4}{0.7}{0.65}{0.5}{1}{3}{1}

  \newarray\aCaption
  \readarray{aCaption}{$q_2$ & $q_0$ & $q_1$ & $q_2$ & $q_0$ & $q_3$}
  \TermZrCapState{0}{5}{4}{0.7}{0.65}{0.5}
  \readarray{aCaption}{$G_3$ & $G_1$ & $G_2$ & $G_3$ & $G_4$}
  \TermZrCapGraph{1}{5}{4}{0.7}{0.65}{0.5}

  \TermZrStateElem{0}{1}{\bot}
  \TermZrStateElem{0}{2}{r}
  \TermZrStateElem{0}{3}{\bot}
  \TermZrStateElem{0}{4}{l}

  \TermZrStateElem{1}{1}{r}
  \TermZrStateElem{1}{2}{\bot}
  \TermZrStateElem{1}{3}{\bot}
  \TermZrStateElem{1}{4}{l}

  \TermZrStateElem{2}{1}{\bot}
  \TermZrStateElem{2}{2}{\bot}
  \TermZrStateElem{2}{3}{r}
  \TermZrStateElem{2}{4}{l}

  \TermZrStateElem{3}{1}{\bot}
  \TermZrStateElem{3}{2}{r}
  \TermZrStateElem{3}{3}{\bot}
  \TermZrStateElem{3}{4}{l}

  \TermZrStateElem{4}{1}{r}
  \TermZrStateElem{4}{2}{\bot}
  \TermZrStateElem{4}{3}{\bot}
  \TermZrStateElem{4}{4}{l}

  \TermZrStateElem{5}{1}{\bot}
  \TermZrStateElem{5}{2}{\bot}
  \TermZrStateElem{5}{3}{rl}
  \TermZrStateElem{5}{4}{\bot}

  \TermZrEdgeFWLab{1}{2}{1}{\tiny$-1$}
  \TermZrEdgeBWLab{1}{4}{4}{\tiny$0$}

  \TermZrEdgeFWLab{2}{1}{3}{\tiny$0$}
  \TermZrEdgeBWLab{2}{4}{4}{\tiny$0$}

  \TermZrEdgeFWLab{3}{3}{2}{\tiny$0$}
  \TermZrEdgeBWLab{3}{4}{4}{\tiny$0$}

  \TermZrEdgeFWLab{4}{2}{1}{\tiny$-1$}
  \TermZrEdgeBWLab{4}{4}{4}{\tiny$0$}

  \TermZrEdgeFWLab{5}{1}{3}{\tiny$0$}
  \TermZrEdgeBWLab{5}{3}{4}{\tiny$0$}

  \TermGridBind{4}{2}{4}{4}{}
  \TermGridBind{3}{3}{3}{4}{}
  \TermGridBind{2}{1}{2}{4}{}

\end{tikzpicture}}
\end{minipage}}
&
\mbox{\begin{minipage}{5.5cm}
\scalebox{1.0}{\begin{tikzpicture}
  \scriptsize
        \TermGridGen{0.0}{0.8}{6}{0.0}{0.4}{4}{0.7}{0.5}{0}
        \TermGridEdgeC{1}{2}{2}{1}{$-\!1$}{above}
        \TermGridEdgeC{2}{1}{3}{3}{$0$}{above}
        \TermGridEdgeC{3}{3}{4}{2}{$0$}{above}
        \TermGridEdgeC{4}{2}{5}{1}{$-\!1$}{above}
        \TermGridEdgeC{5}{1}{6}{3}{$0$}{above}
        \TermGridEdgeC{6}{3}{5}{4}{$0$}{above}
        \TermGridEdgeC{5}{4}{4}{4}{$0$}{above}
        \TermGridEdgeC{4}{4}{3}{4}{$0$}{above}
        \TermGridEdgeC{3}{4}{2}{4}{$0$}{above}
        \TermGridEdgeC{2}{4}{1}{4}{$0$}{above}
\end{tikzpicture}}
\end{minipage}}
\\
(a) An accepting run $\pi=\mu.\lambda.\mu'$ in $\mathcal{A}_{24}$.
&
(b) The graph $\mathcal{H}_\pi$.
\end{tabular}
\end{center}
\caption{
Constructing the ranking function for a relation (see also Fig.\ \ref{fig:rf:zigzag}) $R(\x,\x') \iff x_2\mi x'_1\leq -1 \wedge x_3\mi x'_2\leq 0 \wedge x_1\mi x'_3\leq 0 \wedge x'_4\mi x_4\leq 0 \wedge x'_3\mi x_4\leq 0$. 
Figure (a) shows a~run $\pi$ that accepts word $\gamma=G_3.G_1.G_2.G_3.G_4$. Figure (b) shows $\mathcal{G}_\pi$, obtained by concatenating the symbols (graphs) of $\gamma$. $\mathcal{G}_\pi$ contains a~single path $\rho$ from $\xxki{0}{2}$ to $\xxki{0}{4}$.
}
\label{fig:bounding:rf}
\end{figure}
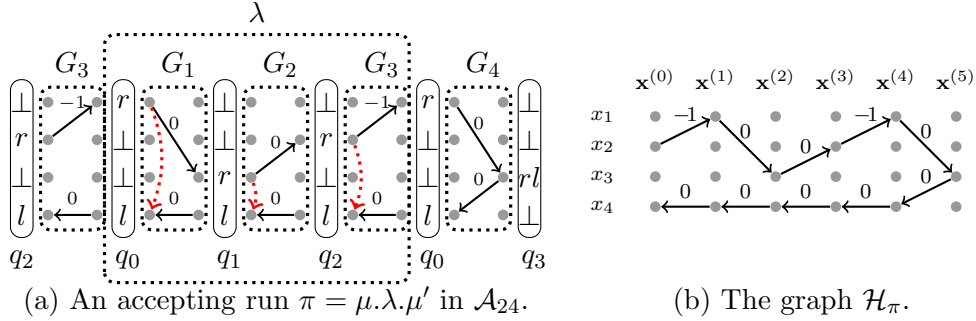

\begin{exa} (Ex.\ \ref{ex:dbm} ctd.) \label{example:rf:decr}
We illustrate the construction of a linear decreasing function for a~well-founded relation $R(\x,\x') \equiv x_2\mi x'_1\leq -1 \wedge x_3\mi x'_2\leq 0 \wedge x_1\mi x'_3\leq 0 \wedge x'_4\mi x_4\leq 0 \wedge x'_3\mi x_4\leq 0$ (see also Figure \ref{fig:rf:zigzag}). By Lemma \ref{lem:core:results:db}, there exists an accepting run $\mu.\lambda.\mu'$ in a~zigzag automaton where $\lambda$ is a cycle such that $\omega(\lambda)<0$. Figure \ref{fig:bounding:rf} depicts such a~run in $\mathcal{A}_{2,4}$ where $\mu$, $\lambda$, and $\mu'$ are labeled with words $G_3$, $G_1.G_2.G_3$, and $G_4$, respectively. We have $\omega(\lambda)=-1$. We follow the construction from Lemma \ref{decreasing} and sum the edges that are present in $\lambda$ (see the solid edges in $G_1$, $G_2$, and $G_3$ in Figure \ref{fig:bounding:rf}). We obtain 
\[(x_1 - x'_3) + (x_3 - x_2') + (x_2 - x'_1) + (x'_4 - x_4) + (x'_4 - x_4) + (x'_4 - x_4) \leq -1 \]
which simplifies to $(x_1+x_2+x_3-3x_4)-(x'_1+x'_2+x'_3-3x'_4)\leq -1$. Letting $f(\x) = -(x_1+x_2+x_3-3x_4)$, we have that $R(\x,\x') \Rightarrow f(\x) > f(\x')$. 
\qed\end{exa}

Next, we prove that all functions of Lemma \ref{decreasing} are bounded, concluding that they are indeed ranking functions. Each run $\pi$ of length $n\geq 1$ in the zigzag automaton $\mathcal{A}_{ij}$, $1\leq i,j\leq N$, recognizes a~word $w=G_0.G_1\dots G_{n-1}$ where $G_0,\dots,G_{n-1}\in\Sigma_R$. Assuming that $E_\ell$ is the set of edges in $G_\ell$ for each $0\leq \ell<n$, we define the concatenation of graphs $G_0,\dots,G_{n-1}$ as $\mathcal{H}_\pi=(V,E)$ where $V=\bigcup_{\ell=0}^n \xk{\ell}$ and
\[\begin{array}{lcl}
  \xxki{\ell}{p}\arrow{c}{}\xxki{\ell+1}{q} \in E & \textrm{iff} & x_p\arrow{c}{}x_q' \in E_\ell \\
  \xxki{\ell+1}{p}\arrow{c}{}\xxki{\ell}{q} \in E & \textrm{iff} & x_p'\arrow{c}{}x_q \in E_\ell
\end{array}\]
for all $0\leq \ell<n$ and $1\leq i,j\leq N$. See Figure~\ref{fig:bounding:rf} for an illustration. Supposing that $\pi$ traverses a cycle $\lambda$ in $\mathcal{A}_{ij}$ (see the cycle $\lambda$ in Figure~\ref{fig:bounding:rf}), $\pi$ can be decomposed into a prefix, the cycle itself and a~suffix. By the definition of zigzag automata, $\mathcal{H}_\pi$ contains exactly one path\footnote{Moreover, this path is acyclic if $i\neq j$ or an elementary cycle if $i=j$.} $\rho$ from $\xxki{0}{i}$ to $\xxki{0}{j}$ and a~(possibly empty) set of elementary cycles $\{\nu_1,\dots,\nu_p\}, p\geq 0$. For instance, $\mathcal{H}_\pi$ from Figure~\ref{fig:bounding:rf} contains a single path $\rho$. The paths $\{\rho,\nu_1,\dots,\nu_p\}$ may traverse the cycle $\lambda$ several times, however each exit point from the cycle must match a~subsequent entry point (the dotted edges in Figure~\ref{fig:bounding:rf}(a) mark such a~matching). These paths from the exit to the corresponding entries give the lower bound on $f(\x)$, formally: $R^n(\x,\x') \Rightarrow f(\x)\geq h$ for some $h\in \zed$ and sufficiently large $n\geq 1$ (Proposition \ref{prop:bijection:existence}). In fact, these paths appear already on graphs $\mathcal{G}^i_R$ for every $i \!\geq\! N^2$ (Lemma \ref{unp:dif:bounded}) and the ``sufficiently large $n$`` can be thus bounded by $N^2$. Hence the need for a~strengthened witness $\relRFc$, as $R$ alone is not enough for proving boundedness of $f(\x)$. Lemma \ref{bounded} combines all these results to prove the existence of a~ranking function.

\begin{prop} \label{prop:bijection:existence}
Let $R(\x,\x')$ be a~difference bounds constraint, let $\pi=q_0\arrow{G_0}{}q_1\arrow{G_1}{}\dots\arrow{}{}q_{n-1}\arrow{G_{n-1}}{}q_n$, for some $n\geq 1$, be an accepting run of a~zigzag automaton $\mathcal{A}_{gh}$ for some $1\leq g,h\leq N$, and $k\in\{0,\dots,n-1\}$ be a constant. Then, there exists a bijection
\[ \beta : \{ j ~|~ (q_k)_j=r \} \rightarrow \{ j ~|~ (q_k)_j=\ell \} \]
such that for every $i \in \{ j ~|~ (q_k)_j=r \}$, the following formula is valid:
\[ \exists b ~.~ \forall \x ~.~ (\exists \x' ~.~ R^n(\x,\x')) ~\Rightarrow~ x_{\beta(i)}-x_i \geq b \]
\end{prop}
\proof{
We define a~shift operator that for every path $\rho$ in $\mathcal{G}_R^m$, $m\geq 1$, of the form $\rho=\xxki{j_1}{i_1}\arrow{c_1}{}\xxki{j_2}{i_2}\arrow{c_2}{}\dots\arrow{c_{p-1}}{}\xxki{j_p}{i_p}$, $p>1$, and every $k\in\zed$, returns the path $\rho^{\rightarrow k}$ defined as:
\[
\rho^{\rightarrow k} \stackrel{def}{=}
\xxki{j_1+k}{i_1}\arrow{c_1}{}\xxki{j_2+k}{i_2}\arrow{c_2}{}\dots\arrow{c_{p-1}}{}\xxki{j_p+k}{i_p}
\]

Let us assume that $G_k=(\x\cup\x',E_k)$ for each $0\leq k<n$ and let us denote by $w$ the word $G_0.G_1\dots G_{n-1}$ accepted by $\pi$. Given a~path $\rho$ in $\mathcal{G}_R^n$, let $V_\rho$ denote the set of all vertices traversed by $\rho$. It follows from the definition of zigzag automata that $\mathcal{H}_\pi$ contains one path $\nu_0$ that starts in $\xxki{0}{g}$ and ends in $\xxki{0}{h}$. $\mathcal{H}_\pi$ may also contain a~(possibly empty) set of elementary cycles $\{\nu_1,\dots,\nu_s\}$ for some $s\geq 0$. By the definition of zigzag automata, the sets of vertices $V_{\nu_0},\dots,V_{\nu_p}$ are pairwise disjoint.
By the definition of zigzag automata, we have:
\[
  |\{j ~|~ (q_k)_j=r\}| = |\{j ~|~ (q_k)_j=\ell \}|
\]

Clearly, for every $1\leq i\leq N$ such that $(q_k)_i=r$, there exists $\nu\in\{\nu_0,\dots,\nu_s\}$ such that $\xxki{k}{i}\in V_\nu$. Since $(q_k)_i=r$, $\nu$ goes to the right from $\xxki{k}{i}$, but it must eventually turn left and reach $\xxki{k}{j}$ such that $(q_k)_j=\ell$ for some $1\leq j\leq N$, either in order to reach $\xxki{0}{h}$ (if $\nu=\nu_0$) or in order to reach $\xxki{k}{i}$ again (if $\nu\neq\nu_0$ is a cycle). Without loss of generality, let $\xxki{k}{j}$ be the first such vertex reachable from $\xxki{k}{i}$ and let us define $\beta(i)\stackrel{def}{=}j$. Clearly, $\beta$ is a~bijection from $\{ j ~|~ (q_k)_j=r \}$ to $\{ j ~|~ (q_k)_j=\ell \}$.
Since $\xxki{k}{j}$ was chosen as the first vertex reachable from $\xxki{k}{i}$ such that $(q_k)_j=\ell$, it follow that the subpath $\rho$ of $\nu$ from $\xxki{k}{i}$ to $\xxki{k}{j}$ traverses only vertices from $\bigcup_{m=k}^{n} \xk{m}$ (since to reach some vertex from $\bigcup_{m=0}^{k-1} \xk{m}$, the path would have to cross some component $(q_k)_t$, $1\leq t\leq N$, such that $(q_k)_t=\ell$). Hence, $\rho$ can be shifted by $-k$ and we obtain a path $\rho'=\rho^{\rightarrow (-k)}$ that starts in $\xxki{0}{i}$ and ends in $\xxki{0}{j}$. Since $\rho'$ is a path in $\mathcal{G}_R^n$, then $R^n \Rightarrow x_i-x_j\leq \omega(\rho')$, by \eqref{dbm-min-paths}. Hence, $R^n(\x,\x') \Rightarrow x_{\beta(i)}-x_i\geq -\omega(\rho')$ is valid. As an immediate consequence, the following formulas are valid too:
\[\begin{array}{rcl}
 \forall \x ~.~ (\exists \x' ~.~ R^n(\x,\x')) & \Rightarrow & x_{\beta(i)}-x_i\geq -\omega(\rho') \\
 \exists b ~.~ \forall \x ~.~ (\exists \x' ~.~ R^n(\x,\x')) &
 \Rightarrow & x_{\beta(i)}-x_i \geq b\rlap{\hbox to 122 pt{\hfill\qEd}}
\end{array}\]
}

The next lemma proves, for any two unprimed variables $x_i,x_j$, that if the difference $x_i-x_j$ is bounded in $R^n(\x,\x')$ for some $n\geq 1$, it is bounded in $R^{N^2}(\x,\x')$ too.

\begin{lem}\label{unp:dif:bounded}
Let $R(\x,\x')$ be a~difference bounds constraint. Then, for each $1\leq i,j\leq N, i\neq j$ and for each $n \geq 1$, the following is a valid formula:
\[\begin{array}{l}
\exists h ~.~ \forall \x ~.~ (\exists \x' ~.~ R^n(\x,\x')) \Rightarrow (x_i - x_j \leq h) 
  \\ \Rightarrow
  \\ \exists h ~.~ \forall \x ~.~ (\exists \x' ~.~ R^{N^2}(\x,\x')) \Rightarrow (x_i - x_j \leq h) 
\end{array}\]
\end{lem}
\proof{
Let us first define, for each $n\geq 1$:
\[ B_n \stackrel{def}{=} \{ (i,j) ~|~ 1\leq i,j\leq N \textrm{ and there is a path from $\xxki{0}{i}$ to $\xxki{0}{j}$ in $\mathcal{G}_R^n$ } \} \]
Clearly, for each $n\geq 1$, $\mathcal{G}_R^n$ is a subgraph of $\mathcal{G}_R^{n+1}$ and hence $B_n \subseteq B_{n+1}$. Next observe that for every $n\geq 1$, every path $\rho$ from $\xxki{0}{i}$ to $\xxki{0}{j}$ in $\mathcal{G}_R^{n+1}$ can be written as $\rho=\tau_0.\nu_1.\tau_1\dots \nu_p.\tau_p$ for some $p\geq 0$ such that $\tau_0,\dots,\tau_p$ traverse only nodes from $\xk{0}\cup\xk{1}$ and $\nu_1,\dots,\nu_p$ traverse only nodes from $\xk{1}\cup\dots\cup\xk{n+1}$. Clearly, if $\nu_k$, $1\leq k\leq p$, is a path from $\xxki{1}{i}$ to $\xxki{1}{j}$ for some $1\leq i,j\leq N$, then there also exists a path from $\xxki{0}{i}$ to $\xxki{0}{j}$ in $\mathcal{G}_R^n$ and consequently, $(i,j)\in B_n$. Hence, we have for all $n\geq 1$:
\[
B_{n+1} = B_{n} \cup \{ (i,j) ~|\!
\begin{array}{l}
1\leq i,j\leq N, \exists 1\leq k_1,\dots,k_p\leq N ~.~ (k_1,k_2),(k_3,k_4),\ldots\in B_{n} \textrm{ and } \mathcal{G}_R  \\
\textrm{ has paths } \xxki{0}{i}\rightarrow^+\xxki{1}{k_1}, \xxki{1}{k_2}\rightarrow^+\xxki{1}{k_3},\xxki{1}{k_4}\rightarrow^+\xxki{1}{k_5},\dots,\xxki{1}{k_p}\rightarrow^+\xxki{0}{j} 
\end{array}
\}
\]
Hence, $B_{n+1}$ is a function of $B_n$ and $\mathcal{G}_R$. Consequently, if $B_n=B_{n+1}$ for some $n\geq 1$, then $B_m=B_n$ for all $m\geq n$. Clearly, $|B_n|\leq N^2$ for any $n\geq 1$. Hence, the sequence $\{B_n\}_{n\geq 1}$ stabilizes after at most $N^2$ steps, formally: $B_n=B_{N^2}$ for all $n\geq N^2$. Consequently, the implication 
\begin{equation}\label{eq:stability}
(i,j)\in B_{n} \Rightarrow (i,j)\in B_{N^2}
\end{equation}
holds for all $n\geq N^2$. In fact, it is also valid for all $1\leq n<N^2$, since we have $B_{n}\subseteq B_{N^2}$ in this case. Hence, \eqref{eq:stability} holds for all $n\geq 1$.
Next, observe that:
\[ \begin{array}{rcll}
 (i,j)\in B_n & \textrm{iff} & \textrm{there exists a path $\rho$ from $\xxki{0}{i}$ to $\xxki{0}{j}$ in $\mathcal{G}_R^n$} \\
 & \textrm{iff} & R^n(\x,\x') \Rightarrow (x_i-x_j\leq \omega(\rho)) \textrm{ is valid} & \textrm{(by \eqref{dbm-min-paths})} \\
 & \textrm{iff} & (\exists \x' ~.~ R^n(\x,\x')) \Rightarrow (x_i-x_j\leq \omega(\rho)) \textrm{ is valid} \\
 & \textrm{iff} & \exists h ~.~ \forall \x ~.~ (\exists x' ~.~ R^n(\x,\x')) \Rightarrow (x_i-x_j\leq h) \textrm{ is valid}
\end{array} \]
Finally, we combine the above with \eqref{eq:stability} and conclude that for all $n\geq 1$ and for all $1\leq i,j\leq N$, we have:
\[ \begin{array}{ccl}
 (i,j) \in B_n & \Leftrightarrow & \exists h ~.~ \forall \x ~.~ (\exists x' ~.~ R^n(\x,\x')) \Rightarrow (x_i-x_j\leq h) \\
 \Downarrow \\
 (i,j) \in B_{N^2} & \Leftrightarrow & \exists h ~.~ \forall \x ~.~ (\exists x' ~.~ R^{N^2}(\x,\x')) \Rightarrow (x_i-x_j\leq h) \\ 
\end{array} \]
Hence, the lemma holds.
\qed}

Finally, we show that each decreasing function of Lemma \ref{decreasing} is also bounded, concluding that it is a~linear ranking function.

\begin{lem} \label{bounded}
Let $R(\x,\x')$ be a~difference bounds constraint defining a well-founded relation $R\subseteq\zed^\x\times\zed^\x$ such that $R^{5^N}(\x,\x')$ is consistent. Then, there exists a~linear ranking function for $\relRFc$.
\end{lem}
\proof{
Let $\mu.\lambda.\mu'$ be an accepting run from Lemma \ref{decreasing} where $\lambda$ is a~negative-weight cycle of the form $\lambda = q_0\arrow{G_0}{}q_1\arrow{G_1}{}q_2\dots q_{p-1}\arrow{G_{p-1}}{} q_0$ where $p=|\lambda|$. Let $n=|\mu.\lambda.\mu'|$. Further, let $f(\x)$ be the the corresponding linear decreasing function constructed in Lemma \ref{decreasing} from $\lambda$. Recall that $f(\x)$ denotes the negated sum of all unprimed terms in 
\[
\sum\limits_{\substack{0\leq j<p \\ (q_j)_i=r}} (-x'_i+x_i) ~+
\sum\limits_{\substack{0\leq j<p \\ (q_j)_i=\ell}} (-x_i+x'_i)
\]
Hence, for each $0\leq j<p$ and $1\leq i\leq N$, $(q_j)_i$ contributes to $f(\x)$ with terms:
\[\begin{array}{cl}
 \{-x_i\} & \textrm{if $(q_j)_i=r$,} \\
 \{+x_i\} & \textrm{if $(q_j)_i=\ell$,} \\
 \emptyset & \textrm{otherwise.}
\end{array}\]
Let $n \stackrel{def}{=} |\mu.\lambda.\mu'|$. By Proposition \ref{prop:bijection:existence}, for each $0\leq j<p$, there exists a~bijection
\[ \beta_j: \{i ~|~ (q_j)_i=r\} \rightarrow \{i ~|~ (q_j)_i=\ell\} \]
such that, for each $k \in \{i ~|~ (q_j)_i=r\}$:
\[ \exists h ~.~ \forall \x ~.~ (\exists \x' ~.~ R^n(\x,\x')) \Rightarrow (x_{\beta_j(k)} - x_{k}\geq h) \]
By Lemma \ref{unp:dif:bounded}, we then have:
\[ \exists h ~.~ \forall \x ~.~ (\exists \x' ~.~ R^{N^2}(\x,\x')) \Rightarrow (x_{\beta_j(k)} - x_{k}\geq h) \]
Clearly:
\[ f(\x)=\sum_{0\leq j<p}\sum_{ \substack{1\leq k\leq N \\ (q_j)_k=r} } (x_{\beta_j(k)}-x_{k}) \] 
Thus, since each term $x_{\beta_j(i)}-x_{i}$ in the above sum is bounded in $\exists \x' ~.~ R^{N^2}(\x,\x')$, it follows that the sum of these terms is bounded too:
\begin{equation}\label{eq:rf:part1}
 \exists h ~.~ \forall \x ~.~ (\exists \x' ~.~ R^{N^2}(\x,\x')) \Rightarrow f(\x)\geq h 
\end{equation}
By Lemma \ref{decreasing}, we have:
\begin{equation}\label{eq:rf:part2}
 \forall \x,\x' ~.~ R(\x,\x') \Rightarrow f(\x)>f(\x')
\end{equation}
Since strengthening the hypothesis of any implication preserves its validity, we can infer from \eqref{eq:rf:part1} and \eqref{eq:rf:part2} that:
\[ \exists h ~.~ \forall \x,\x' ~.~ \relRFcBr ~\Rightarrow~ f(\x)>f(\x') ~\wedge~ f(\x)\geq h \] 
Thus, $f(\x)$ is a linear ranking function for $\relRFc$.
\qed}

\begin{exa} (Ex.\ \ref{example:rf:decr} ctd.) \label{example:rf:bnd}
We illustrate the boundedness of $f\!=\!-(x_1 \pl x_2 \pl x_3 \mi 3x_4)$, by following the arguments of Lemma \ref{bounded} and Proposition \ref{prop:bijection:existence}.
The cycle $\lambda$ traverses control states $q_0,q_1,q_2$ (see Figure~\ref{fig:bounding:rf}). Let us consider the following bijections $\beta_0,\beta_1,\beta_2$:
\[ \beta_0=\{(1,4)\}, \beta_1=\{(3,4)\}, \beta_2=\{(2,4)\} \]
(the dotted edges in Figure~\ref{fig:bounding:rf}(a) mark these bijections).
Next, we define the paths $\rho_0,\rho_1,\rho_2$ as subpaths of $\rho$ from Figure \ref{fig:bounding:rf}(b)
\[\begin{array}{lcc}
  \rho_0 & \stackrel{def}{=} & \xxki{1}{1}\arrow{0}{}\xxki{2}{3}\arrow{0}{}\xxki{3}{2}\arrow{-1}{}\xxki{4}{1}\arrow{0}{}\xxki{5}{3}\arrow{0}{}\xxki{4}{4}\arrow{0}{}\xxki{3}{4}\arrow{0}{}\xxki{2}{4}\arrow{0}{}\xxki{1}{4}
\\ \rho_1 & \stackrel{def}{=} & \xxki{2}{3}\arrow{0}{}\xxki{3}{2}\arrow{-1}{}\xxki{4}{1}\arrow{0}{}\xxki{5}{3}\arrow{0}{}\xxki{4}{4}\arrow{0}{}\xxki{3}{4}\arrow{0}{}\xxki{2}{4}
\\ \rho_2 & \stackrel{def}{=} & \xxki{3}{2}\arrow{-1}{}\xxki{4}{1}\arrow{0}{}\xxki{5}{3}\arrow{0}{}\xxki{4}{4}\arrow{0}{}\xxki{3}{4}
\end{array}\]
Note that
\[\begin{array}{lcllllllllllllll}
\rho_0 & = & \xxki{1}{1}\arrow{}{}\dots\arrow{}{}\xxki{1}{\beta(1)}
 & V_{\rho_0}\subseteq\bigcup_{\ell=1}^{5}\xk{\ell}
\\ {\rho_0}^{\rightarrow(-1)} & = & \xxki{0}{1}\arrow{}{}\dots\arrow{}{}\xxki{0}{\beta(1)} 
 & V_{\rho_0^{\rightarrow(-1)}}\subseteq\bigcup_{\ell=0}^{5}\xk{\ell}
\end{array}\]
According to Equation \eqref{dbm-min-paths}, existence of the path ${\rho_0}^{\rightarrow(-1)}$ implies that $R^4(\x,\x') \Rightarrow (x_1-x_{\beta_0(1)}) \leq \omega({\rho_0}^{\rightarrow(-1)})=-1$. Clearly, it follows that $(\exists \x' ~.~ R^4(\x,\x')) \Rightarrow (x_{\beta_0(1)}-x_1) \geq 1$. The bijection $\beta_0$ therefore satisfies the required properties. Next, we apply Proposition \ref{unp:dif:bounded} and infer that $(\exists \x' ~.~ R^{N^2}(\x,\x')) \Rightarrow (x_{\beta_0(1)}-x_1) \geq c_0$ for some $c_0\in\zed$. By analogical reasoning, we infer that
\[\begin{array}{lcl}
(\exists \x' ~.~ R^{N^2}(\x,\x')) \Rightarrow (x_{\beta_1(3)}-x_3) \geq c_1 & \textrm{for some $c_1\in\zed$}
\\
(\exists \x' ~.~ R^{N^2}(\x,\x')) \Rightarrow (x_{\beta_2(2)}-x_2) \geq c_2 & \textrm{for some $c_2\in\zed$}
\end{array}\]
Then, we infer:
\[\begin{array}{cl}
 & (x_{\beta_0(1)}-x_1) \geq c_0 ~\wedge~ (x_{\beta_1(3)}-x_3) \geq c_1 ~\wedge~ (x_{\beta_2(2)}-x_2) \geq c_2
\\ \iff & (x_4-x_1) \geq c_0 ~\wedge~ (x_4-x_3) \geq c_1 ~\wedge~ (x_4-x_2) \geq c_2
\\ \Rightarrow & (x_4-x_1) + (x_4-x_3) + (x_4-x_2) \geq c_0 + c_1 + c_2
\\ \iff & f(\x) \geq c_0 + c_1 + c_2
\end{array}\]
Hence, $(\exists \x' ~.~ R^{N^2}(\x,\x')) \Rightarrow f(\x) \geq c_0+c_1+c_2$ and thus, $f(\x)$ is bounded. Example \ref{example:rf:decr} demonstrated that $f(\x)$ is decreasing. We conclude that $f(\x)$ is a~ranking function.


As an experiment, we have tried the \textsc{iRankFinder}
\cite{ben-amram-genaim-popl13} tool (complete for integer linear
ranking functions), which failed to discover a~ranking function on
this example. This comes with no surprise, since no linear decreasing
function that is bounded after the first iteration exists. However,
\textsc{iRankFinder} finds a~linear ranking function for the witness
relation $\relRFc$ instead. Interestingly, the linear ranking function
found by \textsc{iRankFinder} differs from the one computed in this
example only by a constant. \qed\end{exa}

\subsubsection{Linear Ranking Functions for Octagonal Relations}
\label{sec:rf:oct}

In the rest of this section, let us fix the sets of variables $\x=\{x_1,\dots,x_N\}$ and $\y=\{y_1,\dots,y_{2N}\}$ for some constant $N\geq 1$. The following proposition gives a~way to construct a~linear ranking function for an octagonal relation $R(\x,\x')$ from any linear ranking function for its difference bounds representation $\overline{R}(\y,\y')$.

\begin{prop} \label{prop-rf-db-to-oct-1}
  Let $R(\x,\x')$ be an octagonal constraint, $\overline{R}(\y,\y')$ be its difference bounds encoding and let $\overline{f}(\y)$ be a~linear ranking function for $\overline{R}(\y,\y')$. Then, the function $f(\x) \stackrel{def}{=} \overline{f}(\y)[x_i/y_{2i-1},-x_i/y_{2i}]_{i=1}^{N}$, is a~linear ranking function for $R(\x,\x')$.
\end{prop}
\proof{
Clearly, $f(\x)$ is linear by definition. We have the following equivalences: 
\[ \begin{array}{rl}
R(\x,\x') \iff \overline{R}(\y,\y')[x_i/y_{2i-1},-x_i/y_{2i}]_{i=1}^{N} & \textrm{(by Equation \eqref{dbc-to-oct})} \\
f(\x)=\overline{f}(\y)[x_i/y_{2i-1},-x_i/y_{2i}]_{i=1}^{N} & \textrm{(by definition of $f(\x)$)}
\end{array} \]
Since $\overline{f}(\y)$ is a linear ranking function for $\overline{R}(\y,\y')$, the following formula is valid:
\[ \exists h ~.~ \forall \y,\y' ~.~ \overline{R}(\y,\y') ~\Rightarrow~ \overline{f}(\y) > \overline{f}(\y') ~\wedge~ \overline{f}(\y) \geq h \]
Clearly, its validity is preserved under the substitution $[x_i/y_{2i-1},-x_i/y_{2i}]_{i=1}^{N}$ and thus
\[ \exists h ~.~ \forall \x,\x' ~.~ \forall R(\x,\x') ~\Rightarrow~ f(\x) > f(\x') ~\wedge~ f(\x) \geq h \]
is valid too. Hence, $f(\x)$ is a linear ranking function for $R(\x,\x')$.
\qed}

The next proposition generalizes Proposition \ref{prop-rf-db-to-oct-2} and shows how to construct a~linear ranking function for an octagonal relation $R(\x,\x') \wedge \exists \x' . R^n(\x,\x')$ from any linear ranking function for the difference bounds relation $\overline{R}(\y,\y') \wedge \exists \y' ~.~ \overline{R}^{\,n}(\y,\y')$.

\begin{prop} \label{prop-rf-db-to-oct-2}
Let $R(\x,\x')$ be an~octagonal constraint, $\overline{R}(\y,\y')$ be its difference bounds encoding and let $\overline{f}(\y)$ be a~linear ranking function for $\overline{R}(\y,\y') \wedge \exists \y' ~.~ \overline{R}^{\,n}(\y,\y')$, for a~fixed $n\geq 1$. Then, $f(\x) \stackrel{def}{=} \overline{f}(\y)[x_i/y_{2i-1},-x_i/y_{2i}]_{i=1}^{N}$ is a~linear ranking function for $R(\x,\x') \wedge \exists \x' . R^n(\x,\x')$.
\end{prop}
\proof{
Let us first define the following substitution 
\[ \sigma \stackrel{def}{=} [\xxki{0}{i}/\yyki{0}{2i-1},-\xxki{0}{i}/\yyki{0}{2i},\xxki{n}{i}/\yyki{n}{2i-1},-\xxki{n}{i}/\yyki{n}{2i-1}]_{i=1}^{2N} \]
Next, observe that (the second equivalence is by Proposition \ref{odbc:qelim})
\[
\begin{array}{rcll}
R^n(\xk{0},\xk{n}) & \iff & \exists \xk{1},\dots,\xk{n-1} ~.~ \bigwedge_{i=0}^{n-1}R(\xk{i},\xk{i+1}) \\
  & \iff & \big[\exists \yk{1},\dots,\yk{n-1} ~.~ \bigwedge_{i=0}^{n-1}\overline{R}(\yk{i},\yk{i+1})\big][\sigma] \\
  & \iff & \overline{R}^{\,n}(\yk{0},\yk{n})[\sigma]
\end{array}
\]
Consequently, we have:
\begin{equation}\label{eq:rf:oct:0}
\overline{R^n}(\y,\y') \iff \overline{R}^{\,n}(\y,\y')
\end{equation}
Observe that 
\begin{equation}\label{eq:rf:oct:1}
\begin{array}{rcll}
 \overline{\exists \x' ~.~ R^n(\x,\x')} & \iff &
  \overline{ \big(\exists \y' ~.~ \overline{R^n}(\y,\y')\big)[x_i/y_{2i-1},-x_i/y_{2i}]_{i=1}^{N} } 
  & \textrm{(by Proposition \ref{odbc:qelim})} \\
  & \iff & \exists \y' ~.~ \overline{R^n}(\y,\y')
  \\
  & \iff & \exists \y' ~.~ \overline{R}^{\,n}(\y,\y')
  & \textrm{(by Equation \eqref{eq:rf:oct:0})}
\end{array}
\end{equation}
Consequently, we have:
\[
\begin{array}{rcll}
 \overline{R(\x,\x') \wedge \exists \x' ~.~ R^n(\x,\x')} & \iff &
  \overline{R}(\y,\y') \wedge \overline{ \exists \x' ~.~ R^n(\x,\x')}
  \\
  & \iff & \overline{R}(\y,\y') \wedge \exists \y' ~.~ \overline{R}^{\,n}(\y,\y')
  & \textrm{(by Equation \eqref{eq:rf:oct:1})}
\end{array}
\]
Thus, since $\overline{f}(\y)$ is a~linear ranking function for $\overline{R}(\y,\y') \wedge \exists \y' ~.~ \overline{R}^n(\y,\y')$, then $f(\x)$ is a~linear ranking function for $R(\x,\x') \wedge \exists \x' ~.~ R^n(\x,\x')$, by Proposition \ref{prop-rf-db-to-oct-1}.
\qed}

Finally, we can combine the above results into the main theorem.

\begin{thm}\label{theorem:rf}
Let $R\subseteq\zed^\x\times\zed^\x$ be a relation defined by an octagonal constraint $R(\x,\x')$ and let $V\subseteq\zed^\x\times\zed^\x$ be a relation defined by
\[ 
 V(\x,\x') ~\equiv~ \relRFd
\]
Then, $R$ is well founded if and only if $V$ is well founded if and only if $V(\x,\x')$ has a~linear ranking function. Moreover, both $V(\x,\x')$ and the linear ranking function are computable in polynomial time.
\end{thm}
\proof{
The fact that $R(\x,\x')$ is well founded if and only if $V(\x,\x')$ is well founded follows from Lemma \ref{strenghten}. Thus, if $R(\x,\x')$ is not well founded, neither is $V(\x,\x')$ and hence, $V(\x,\x')$ has no (linear) ranking function. In the rest of the proof, we show that if $R(\x,\x')$ is well founded, then there exists a~linear ranking function for $V(\x,\x')$. As a~first subcase, suppose that $R^{5^{2N}}(\x,\x')$ is inconsistent. Then clearly, $V(\x,\x')$ is inconsistent too and, trivially, $V(\x,\x')$ has a~linear ranking function. As a~second subcase, suppose that $R^{5^{2N}}(\x,\x')$ is consistent. By Proposition \ref{oct:to:dbm:consistency}, $\overline{R}^{5^{2N}}(\y,\y')$ is consistent too. Since $R$ is well founded, $\overline{R}$ is well founded too, by Lemma \ref{lem:core:results}. Then, by Lemma \ref{bounded}, there exists a~linear ranking function $\overline{f}$ for $\relRFcDbOct$. By Proposition \ref{prop-rf-db-to-oct-2}, the function defined as $f \stackrel{def}{=} \overline{f}[x_i/y_{2i-1},-x_i/y_{2i}]_{i=1}^{N}$ is a~linear ranking function for $\relRFcOct$, formally:
\begin{equation}\label{th:rf:eq:1}
\exists h ~.~ \forall \x,\x' ~.~ \relRFcOctBr ~\Rightarrow~ f(\x) > f(\x') ~\wedge~ f(\x) \geq 0
\end{equation}
Since $4N^2 < 5^{2N}$ for all $N\geq 1$, then $\pre_R^{4N^2}(\zed^\z) \supseteq \pre_R^{5^{2N}}(\zed^\x)$, by Proposition \ref{prop:pre:properties}. Consequently, $\exists \x' ~.~ R^{5^{2N}}(\x,\x') \Rightarrow \exists \x' ~.~ R^{4N^2}(\x,\x')$ and therefore
\begin{equation}\label{th:rf:eq:2}
\relRFdBr \Rightarrow \relRFcOctBr
\end{equation}
Combining \eqref{th:rf:eq:1} with \eqref{th:rf:eq:2}, we infer that $f(\x)$ is a linear ranking function for $\relRFd$.

By Lemma \ref{lem:fast:power}, $V$ can be computed in at most $\mathcal{O}(N^4\cdot(N+\log_2 \maxcoef{R}))$ time and moreover, $\maxcoef{V}$ is of the order $\mathcal{O}(\maxcoef{R} \cdot N \cdot 2^{N})$. Consistency of $V(\x,\x')$ can then be checked in at most $\mathcal{O}(N^3 \cdot (N+ \log_2\maxcoef{V})) = \mathcal{O}(N^3 \cdot (N+ \log_2\maxcoef{R}))$ time, by Corollary \ref{oct:consistency:test}. If $V \iff \textbf{false}$, one can return an arbitrary linear function $f(\x)$. Otherwise, if $V \not\iff \textbf{false}$, one can compute $\overline{V} \equiv \relRFcDbOct$, again in at most $\mathcal{O}(N^3 \cdot (N+ \log_2\maxcoef{R}))$ time, as a consequence of Proposition \ref{odbc:qelim}, Proposition \ref{oct-eq}, and Corollary \ref{oct:consistency:test}. Then, a~linear ranking function for $\overline{R}$ can be computed in time that is polynomial in the bit-size of $V(\x,\x')$, as proved in \cite{ben-amram-genaim-popl13} (see Corollary~4.8 in Section~4.1). 
It follows easily from Definition \ref{odbc} that $V(\x,\x')$ can be represented using $\mathcal{O}(\log_2(\maxcoef{V})\cdot 3\cdot(2N)^2) = \mathcal{O}(N^2\cdot (\log_2 N + \log_2 \maxcoef{R}))$ bits. Thus, the time needed to compute $\overline{f}$ is polynomial in $\mu_R$ and $N$. Finally, one computes $f \stackrel{def}{=} \overline{f}[x_i/y_{2i-1},-x_i/y_{2i}]_{i=1}^{N}$, again in polynomial time.
\qed}

\section{Linear Affine Relations}
\label{sec:termination:affine}

The previous section was concerned with computing weakest
non-termination preconditions for non-deterministic integer relations
(octagonal relations). Here, we present linear affine relations which
are a~general model of deterministic transition relations. Linear
affine relations are conjunctions of equalities of the form $x' =
a_1x_1 + \ldots + a_nx_n + b$, where $a_1,\ldots,a_n \in \zed$ are
integer coefficients, and Presburger definable conditions on the
unprimed variables $x_1,\ldots,x_n$. First, we show that the weakest
recurrent set of a linear affine relation $R$ can be computed as the
limit of a descending Kleene sequence $pre_R(\zed^{\vec{x}}) \supseteq
pre^2_R(\zed^{\vec{x}}) \supseteq \ldots$. Second, this set can be
defined in Presburger arithmetic for a~subclass of affine relations
with the {\em finite monoid property} (Section
\ref{sec:wnt:monoids}). Finally, we relax the finite monoid condition
and describe a~method for generating sufficient termination
conditions, i.e.\ sets $S \in \zed^{\vec{x}}$ such that $S \cap \wrs(R) =
\emptyset$, for the class of {\em polynomially bounded} affine
relations (Section \ref{sec:termination:poly:bounded}).

\begin{defi} \label{def:affine:rel}
Let $\vec{x}=\langle x_1,\ldots,x_N \rangle$ be a~vector of variables
ranging over $\zed$. A relation $R \subseteq \zed^\x \times \zed^\x$
is said to be an \emph{affine relation} if it can be defined by
a~formula $R(\x,\x')$ of the form:
\begin{equation}\label{affine-transform}
  R(\x,\x') ~\iff~ \vec{x'} = A\times \vec{x} + \vec{b} ~\wedge~ \phi(\vec{x})
\end{equation}
where $A \in \zed^{N \times N}$, $\vec{b} \in \zed^N$, and $\phi$ is
a~ quantifier-free Presburger formula over unprimed variables only,
called the \emph{guard} of $R$. The formula $\vec{x'} = A\times
\vec{x} + \vec{b}$, defining a~ linear transformation, is called the
\emph{update} of $R$.
\end{defi}

\subsection{\bf Background on Linear Algebra}

We first recall several notions of linear algebra, needed in the
following. For a comprehensive textbook on linear algebra, we refer to
\cite{schrijver}. A complex number $r$ is said to be a~\emph{root of
  the unity} if $r^d = 1$ for some integer $d > 0$. If $A \in \zed^{n
  \times n}$ is a~square matrix, and $\vec{v} \in \zed^n$ is a~column
vector of integer constants, then any complex number $\lambda \in
\complex$ such that $A \vec{v} = \lambda \vec{v}$, for some complex
vector $\vec{v} \in \complex^n$, is called an \emph{eigenvalue} of
$A$. The vector $\vec{v}$ in this case is called an \emph{eigenvector}
of $A$. It is known that the eigenvalues of $A$ are the roots of the
\emph{characteristic polynomial} $P_A(x)=\mbox{det}(A - x I_n) = 0$,
which is an effectively computable univariate polynomial. The
\emph{minimal polynomial} of $A$ is the polynomial $\mu_A$ of lowest
degree such that $\mu_A(A) = 0$. By the Cayley-Hamilton Theorem, the
minimal polynomial always divides the characteristic polynomial,
i.e.\ the roots of the former are root of the latter.

If $\lambda_1,\ldots,\lambda_m$ are the eigenvalues of $A$, then
$\lambda_1^p, \ldots, \lambda_m^p$ are the eigenvalues of $A^p$, for
all integers $p > 0$. A~matrix is said to be \emph{diagonalizable} if
and only if there exists a~non-singular matrix $U \in \complex^{N
  \times N}$ and a~diagonal matrix with the eigenvalues
$\lambda_1,\ldots,\lambda_m$ occurring on the main diagonal, such that
$A = U \times D \times U^{-1}$. This is the case if and only if
$\mu_A$ has only roots of multiplicity one.\footnote{See e.g. Thm 8.47
  in \cite{boigelot-thesis}.}

\subsection{\bf Termination Preconditions for Deterministic Relations}
First, we show that the pre-image function of a deterministic relation
is $\cap$-continuous. Since affine transformations are deterministic,
this means that their weakest non-termination preconditions can be
computed as limits of descending Kleene sequences. Let $\vec{x}$ be a
set of variables in the following.

\begin{lem} \label{join-continuous}
  Let $R \subseteq \zed^{\vec{x}} \times \zed^{\vec{x}}$ be
  a~deterministic relation. Then, $\pre_R$ is $\cap$-continuous.
\end{lem}
\proof{ Let $I=\{0,\dots,d\}$, $d\in\nat_\infty$, and $\{S_i \subseteq
  \zed^{\vec{x}}\}_{i\in I}$ be a~potentially infinite collection of
  sets. We prove that:
  \[ \begin{array}{c} 
    \pre_R(\bigcap_{i\in I} S_i) = \bigcap_{i\in I} \pre_R(S_i)
    \textrm{.} 
  \end{array} \]
  ``$\subseteq$'' By the monotonicity of $\pre_R$ (Proposition
  \ref{prop:pre:properties}), we have $\pre_R(\bigcap_{i\in I} S_i)
  \subseteq \pre_R(S_i)$ for all $i\in I$ and hence,
  $\pre_R(\bigcap_{i\in I} S_i) \subseteq \bigcap_{i\in I}
  \pre_R(S_i)$.  

  \smallskip\noindent ``$\supseteq$'' Let $v\in \bigcap_{i\in I}
  \pre_R(S_i)$. Then, there exists $v_i\in S_i$ such that $(v,v_i) \in
  R$ for all $i\in I$. Since $R$ is deterministic, then $v_0=v_i$ for
  all $i\in I$ and hence $v_0\in \bigcap_{i\in I} S_i$.  Consequently,
  $v\in \pre_R(\bigcap_{i\in I} S_i)$.  \qed}

For the rest of this section, we extend the notion of {\em closed
  form} (Definition \ref{closed-form-def}) from sequences of sets $S
\subseteq \zed^\x$ to sequences of powers of relations $R \subseteq
\zed^\x \times \zed^\x$.

\begin{defi}\label{closed-form-rel}
  Let $R \subseteq \zed^\x \times \zed^\x$ be a relation. The {\em
    closed form} of $R$ is a formula $\widehat{R}(k,\x,\x')$ such
  that, for all $n \geq 1$ and all $\nu,\nu' \in \zed^\x$:
  $$(\nu,\nu') \in R^n \iff (\nu,\nu') \models \widehat{R}[n/k]$$
\end{defi}

Next, we prove that the closed form of a deterministic relation can be
defined in Presburger arithmetic whenever the closed form of its
update can be defined in Presburger arithmetic. Concretely, whenever
the logical definition of a relation $R$ can be split into a guard and
a deterministic update, and the closed form of $R$ can be computed
based on the closed form of the update.
\begin{lem}\label{det-closed-form}
  Let $R \subseteq \zed^\x \times \zed^\x$, $\x = \{x_1,\ldots,x_N\}$,
  be a deterministic relation and $\varphi(\x)$ be a~guard. Then the
  closed form of the relation defined by the formula $R(\x,\x') \wedge
  \varphi(\x)$ is:
  $$(\widehat{R \wedge \varphi})(k, \vec{x},\vec{x'}) \iff
  \widehat{R}(k,\x,\x') \wedge \forall 1 \leq \ell < k ~\exists \y ~.~
  \widehat{R}(\ell,\x,\y) \wedge \varphi(\y)$$ where $\widehat{R}$ is
  the closed form of $R$ and $\y = \{y_1,\ldots,y_N\}$.
\end{lem}
\proof{``$\Rightarrow$'' Let $\nu,\nu' \in \zed^{\vec{x}}$ be a~pair
  of valuations, such that $(\nu,\nu') \models (R \wedge \varphi)^n$,
  for some integer $n \geq 1$. Then we also have $(\nu,\nu') \models
  (\widehat{R \wedge \varphi})[n/k]$. Consequently, there exists
  a~sequence of valuations $\nu=\nu_0,\nu_1,\ldots,\nu_n=\nu' \in
  \zed^{\vec{x}}$, such that $(\nu_i,\nu_{i+1}) \models R \wedge
  \varphi$. By Definition \ref{closed-form-rel}, we have that $(\nu_0,
  \nu_n) \models \widehat{R}[n/k]$ and $(\nu_0,\nu_i) \models
  (\widehat{R} \wedge \varphi)[i/k]$, for all $i = 0,\ldots,n-1$.

  \smallskip\noindent ''$\Leftarrow$'' Let $\nu, \nu' \in
  \zed^{\vec{x}}$ be two valuations such that:
  \begin{itemize}
  \item $(\nu,\nu') \models \widehat{R}[n/k]$ for some
    $n \geq 1$ and, 
  \item for all $i = 0, \ldots, n-1$ there exists a valuation $\nu_i
    \in \zed^\x$ such that $(\nu,\nu_i)\models \widehat{R}[i/k]$ and
    $\nu_i \models \varphi$. 
  \end{itemize}
  Since $\widehat{R}[n/k]$ defines $R^n$, by Definition
  \ref{closed-form-rel}, there exists a~ sequence of valuations
  $\nu=\nu'_0,\nu'_1,\ldots,\nu'_n=\nu' \in \zed^{\vec{x}}$ such that
  $(\nu'_i,\nu'_{i+1}) \models R$. By the fact that $R$ was assumed to
  be deterministic, we have $\nu_i = \nu'_i$ for all $i=0,\ldots,n-1$,
  hence $\nu'_i \models \varphi$, for all $i=0,\ldots,n-1$. Clearly
  then $(\nu,\nu') \models (\widehat{R \wedge \varphi})[n/k]$. \qed}

Since linear affine relations are deterministic (Definition
\ref{def:affine:rel}), by Lemma \ref{join-continuous} they are also
$\cap$-continuous, and the weakest recurrent set of an arbitrary
linear affine relation $R$ can be computed as $\wrs(R)=\bigcap_{m\geq
  0} \pre^m_R(\zedToX)$, by Lemma \ref{wrs-kleene}. Hence, the weakest
recurrent set can be defined using the closed form of $R$:
\[ (\wrs(R))(\vec{x}) \iff \forall k \geq 1 ~.~ 
\exists \vec{x'} ~.~ \widehat{R}(k,\vec{x},\vec{x'}) \]
Considering that the formula defining $R$ is of the form $R_u(\x,\x')
\wedge \varphi(\x)$ where $R_u(\x,\x')$ is a~deterministic update and
$\varphi(\x)$ is a~Presburger guard, we can write the closed form of
$R$ as:
\[ 
  \widehat{R}(k,\x,\x') \iff \widehat{R}_u(k,\vec{x},\vec{x'}) \wedge
  \forall 1 \leq \ell < k ~\exists \vec{y} ~.~
  \widehat{R}_u(\ell,\vec{x},\vec{y}) \wedge \varphi(\vec{y})
\]
by Lemma \ref{det-closed-form}. Then, the definition of the weakest
recurrent set of a~linear affine relation is (after the elimination of
the trailing existential quantifier and renaming $\ell$ with $k$ and
$\y$ with $\x'$):
\begin{equation}\label{lin-wrs}
  (\wrs(R))(\vec{x}) ~\iff~ \forall k \geq 1 ~.~ \exists \vec{x}' ~.~
  \widehat{R}_u(k,\vec{x},\vec{x}') \wedge \varphi(\vec{x}')
\end{equation}

\subsection{Finite Monoid Affine Relations}
\label{sec:wnt:monoids}

The class of finite monoid affine relations was the first class of
integer relations for which the transitive closure has been shown to
be Presburger definable, by Boigelot
\cite{boigelot-thesis}. Informally, an affine relation is a~finite
monoid relation if the set of powers of its transformation matrix is
finite. Originally, Boigelot characterized this class by two decidable
conditions in \cite{boigelot-thesis} (we report on these conditions in
Theorem \ref{finite-monoid-conditions}). Later, Finkel and Leroux
noticed in \cite{finkel-leroux} that Boigelot's conditions correspond
to the finite monoid property, which is also known to be decidable
\cite{mandel77}.

Given a vector $\vec{x} = \langle x_1,\ldots,x_N \rangle$ of
variables, an affine transformation
$$R(\x,\x') ~\iff~ \x' = A\times \x + \vec{b} ~\wedge~ \varphi(\x)$$
where $A \in \zed^{N \times N}$, $\vec{b} \in \zed^N$, is said to have
the \emph{finite monoid property} \cite{boigelot-thesis,finkel-leroux}
if the monoid of powers of $A$, denoted as $\langle \mathcal{M}_A,
\times \rangle$, where $\mathcal{M}_A = \{A^i ~|~ i \geq 0\}$, is
finite. Here ${A}^0 = I_N$ and ${A}^i = A\times {A}^{i-1}$, for $i >
0$. It has been shown in \cite{finkel-leroux} that the finite monoid
property can be equivalently characterized by the following two
conditions.

\begin{thm}[\cite{boigelot-thesis,finkel-leroux}]\label{finite-monoid-conditions} 
An affine transformation $R(\x,\x') \iff A\times \vec{x} + \vec{b}
\wedge \varphi(\x)$, where $A \in \zed^{N \times N}$ and $\vec{b} \in
\zed^N$, has the finite monoid property if and only if there exists $p
> 0$ such that the following hold:
\begin{enumerate}
\item every eigenvalue of $A^p$ belongs to the set $\{0,1\}$, and
\item the minimal polynomial $\mu_{A^p}(x)$ of $A^p$ belongs to the
  set $\{0,x,x-1,x(x-1)\}$ (or, equivalently, $A^p$ is diagonalizable).
\end{enumerate}
\end{thm}
Both conditions in the above theorem are decidable
\cite{boigelot-thesis,mandel77}. It was shown in
\cite{boigelot-thesis,finkel-leroux,cav10} that the closed form of
(the update part of) a linear affine transformation with the finite
monoid property is Presburger definable. This entails the decidability
of the universal termination problem for finite monoid affine relations.

\begin{thm}\label{theorem:term:monoids}
  The weakest non-termination precondition of a~finite monoid affine
  relation is Presburger definable and effectively
  computable. Consequently, the termination problem is decidable for
  finite monoid affine relations.
\end{thm}
\proof{Let $R \subseteq \zed^\x \times \zed^\x$ be a~finite monoid
  affine relation defined by a formula $R_u(\x,\x') \wedge
  \varphi(\x)$. By Equation (\ref{lin-wrs}) we have:
  $$(\wrs(R))(\x) \iff \forall k \geq 1 ~.~ \exists \vec{x}' ~.~
  \widehat{R}_u(k,\vec{x},\vec{x}') \wedge \varphi(\vec{x}')$$ Since
  both $\widehat{R}_u(k,\vec{x},\vec{x}')$ and $\varphi(\x')$ are
  Presburger formulas, $\wrs(R)(\x)$ is a Presburger formula as well.
  Since Presburger arithmetic is decidable \cite{presburger29}, the
  termination problem can be decided by checking whether $\wrs(R) =
  \emptyset$. \qed}

\subsection{Polynomially Bounded Affine Relations}
\label{sec:termination:poly:bounded}
In the following, we study another subclass of affine relations with
linear guards and transformation matrices whose eigenvalues are either
zero or roots of the unity.
\begin{defi} \label{def:poly:bounded}
If $\vec{x}=\langle x_1,\ldots,x_N \rangle$ is a~vector of variables
ranging over $\zed$, a {\em polynomially bounded affine relation} is
a~relation defined by a formula of the form:
\begin{equation}\label{poly-bounded-affine-transform}
  R(\x,\x') ~\iff~ \vec{x'} = A\times \vec{x} + \vec{b} ~\wedge~
  C\vec{x} \geq \vec{d}
\end{equation}
where $A \in \zed^{N \times N}$, $C \in \zed^{P \times N}$ are
matrices, and $\vec{b} \in \zed^N$, $\vec{d} \in \zed^P$ are column
vectors of integer constants, for some $P > 0$, and moreover, all
eigenvalues of $A$ are either zero or roots of the unity.
\end{defi}
\noindent
Note that, if $A$ is a~finite monoid matrix, then all eigenvalues of
$A$ are either zero or roots of the unity. Thus, the condition on $A$
is weaker for polynomially bounded affine relations.  However, since
the guard of finite monoid relations is more general (Presburger), the
two classes are incomparable.

The closed form of polynomially bounded affine relations cannot be
defined in Presburger arithmetic\footnote{The closed form
  $\widehat{R}(k,\x,\x')$ of a polynomially bounded affine relation is
  defined by polynomial functions in $k$, of arbitrary degrees. It is
  possible to show that a polynomial function of degree greater than
  one is not Presburger definable \cite{GS1966}) .}, thus we renounce
defining $\wrs(R)$ precisely, and content ourselves with the discovery
of {\em sufficient conditions for termination}. Basically, given
a~linear affine relation $R$, we aim at finding a~disjunction
$\phi(\vec{x})$ of linear constraints on $\vec{x}$, such that $\phi
\wedge \wrs(R)$ is inconsistent without explicitly computing
$\wrs(R)$. For this, we use several existing results from linear
algebra (see, e.g., \cite{everest03}).  In the following, it is
convenient to work with the equivalent homogeneous form:
\begin{equation}\label{lin-affine-h}
\begin{array}{c}
R(\vec{x},\vec{x'}) \equiv C_h\vec{x}_h \geq \vec{0} ~\wedge~
\vec{x'}_h = A_h \vec{x}_h ~\wedge~ x_{N+1} = 1 \\\\
A_h = \left(\begin{array}{cc}
    A~& \vec{b} \\
    0 & 1 \\
    \end{array}\right)~ 
C_h = \left(\begin{array}{cc}
    C & -\vec{d} 
    \end{array}\right)~ 
\vec{x}_h = \left(\begin{array}{c} \vec{x} \\
           x_{N+1} 
           \end{array}\right)
\end{array}
\end{equation}
The weakest recurrent set of $R$ can be then defined as:
\begin{equation}\label{lin-wrs-poly-bounded}
(\wrs(R))(\vec{x}) ~\equiv~ \exists x_{N+1} ~.~ \forall k \geq 0 ~.~ 
C_h A_h^k \vec{x}_h \geq \vec{0} ~\wedge~ x_{N+1} = 1
\end{equation}

\begin{defi}
A function $f : \nat \rightarrow \complex$ is said to be a~{\em
  C-finite recurrence} if and only if:
$$f(n+d) = a_{d-1}f(n+d-1) + \ldots + a_1f(n+1) + a_0f(n), ~\forall
n \geq 0$$ for some $d \in \nat$ and
$a_0,a_1,\ldots,a_{d-1} \in \complex$, with $a_{d-1}
\neq 0$. The polynomial $x^d - a_{d-1}x^{d-1} - \ldots a_1x - a_0$ is
called the {\em characteristic polynomial} of $f$.
\end{defi}
A C-finite recurrence always admits a~closed form.
\begin{thm}[\cite{everest03}]\label{c-finite-closed}
The closed form of a~C-finite recurrence is:
$$f(n) = p_1(n)\lambda_1^n + \ldots + p_s(n)\lambda_s^n$$ where
$\lambda_1,\ldots,\lambda_s \in \complex$ are non-zero distinct roots
of the characteristic polynomial of $f$, and $p_1,\ldots,p_s \in
\complex[n]$ are polynomials of degree less than the multiplicities of
$\lambda_1, \ldots, \lambda_s$, respectively.
\end{thm}
Next, we define the closed form for the sequence of powers of $A$. 
\begin{cor}\label{matrix-powers-closed}
Given a~square matrix $A \in \zed^{N \times N}$, we have, for all $n >
0$: $$(A^n)_{i,j} = p_{1,i,j}(n) \lambda_1^n + \ldots + p_{s,i,j}(n)
\lambda_s^n$$ where $\lambda_1,\ldots,\lambda_s \in \complex$ are
non-zero distinct eigenvalues of $A$, and $p_{1,i,j},\ldots,p_{s,i,j}
\in \complex[n]$ are polynomials of degree less than the
multiplicities of $\lambda_1,\ldots,\lambda_s$, respectively.
\end{cor}
\proof{If $\mbox{det}(A - xI_n) = x^d - a_{d-1}x^{d-1} - \ldots - a_1x
  - a_0$ is the characteristic polynomial of $A$, then we have $$A^d -
  a_{d-1}A^{d-1} - \ldots - a_1A - a_0 = 0$$ by the Cayley-Hamilton
  Theorem. If we define $f_{i,j}(n) = (A^n)_{i,j}$, for all $n > 0$,
  by multiplying the above equality with $A^n$, we obtain:
  \[\begin{array}{rcl}
  A^{n+d} & = & a_{d-1}A^{n+d-1}+ \ldots +a_1A^{n+1}+a_0A^n \\
  f_{i,j}(n+d) & = & a_{d-1}f_{i,j}(n+d-1)+\ldots+a_1f_{i,j}(n+1)+a_0f_{i,j}(n)  
  \end{array}\]
  By Theorem \ref{c-finite-closed}, we have that
  $$(A^n)_{i,j} = p_{1,i,j}(n) \lambda_1^n + \ldots + p_{s,i,j}(n)
  \lambda_s^n$$ for some polynomials $p_{1,i,j},\ldots,p_{s,i,j} \in
  \complex[n]$ of degrees less than the multiplicities of
  $\lambda_1,\ldots,\lambda_s$, respectively. \qed}

\begin{lem}\label{rational-poly}
Given a~square matrix $A \in \zed^{N \times N}$, whose non-zero
eigenvalues are all roots of the unity. Then
$(A^n)_{i,j} \in \rat[n]$, for all $1 \leq i,j \leq N$, are
effectively computable polynomials with rational coefficients.
\end{lem}
\proof{ Assume from now on that all non-zero eigenvalues
  $\lambda_1,\ldots,\lambda_s$ of $A$ are such that $\lambda_1^{d_1} =
  \ldots = \lambda_s^{d_s} = 1$, for some integers $d_1,\ldots,d_s >
  0$. The method given in \cite{boigelot-thesis} for testing the
  finite monoid condition for $A$ gives also bounds for
  $d_1,\ldots,d_s$. Then we have $\lambda_1^L = \ldots \lambda_s^L =
  1$, where $L=\mbox{lcm}(d_1, \ldots, d_s)$. As $d_1,\ldots,d_s$ are
  effectively bounded, so is $L$. By Corollary
  \ref{matrix-powers-closed}, we have that, if $n$ is a~multiple of
  $L$, then $(A^n)_{i,j}=p_{i,j}(n)$ for some effectively computable
  polynomial $p_{i,j} \in \complex[n]$, of degree $d_{ij} > 0$,
i.e.\ for $n$ multiple of $L$, $A^n$ is polynomially definable. But since
  $p_{i,j}(n)$ assumes real values in an infinity of points $n = kL,~
  k > 0$, it must be that its coefficients are all real numbers,
  i.e.\ $p_{i,j} \in \real[n]$. Moreover, these coefficients are the
  solutions of the integer system:
\[\left\{\begin{array}{ccc}
p_{i,j}(L) & = & (A^L)_{i,j} \\ & \ldots & \\ p_{i,j}((d_{ij}+1)L) & =
& (A^{(d_{ij}+1)L})_{i,j}
\end{array}\right.\] 
Clearly, since $A \in \zed^{N \times N}$, $A^p \in \zed^{N \times N}$,
for any $p > 0$. Hence $p_{i,j} \in \rat[n]$.  \qed}

We turn now back to the problem of defining $\wrs(R)$ for linear
affine relations $R$ of the form (\ref{lin-wrs-poly-bounded}). First
notice that, if all non-zero eigenvalues of $A$ are roots of the
unity, then the same holds for $A_h$ (\ref{lin-affine-h}). By Lemma
\ref{rational-poly}, one can find rational polynomials $p_{i,j}(k)$
defining $(A_h^k)_{i,j}$, for all $1 \leq i,j \leq N$. The condition
(\ref{lin-wrs-poly-bounded}) becomes a~conjunction of the form:
\begin{equation}\label{poly-system}
(\wrs(R))(\vec{x}) \equiv \bigwedge_{i=1}^n \forall k \geq 1 ~.~ P_i(k,
\vec{x}) \geq 0
\end{equation}
where each $P_i = a_{i,d}(\vec{x}) \cdot k^d + \ldots +
a_{i,1}(\vec{x}) \cdot k + a_{i,0}(\vec{x})$ is a~polynomial in $k$
whose coefficients are the linear combinations $a_{i,d} \in
\rat[\vec{x}]$. We are looking for a~sufficient condition for
termination, which is, in this case, any set of valuations of
$\vec{x}$ that would invalidate (\ref{poly-system}).  The following
proposition gives sufficient invalidating clauses for each conjunct
above. By taking the disjunction of all these clauses we obtain
a~sufficient termination condition for $R$.

\begin{lem}\label{poly-neg}
Given a~polynomial $P(k,\vec{x}) = a_d(\vec{x}) \cdot k^d + \ldots +
a_1(\vec{x}) \cdot k + a_0(\vec{x})$, for each valuation $\nu \in
\zed^\x$ there exists an integer $n > 0$ such that $P(n,\nu(\vec{x}))
< 0$ if, for some $i = 0,1,\ldots,d$, we have $a_{d-i}(\nu(\vec{x})) <
0$ and $a_d(\nu(\vec{x})) = a_{d-1}(\nu(\vec{x})) = \ldots =
a_{d-i+1}(\nu(\vec{x})) = 0$.
\end{lem}
\proof{ Assuming that:  
  $$a_{d-i}(\nu(\vec{x})) < 0 ~\mbox{and}~ a_d(\nu(\vec{x})) =
  a_{d-1}(\nu(\vec{x})) = \ldots = a_{d-i+1}(\nu(\vec{x})) = 0$$ for
  some $0 \leq i \leq d$, we have $P(k,\nu(\vec{x})) =
  a_{d-i}(\nu(\vec{x})) \cdot k^d + \ldots + a_1(\nu(\vec{x})) \cdot k
  + a_0(\nu(\vec{x}))$. Since the dominant coefficient
  $a_{d-i}(\nu(\vec{x}))$ is negative, the polynomial will assume only
  negative values, from some point on. \qed}

\begin{exa}
Consider the following program \cite{byron}, and its linear
transformation matrix $A$.
\begin{center}
\begin{minipage}{5cm}
\begin{tabbing}
while \= ($x \geq 0$)  \\
\> $x' = x + y$ \\
\> $y' = y + z$
\end{tabbing}
\end{minipage}
\begin{minipage}{5cm}
\[\begin{array}{c}
A = 
\left(\begin{array}{ccc}
1 & 1 & 0 \\
0 & 1 & 1 \\
0 & 0 & 1
\end{array}\right)
\end{array}\]
\end{minipage}
\begin{minipage}{5cm}
\[\begin{array}{c}
A^k = 
\left(\begin{array}{ccc}
1 & k & \frac{k(k-1)}{2} \\
0 & 1 & k \\
0 & 0 & 1 
\end{array}\right)
\end{array}\]
\end{minipage}
\end{center}
The characteristic polynomial of $A$ is $\mbox{det}(A - \lambda I_3) =
(1-\lambda)^3$, hence the only eigenvalue is $1$, with multiplicity
$3$. Then we compute $A^k$ (see above), and $x' = x + k \cdot y +
\frac{k(k-1)}{2} z$ gives the value of $x$ after $k$ iterations of the
loop. Since only $x$ occurs within the guard of the loop, the weakest
non-termination precondition is: $\forall k \geq 1 ~.~ \frac{z}{2}
\cdot k^2 + (y - \frac{z}{2}) \cdot k + x \geq 0$. Lemma
\ref{poly-neg} gives a~ sufficient condition for termination: $(z < 0)
\vee (z = 0 \wedge y < 0) \vee (z = 0 \wedge y = 0 \wedge x < 0)$.
\end{exa}

We can generalize this method further to the case where all
eigenvalues of $A$ are of the form $q \cdot r$, with $q \in \real$ and
$r \in \complex$ being a~ root of the unity\footnote{A complex number
  $r = \cos(\theta) + i\sin(\theta)$, of absolute value $|r|=1$, is a
  root of the unity if and only if $\theta = \frac{a\pi}{b}$, for some
  $a,b \in \nat$, $b \neq 0$.}. The main reason for not using this
condition from the beginning is that we are, to this point, unaware of
its decidability status. With this condition instead, it is sufficient
to consider only the eigenvalues with the maximal absolute value, and
the polynomials obtained as sums of the polynomial coefficients of
these eigenvalues. The result of Lemma \ref{rational-poly} and the
sufficient condition of Lemma \ref{poly-neg} carry over when using
these polynomials instead.

\section{Termination Analysis of Integer Programs}
\label{sec:intprograms}

In this section, we extend the computation of weakest non-termination
preconditions from simple conjunctive loops to programs with possibly
nested loops. The method described here applies the {\em transition
  invariants} technique, initially developed for proving program
termination \cite{transition-invariants}, to the computation of
termination preconditions.

The method can be summarized as follows. Suppose that $R$ is the
(possibly disjunctive) transition relation of a program. Our method
first computes (1) a~{\em reachability relation}, defined as an
over-approximation of a restriction of the transitive closure of the
transition relation $R^+$ to a set $Init$ of initial program
configurations, formally $Reach \supseteq \{ (\nu,\nu') ~|~ (\nu,\nu')
\in R^+, \nu \in Init\}$, and (2) a~{\em transition invariant},
defined as an over-approximation of the transitive closure of $R$
restricted to states reachable from the set of initial configurations,
formally $TInv \supseteq \{ (\nu,\nu') ~|~ (\nu,\nu') \in R^+, \nu \in
R^*(Init) \}$. Then, $TInv$ is over-approximated with a union $R_1
\cup \dots \cup R_m$, $m\geq 1$, of octagonal relations. Next, the
weakest non-termination precondition $\wnt(R_i)$, $1\leq i\leq m$, can be
computed using techniques from Sections~\ref{sec:termination:octagons}
and \ref{sec:termination:affine}. The weakest non-termination
precondition of the program is then over-approximated by the pre-image
of $\wnt(R_1) \cup \ldots \cup \wnt(R_m)$ via the reachability
relation, formally $Reach^{-1}(\wnt(R_1) \cup \ldots \cup \wnt(R_m))$,
or equivalently, $\bigcup_{i=1}^m Reach^{-1}(\wnt(R_i))$. The
complement of this set is then a~valid termination precondition.

The technique presented in this section can be further applied to
programs with (recursive) procedure calls, by using the program
transformation described in \cite{cook-podelski-rybalchenko-fmsd09},
which turns a~program $P$ with recursive procedure calls into
a~program $P'$ without procedures such that
$\wrs(P)\subseteq\wrs(P')$. The main ingredient of this technique is
the {\em summarization} of procedures, i.e.\ computing (an
over-approximation of) the relation between the values of the input
parameters and the values returned by the procedure.

\subsection{\bf Example}\label{int:programs:motivation}

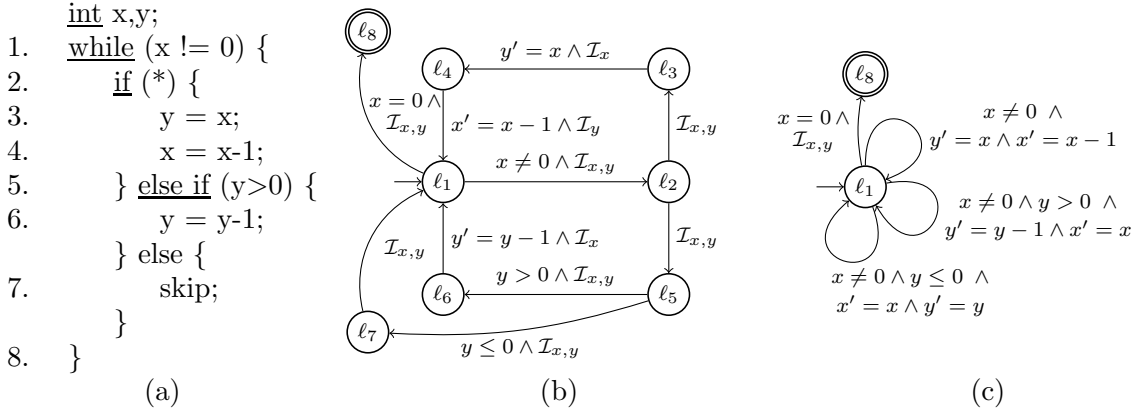
\begin{figure}[h]
\begin{tabular}{ccc}
\begin{minipage}{4cm}
\begin{tabbing} 
  xxxx\=xxx\=xxx\=xxx\=xxx\=xxx\=xxx\=\kill
     \> \key{int} x,y; \\
  1. \> \key{while} (x != 0) \{ \\
  2. \> \> \key{if} (*) \{ \\
  3. \> \> \> y = x; \\
  4. \> \> \> x = x-1; \\
  5. \> \> \} \key{else if} (y$>$0) \{ \\
  6. \> \> \> y = y-1; \\
     \> \> \} else \{ \\
  7. \> \> \> skip; \\
     \> \> \}  \\
  8. \> \}
\end{tabbing}
\end{minipage}
&
\mbox{\begin{minipage}{5.7cm}
\mbox{\scalebox{1.0}{\begin{tikzpicture}
\ttfamily
  \scriptsize

  \tikzset{
    sState/.style={draw=black,circle,inner sep=2pt,semithick},
    sInitial/.style={initial,initial text=}
  }

  \node[sState,sInitial] (l1) at (0mm,15mm) {$\ell_1$};
  \node[sState] (l2) at (30mm,15mm) {$\ell_2$};
  \node[sState] (l3) at (30mm,30mm) {$\ell_3$};
  \node[sState] (l4) at (0mm,30mm) {$\ell_4$};
  \node[sState] (l5) at (30mm,0mm) {$\ell_5$};
  \node[sState] (l6) at (0mm,0mm) {$\ell_6$};
  \node[sState] (l7) at (-10mm,-5mm) {$\ell_7$};
  \node[sState,accepting] (l8) at (-10mm,35mm) {$\ell_8$};

  \path[->]
    (l1) edge [bend angle=40,bend left] (l8)
    (l1) edge node [above] {$x\neq 0 \wedge \mathcal{I}_{x,y}$} (l2)
    (l2) edge node [right] {$\mathcal{I}_{x,y}$} (l3)
    (l3) edge node [above] {$y'=x \wedge \mathcal{I}_x$} (l4)
    (l4) edge node [right] {$x'=x-1 \wedge \mathcal{I}_y$} (l1)
    (l2) edge node [right] {$\mathcal{I}_{x,y}$} (l5)
    (l5) edge node [above] {$y>0 \wedge \mathcal{I}_{x,y}$} (l6)
    (l6) edge node [right] {$y'=y-1 \wedge \mathcal{I}_x$} (l1)
    (l5) edge [bend angle=10,bend left] node [below] {$y\leq0 \wedge \mathcal{I}_{x,y}$} (l7)
    (l7) edge [bend angle=40,bend left] (l1);

  \node at (-5mm,24mm) {\shortstack{$x=0\,\wedge$\\\,$\mathcal{I}_{x,y}$}};
  \node at (-5mm,6mm) {$\mathcal{I}_{x,y}$};
 
\end{tikzpicture}}}
\end{minipage}} 
&
\mbox{\begin{minipage}{5cm}
\hspace{-4mm}\mbox{\scalebox{1.0}{\begin{tikzpicture}
\ttfamily
  \scriptsize

  \tikzset{
    sState/.style={draw=black,circle,inner sep=2pt,semithick},
    sInitial/.style={initial,initial text=}
  }

  \node[sState,sInitial] (l1) at (0mm,15mm){$\ell_1$};
  \node[sState,accepting] (l8) at (0mm,30mm){$\ell_8$};

  \path[->] 
    (l1) edge [bend angle=10,bend left] node [left] {\shortstack{$x=0\,\wedge$\\$\mathcal{I}_{x,y}$}} (l8)
         edge [loop right,out=90,in=20,looseness=10] (l1)
         edge [loop right,out=10,in=-60,looseness=10] (l1)
         edge [loop right,out=-70,in=-140,looseness=10] (l1);

  \node at (21mm,23mm) {\shortstack{$x\neq 0 ~\wedge$ \\ $y'=x \wedge x'=x-1$}};
  \node at (23mm,11mm) {\shortstack{$x\neq 0 \wedge y>0 ~\wedge$ \\ $y'=y-1 \wedge x'=x$}};
  \node at(6mm,1mm) {\shortstack{$x\neq 0 \wedge y\leq 0 ~\wedge$ \\ $x'=x \wedge y'=y$}};
 
\end{tikzpicture}}}
\end{minipage}}
\\
(a) & (b) & (c)
\end{tabular}
\caption{\label{fig:running}An integer program and its control flow graph}
\end{figure}

Consider the non-deterministic integer program in Figure
\ref{fig:running}(a). If $x=0$ initially, the program does not enter
the main loop, and terminates trivially. Otherwise, the program may
enter an infinite computation. If $y\leq 0$ initially, the program can
iterate the third branch of the main loop infinitely many
times. Otherwise, if $y>0$ initially, the program can iterate the
second branch $y$ times and then iterate the third branch infinitely
many times.

We view programs as control flow graphs labeled with arithmetic
formulas. Figure \ref{fig:running}(b) depicts the control flow graph
of the program in Figure \ref{fig:running}(a). We write
$\mathcal{I}_{x_1,\dots,x_m}$ as a shorthand for $\bigwedge_{i=1}^{m}
x_i'=x_i$. The mechanics of our algorithm computing the weakest
non-termination precondition applied on the above example are described in the
following. First, we reduce the three loops $\ell_1 \arrow{}{} \ell_2
\arrow{}{} \ell_3 \arrow{}{} \ell_4 \arrow{}{} \ell_1$, $\ell_1
\arrow{}{} \ell_2 \arrow{}{} \ell_5 \arrow{}{} \ell_6 \arrow{}{}
\ell_1$ and $\ell_1 \arrow{}{} \ell_2 \arrow{}{} \ell_5 \arrow{}{}
\ell_7 \arrow{}{} \ell_1$ in Figure \ref{fig:running}(b) into
self-loops, obtaining a reduced control flow graph in Figure
\ref{fig:running}(c). Then, we compute the transitive summary relation
induced by all non-trivial runs of the program starting and ending at
$\ell_1$ (this notion is formally defined in the next section). This
relation is given in disjunctive normal form:
\[\small
\begin{array}{lcl}
\sem{P}^+(\ell_1,\ell_1) & \iff & R_1 \vee R_2 \vee R_3 \vee R_4 \vee R_5 \vee R_6 \vee R_7 \\
\\
R_1 & \iff & x \leq -1 \wedge y' \leq x \wedge y'=x'+1 \\
R_2 & \iff & y' \geq 1 \wedge y' \leq x \wedge y'=x'+1 \\
R_3 & \iff & y' \geq 0 \wedge y' \leq y-1 \wedge x'=x \wedge x' \leq -1 \\
R_4 & \iff & x' \geq 1 \wedge x'=x \wedge y' \geq 0 \wedge y' \leq y-1 \\
R_5 & \iff & x'=x \wedge x' \leq -1 \wedge y'=y \wedge y' \leq 0 \\
R_6 & \iff & x' \geq 1 \wedge x'=x \wedge y'=y \wedge y' \leq 0 \\
R_7 & \iff & x' \geq 1 \wedge y' \geq 0 \wedge x' \leq x-1 \wedge y'\leq x'
\end{array}\]
Notice that, since $\ell_1$ is the initial control state of the
program, the set of valuations reached at $\ell_1$ is the universal
set $\zed^\x$. A {\em transition invariant} of the program is the
restriction of the summary relation to the reachable states, which, in
this case, is $\sem{P}^{TInv}(\ell_1, \ell_1) = \sem{P}^+(\ell_1,
\ell_1)$. Next, we compute the weakest non-termination precondition of each
disjunct of the transition invariant, obtaining the formulas
$\wnt(R_1),\dots,\wnt(R_7)$ below:
\[\small\begin{array}{c}
\begin{array}{lcl}
\wnt(R_1) & \iff & x\leq -1 \\
\wnt(R_2) & \iff & \textbf{false} \\
\wnt(R_3) & \iff & \textbf{false} \\
\wnt(R_4) & \iff & \textbf{false} \\
\wnt(R_5) & \iff & x\leq -1 \wedge y\leq 0 \\
\wnt(R_6) & \iff & x\geq 1 \wedge y\leq 0 \\
\wnt(R_7) & \iff & \textbf{false} \\
\end{array}
\end{array}\]
The disjunction of these non-termination precondition defines a set of
configurations of the program, from which infinite runs, starting at
$\ell_1$, are guaranteed to exist:
\[
  \wnt(R_1) \vee \dots \vee \wnt(R_7) \iff 
  (x \leq -1) \vee (x\geq 1 \wedge y\leq 0)
\]
Finally, we compute the pre-image of this set via the (reflexive and
transitive) reachability relation defined as $\sem{P}^*(\ell_1,\ell_1)
= \sem{P}^+(\ell_1,\ell_1) \vee \mathcal{I}_\x$, obtaining thus the
weakest non-termination precondition of the program:
\[\begin{array}{rcl}
 (\sem{P}^*(\ell_1,\ell_1))^{-1}(\wnt(R_1) \vee \dots \vee \wnt(R_7)) & \iff & \\
    (x\geq 1 \wedge y\leq 0) \vee
    (x\geq 1\wedge y\geq 1) \vee
    (x\geq 2) \vee
    (x\leq -1)
   & \iff & x\neq 0 
\end{array}\]
This result matches the intuition. Indeed, the program will terminate
if and only if $x=0$, in which case the {\tt while} loop is never
entered. For $x\neq0$, the program enters the {\tt while} loop and may
get stuck into an infinite loop, for every initial value of $y$.

\subsection{\bf Syntax and Semantics}\label{background:int:syntax:semantics}

In the following, we abstract from specific programming language
constructs and assume that programs are represented by control flow
graphs whose edges are labeled by quantifier-free Presburger
arithmetic formulas defining relations. Formally, an \emph{integer
  program} is a tuple $P = \langle \x, Q,$ $q_{init}, \Delta \rangle$,
where:
\begin{itemize}
\item $\x$ is the set of variables of $P$
\item $Q$ are the \emph{control states} of $P$ 
\item $\Delta$ is a~set of \emph{transition rules} $q
  \arrow{R(\x,\x')}{} q'$, where $q,q' \in Q$ are the source and
  destination states, and $R(\x,\x')$ is a~quantifier-free Presburger
  formula
\item $q_{init}$ is the \emph{initial} control state of $P$
\end{itemize}

\begin{exa}
The program whose control flow graph is shown in Figure
\ref{fig:running}(b) can be formalized as $P=\langle \x, Q, \ell_1,
\Delta \rangle$, where $\x=\{x,y\}$, $Q=\{\ell_1,\dots,\ell_8\}$,
$\Delta = \{ t_1,\dots,t_{10} \}$, and
\[\begin{array}{lcl}
  t_{1} & = & \ell_1\arrow{x\neq 0 ~\wedge~ \mathcal{I}_{x,y}}{}\ell_2 \\
  t_{2} & = & \ell_2\arrow{\mathcal{I}_{x,y}}{}\ell_3 \\
  t_{3} & = & \ell_3\arrow{y'=x ~\wedge~ \mathcal{I}_{x}}{}\ell_4 \\
  t_{4} & = & \ell_4\arrow{x'=x-1 ~\wedge~ \mathcal{I}_{y}}{}\ell_1
\end{array}
\begin{array}{lcl}
  t_{5} & = & \ell_2\arrow{\mathcal{I}_{x,y}}{}\ell_5 \\
  t_{6} & = & \ell_5\arrow{y>0 ~\wedge~ \mathcal{I}_{x,y}}{}\ell_6 \\
  t_{7} & = & \ell_6\arrow{y'=y-1 ~\wedge~ \mathcal{I}_{x}}{}\ell_1 \\\\
\end{array}
\begin{array}{lcl}
  t_{8} & = & \ell_5\arrow{y\leq0 ~\wedge~ \mathcal{I}_{x,y}}{}\ell_7 \\
  t_{9} & = & \ell_7\arrow{\mathcal{I}_{x,y}}{}\ell_1 \\
  t_{10} & = & \ell_1\arrow{x=0 ~\wedge~ \mathcal{I}_{x,y}}{}\ell_8\\
  \rlap{\hbox to 148 pt{\hfill\qEd}}
\end{array}\]
\end{exa}
A \emph{configuration} of a program $P = \langle \vec{x}, Q,q_{init},
\Delta \rangle$ is a~pair $\langle q, \nu \rangle$, where $q \in Q$ is
a~control state and $\nu \in \zed^{\vec{x}}$ is a~valuation of the
variables.  Given two configurations $\langle q,\nu \rangle$ and
$\langle q',\nu' \rangle$ of a~program $P$, the configuration $\langle
q',\nu' \rangle$ is said to be an \emph{immediate successor} of
$\langle q,\nu \rangle$ if and only if $q \arrow{R(\x,\x')}{} q' \in
\Delta$ and $(\nu,\nu') \models R$.  For any $k\geq 0$, a~\emph{run}
of length $k$ of the program $P$ from $q$ to $q'$ is a~finite sequence
$\langle q_0,\nu_0 \rangle \arrow{}{} \langle q_1,\nu_1 \rangle
\arrow{}{} \ldots \arrow{}{} \langle q_k,\nu_k \rangle$, such that
$q=q_0$, $q'=q_k$, and $\langle q_{i+1},\nu_{i+1} \rangle$ is an
immediate successor of $\langle q_i,\nu_i \rangle$, for all $0\leq
i<k$.  Given two configurations $\langle q,\nu \rangle$ and $\langle
q',\nu' \rangle$ of a~program $P$, the configuration $\langle q',\nu'
\rangle$ is said to be a~{\em successor} of $\langle q,\nu \rangle$ if
there exists a run of length $k\geq 0$ from $\langle q,\nu \rangle$ to
$\langle q',\nu' \rangle$.  An \emph{infinite run} of a~program $P$
from a~control state $q$ is an infinite sequence $\langle q_0,\nu_0
\rangle \arrow{}{} \langle q_1,\nu_1 \rangle \arrow{}{} \ldots$ such
that $q=q_0$ and $\langle q_{i+1},\nu_{i+1} \rangle$ is an immediate
successor of $\langle q_i,\nu_i \rangle$ for all $i\geq 0$.  The
transitive closure of the transition relation $\sem{P}^+: (Q\times Q)
\rightarrow 2^{\zed^\x\times\zed^\x}$, the reflexive and transitive
closures of the transition relation $\sem{P}^*: (Q\times Q)
\rightarrow 2^{\zed^\x\times\zed^\x}$, and the weakest non-termination
precondition $\sem{P}^{wnt}: Q \rightarrow 2^{\zed^\x}$ of the program
$P$ are defined for each $q,q'\in Q$ as follows:
\[\begin{array}{lcl}
  \sem{P}^+(q,q') & \stackrel{def}{=} & \!\{ \langle \nu,\nu' \rangle ~|~ \langle q,\nu \rangle\!\arrow{}{}\!\dots\!\arrow{}{}\! \langle q',\nu' \rangle \mbox{ is a~run of $P$ of length $k\!\geq\! 1$} \}
  \\
  \sem{P}^*(q,q') & \stackrel{def}{=} & \!\{ \langle \nu,\nu' \rangle ~|~ \langle q,\nu \rangle\!\arrow{}{}\!\dots\!\arrow{}{}\! \langle q',\nu' \rangle \mbox{ is a~run of $P$ of length $k\!\geq\! 0$} \}
  \\
  \sem{P}^{wnt}(q) & \stackrel{def}{=} & \!\{ \nu ~|~ \langle q,\nu \rangle\!\arrow{}{}\!\dots\! \mbox{ is an infinite run of $P$} \}
\end{array}\]
Note that the set of configurations with control state $q$ that are
reachable from $q_{init}$, can be defined as the post-image of
$\zed^\x$ via $\sem{P}^*(q_{init},q)$,
i.e.\ $\sem{P}^*(q_{init},q)(\zed^\x)$. With this notation, the {\em
  strongest transition invariant} $\sem{P}^{TInv}: (Q\times Q)
\rightarrow 2^{\zed^\x\times\zed^\x}$ of a~program $P$ is defined for
each $q,q'\in Q$ as the restriction of the transitive closure of the transition relation to the
set of reachable configurations:
\[ \sem{P}^{TInv}(q,q') \stackrel{def}{=} \{ \langle\nu,\nu'\rangle \in \sem{P}^+(q,q') ~|~ \nu \in \big(\sem{P}^*(q_{init},q)\big)(\zed^\x) \} \]
When $\sem{P}^+$, $\sem{P}^*$, $\sem{P}^{TInv}$, or $\sem{P}^{wnt}$ is
not computable, one may content oneself with computing the following
over-approximations:
\[\begin{array}{lcllcl}
\sem{P}^+_\sharp &:& (Q\times Q)\rightarrow 2^{\zed^\x\times\zed^\x}\textrm{,}
&
\sem{P}^{TInv}_\sharp &:& (Q\times Q)\rightarrow 2^{\zed^\x\times\zed^\x}\textrm{,}
\\
\sem{P}^*_\sharp &:& (Q\times Q)\rightarrow 2^{\zed^\x\times\zed^\x}\textrm{,}
&
\sem{P}^{wnt}_\sharp &:& Q\rightarrow 2^{\zed^\x\times\zed^\x}\textrm{,}
\end{array}\]
These are arbitrary mappings such that: 
\[\begin{array}{lcllcl}
\sem{P}^+_\sharp(q,q') &\supseteq& \sem{P}^+(q,q')\textrm{,}
&
\sem{P}^{TInv}_\sharp(q,q') &\supseteq& \sem{P}^{TInv}(q,q')\textrm{,}
\\
\sem{P}^*_\sharp(q,q') &\supseteq& \sem{P}^*(q,q')\textrm{,}
&
\sem{P}^{wnt}_\sharp(q) &\supseteq& \sem{P}^{wnt}(q)\textrm{,}
\end{array}\]
for all $q,q'\in Q$. Any set $\sem{P}^{TInv}_\sharp$ that satisfies
the above inclusion is called a {\em transition invariant}.

\subsection{Computing Termination Preconditions for Integer Programs}
 
The following theorem is used to compute a termination precondition of
an integer program, using a set of precomputed transition
invariants. In fact we compute an over-approximation of the weakest
non-termination precondition. The complement of this set is a
termination precondition, i.e.\ a set of initial configurations from
which the program is guaranteed to terminate.
\begin{thm} \label{theorem:wnt:proc}
Let $P = \langle \vec{x}, Q, q_{init}, \Delta \rangle$ be a~program,
$\sem{P}^*_\sharp \supseteq \sem{P}^*$ be an over-approximation of the
reflexive and transitive closure of the transition relation,
$\sem{P}^{TInv}_\sharp \supseteq \sem{P}^{TInv}$ be a transition
invariant and, for each $q\in Q$, let
$R_{q,1},\dots,R_{q,p_q}\subseteq \zed^\x\times\zed^\x$ be relations,
such that $\sem{P}^{TInv}_\sharp(q,q) = \bigcup_{k=1}^{p_q} R_{q,k}$,
for some $p_q\geq 1$. Let
\[\mathcal{N} \stackrel{def}{=} 
\bigcup_{q\in Q} \left( \big(\sem{P}^*_\sharp(q_{init},q)\big)^{-1}
\left( \bigcup_{k=1}^{p_q} \wnt(R_{q,k}) \right)\right)\] Then,
$\sem{P}^{wnt}(q_{init}) \subseteq \mathcal{N}$. Moreover, if
$\sem{P}^{TInv}_\sharp = \sem{P}^{TInv}$ and $\sem{P}^*_\sharp =
\sem{P}^*$, then $\mathcal{N} = \sem{P}^{wnt}(q_{init})$.
\end{thm}
\proof{ 

We first prove that $\sem{P}^{wnt}(q_{init}) \subseteq \mathcal{N}$.
Let $\nu_0 \in \sem{P}^{wnt}(q_{init})$ be a valuation, and let
$\rho_1 = \langle q_{init},\nu_0 \rangle \langle q_1,\nu_1 \rangle
\langle q_2,\nu_2 \rangle \dots$ be an infinite run of $P$ starting
with $\nu_0$. Since the set of control states $Q$ is finite, there
exists $q \in Q$, and infinitely many integers $1\leq \ell_1 < \ell_2
< \ell_3 < \dots$ such that $q = q_{\ell_1} = q_{\ell_2} = q_{\ell_3}
= \dots$ It follows from the definition of $\sem{P}^{TInv}_\sharp$
that $\langle \nu_{\ell_j},\nu_{\ell_{j+1}} \rangle \in
\sem{P}^{TInv}_\sharp(q,q)$ for all $j\geq 1$. Let $\mu_i$ denote
$\nu_{\ell_i}$, for all $i \geq 1$. Then $\rho_2 = \langle
q_{init},\nu_0 \rangle \langle q,\mu_1 \rangle \langle q,\mu_2 \rangle
\dots$ is an infinite subsequence of $\rho_1$.

Since $\sem{P}^{TInv}_\sharp(q,q) = \bigcup_{k=1}^{p_q} R_{q,k}$, it
follows from the definition of $\sem{P}^{TInv}_\sharp$ that for each
$1\leq k < \ell$, there exists $1\leq j\leq p_q$ such that $\langle
\mu_k, \mu_\ell \rangle \in R_{q,j}$. Consequently, there exists
a~function $f: \{ (k,\ell) ~|~ 1\leq k < \ell \} \rightarrow
\{R_{q,1},\dots,R_{q,p_q}\}$ such that $\langle \mu_k, \mu_\ell
\rangle \in f(k,\ell)$ for all $1\leq k < \ell$. Let $\sim_f$ be the
kernel of $f$, i.e.\ the equivalence relation defined as $\langle
k,\ell \rangle \sim_f \langle k',\ell' \rangle$ if and only if
$f(k,\ell) = f(k',\ell')$. Clearly, $\sim_f$ has finite index, since
the range of $f$ is finite. Consequently, by the Ramsey theorem
\cite{ramsey}, there exists an infinite sequence of integers $1\leq
k_1<k_2<k_3<\dots$ and an equivalence class $[(m,n)]_{\sim_f}$ for
some $1\leq m<n$ such that $\langle k_i, k_{i+1} \rangle \sim_f
\langle m, n \rangle$ for all $i\geq 1$. Thus, there exists $1\leq
j\leq p_q$ such that $f(k_i,k_{i+1})=R_{q,j}$ for all $i\geq
1$. Consequently, $\mu_{k_1}\mu_{k_2}\dots$ is an infinite run of
$R_{q,j}$ and hence, $\mu_{k_1}\in \wnt(R_{q,j})$. Since $\langle
\nu_0, \mu_{k_1} \rangle \in \sem{P}^*_\sharp(q_{init},q)$, by the
definition of $\sem{P}^*_\sharp$, it follows that
\[
  \nu_0 \in \left( \sem{P}^*_\sharp(q_{init},q)\right)^{-1}
  \big(\wnt(R_{q,j})\big) \subseteq
  \big(\sem{P}^*_\sharp(q_{init},q)\big)^{-1} \left( \bigcup_{k=1}^{p}
  \wnt(R_{q,k}) \right) \subseteq \mathcal{N}
\]
hence $\nu_0 \in \mathcal{N}$, i.e.\ $\sem{P}^{wnt}(q_{init})
\subseteq \mathcal{N}$. 

Next, we prove that $\sem{P}^{wnt}(q_{init}) \supseteq \mathcal{N}$
under the assumption that $\sem{P}^{TInv}_\sharp = \sem{P}^{TInv}$ and
$\sem{P}^*_\sharp = \sem{P}^*$. Together with the previous point, this
is sufficient to prove that $\sem{P}^{wnt}(q_{init}) = \mathcal{N}$.
Let $\nu\in\mathcal{N}$. By the definition of $\mathcal{N}$ and since
$\sem{P}^*_\sharp = \sem{P}^*$, there exists $q\in Q$,
$\nu_0\in\zed^\x$, and $k\in\{1,\dots,p_q\}$ such that (i) there
exists a run $\rho$ from the configuration $\langle q_{init}, \nu
\rangle$ to the configuration $\langle q, \nu_0 \rangle$, and (ii)
$\nu_0\in\wnt(R_{q,j})$ for some $j \in \{1,\ldots,p_q\}$. Since
$\nu_0\in\wnt(R_{q,j})$, there exist infinitely many valuations
$\nu_1,\nu_2,\dots$ such that $\langle \nu_i, \nu_{i+1} \rangle \in
R_{q,j}$ for all $i\geq 0$. Since $\sem{P}^{TInv}(q,q) =
\sem{P}^{TInv}_\sharp(q,q) = \bigcup_{k=1}^{p_q} R_{q,k}$, we have
that $R_{q,j} \subseteq \sem{P}^{TInv}(q,q) \subseteq \sem{P}^+(q,q)$,
by the definition of the strongest transition invariant
$\sem{P}^{TInv}(q,q)$. But then, for each $i \geq 0$ there exists a
run $\rho_i$ of strictly positive length from $\langle q,\nu_i
\rangle$ to $\langle q,\nu_{i+1} \rangle$. Consequently,
$\rho.\rho_1.\rho_2\dots$ is an infinite run of $P$ and hence,
$\nu\in\sem{P}^{wnt}(q_{init})$. \qed}

Algorithm \ref{alg:wrs:proc} computes a sound over-approximation of
the weakest non-termination precondition of an integer program. It
uses a function \textsc{WNT}$(R)$ to compute the weakest
non-termination precondition of an octagonal, finite monoid or
polynomially bounded affine relation. Based on our previous results,
\textsc{WNT}$(R)$ is precisely the weakest non-termination
precondition, if $R$ is octagonal (Algorithm \ref{alg:recurset}) or
finite monoid affine (Theorems \ref{th:oct:wrs} and
\ref{theorem:term:monoids}, respectively), and \textsc{WNT}$(R)$ is
an over-approximation of the above, if $R$ is a polynomially bounded
affine relation (Equation (\ref{poly-system}) and Lemma
\ref{poly-neg}).
\begin{algorithm}[b]
\begin{algorithmic}[0]
\State {\bf input} A~program $P = \langle \vec{x}, Q, q_{init}, \Delta
\rangle$ \State {\bf output} A non-termination precondition
$\sem{P}^{wnt}_\sharp(q_{init})$
\end{algorithmic}
\begin{algorithmic}[1]
\Function{NT\_PROGRAM}{P}
\State $\mathcal{N} \gets \emptyset$
\For{{\bf each} $q\in Q$}
\If{$q \arrow{R_1}{} \ldots \arrow{R_n}{} q ~\mbox{is the only elementary cycle involving}~ q ~\mbox{in}~ P$}
\label{alg:wrs:proc:N1}
\State $R \leftarrow \exists \x_1 \ldots \exists \x_{n-1} ~.~ R_1(\x,\x_1) \wedge \ldots R_n(\x_{n-1},\x')$
\If{$R ~\mbox{defines an octagonal, fin.\ monoid or poly.\ bounded affine relation}$} 
\State $\mathcal{N} \gets \mathcal{N} \cup {\big(\sem{P}^*_\sharp(q_{init},q)\big)}^{-1}\big(\mbox{\textsc{WNT}}(R)\big)$
\State \textbf{continue}
\EndIf
\EndIf
\State {\bf find} octagonal relations $R'_1, \dots, R'_p$ 
       s.t. $\sem{P}^{TInv}(q,q) \subseteq \big(R'_1 \cup \dots \cup R'_p\big)$
       \label{line:find:tinv}
\State $\mathcal{N} \gets \mathcal{N} \cup {\big(\sem{P}^*_\sharp(q_{init},q)\big)}^{-1}\big(\bigcup_{j=1}^{p}\mbox{\textsc{WNT}}(R'_j)\big)$ \label{line:compute:wnt} \label{alg:wrs:proc:N2}
\EndFor
\State\Return $\mathcal{N}$
\EndFunction
\end{algorithmic}
\caption{Computing a Non-termination Precondition for a
  Program}\label{alg:wrs:proc}
\end{algorithm}
Let $P = \langle \vec{x}, Q, q_{init}, \Delta \rangle$ be an integer
program, for which we would like to compute a non-termination
precondition $\sem{P}^{wnt}_\sharp(q_{init})$. Since the set of
control states of $P$ is finite, any infinite computation of $P$ will
eventually iterate through the same state $q \in Q$ infinitely
often.~Hence we must compute non-termination preconditions for all
states $q \in Q$, i.e.\ sets of configurations from which a
computation iterating $q$ infinitely often is possible. For reasons of
precision, here we distinguish two cases:
\begin{itemize}
\item If $q$ occurs within only one elementary cycle, then every
  infinite run involving $q$ infinitely often must iterate this
  cycle. If, moreover, the composition of the relations on the cycle
  defines an: 
  \begin{itemize}
    \item octagonal relation or a finite monoid affine relation $R$,
      then we can compute $\wnt(R)$ precisely (see Theorems
      \ref{th:oct:wrs} and \ref{theorem:term:monoids}, respectively).
    \item polynomially bounded affine relation $R$, then we can
      compute an over-appro\-xi\-mation of $\wnt(R)$ (see Equation
      (\ref{poly-system}) and Lemma \ref{poly-neg}).
  \end{itemize}
  Notice that equivalence of a formula with an octagonal constraint
  can be decided using integer linear programming \cite{schrijver},
  whereas the finite monoid and polynomial boundedness of an affine
  relation can be decided using Theorem \ref{finite-monoid-conditions}
  and the decidability of its preconditions
  \cite{boigelot-thesis,mandel77}.
\item Otherwise, we compute a transition invariant
  $\sem{P}^{TInv}_\sharp(q,q)$ and over-approximate it with a set of
  octagonal relations $R'_1,\ldots,R'_p$, for some $p \geq 1$. Since
  we can compute $\wnt(R'_i)$ for each such octagonal relation, we can
  apply Theorem \ref{theorem:wnt:proc} to obtain
  $\sem{P}^{wnt}(q_{init})$.
\end{itemize}
Alternatively, one can see the first case above (lines 4-8 of
Algorithm \ref{alg:wrs:proc}) as a special case of Theorem
\ref{theorem:wnt:proc}, in which the transition invariant
$\sem{P}^{TInv}(q,q)$ can be safely replaced by the weakest
non-termination precondition $\wnt(R)$, since $R$ is the only cycle
that can be iterated infinitely often. Since we consider the pre-image
of this set via the reflexive and transitive closure of the
reachability relation $\sem{P}^*_\sharp(q_{init},q)$, we are
guaranteed to iterate this loop only through reachable configurations.

Any procedure for computing transition invariants can be used for the
purposes of this algorithm. For reasons of self-containment, Section
\ref{sec:computing-transition-invariants} describes an algorithm for
computing reflexive and transitive closures of the transition
relations $\sem{P}^*_\sharp(q_{init}, q)$, and transition invariants
$\sem{P}^{TInv}_\sharp(q)$, for every $q\in Q$. A version of this
algorithm was implemented in the \textsc{Flata} tool \cite{flata}, and
is guaranteed to return the exact reflexive and transitive closures of
the transition relations $\sem{P}^*(q_{init}, q)$, and the strongest
transition invariants of the program $\sem{P}^{TInv}(q)$, for a
specific class of programs, called {\em flat} (see Section
\ref{sec:flat-integer-programs}). A formal proof of correctness of
Algorithm \ref{alg:wrs:proc} is given in Section
\ref{sec:flat-integer-programs}.

\subsection{Computing Transition Invariants}
\label{sec:computing-transition-invariants}

The core of the method for computing transition invariants, needed by
the non-termination precondition Algorithm~\ref{alg:recurset}, is a
procedure that computes, for any two control states $q, q' \in Q$ of
an integer program $P = \langle \vec{x}, Q, q_{init}, \Delta \rangle$,
an over-approximation $\sem{P}^+_\sharp(q,q')$ of the transitive
closure $\sem{P}^+(q,q')$. The reflexive and transitive closure
$\sem{P}^*_\sharp(q,q')$ can be computed using the alternative
definition: $\sem{P}^*_\sharp(q,q')=\sem{P}^+_\sharp(q,q')$, if $q
\neq q'$, and $\sem{P}^*_\sharp(q,q)=\sem{P}^+_\sharp(q,q) \cup
\mathcal{I}_\x$. Using the reflexive and transitive closure, one can
compute an over-approximation of the reachable set, at any control
state $q \in Q$, as: $Reach^\sharp_P(q) = \sem{P}^*_\sharp(q_{init},
q)(\zed^\x)$. The transition invariant $\sem{P}^{TInv}_\sharp(q,q')$
given by the transitive closure $\sem{P}^{+}_\sharp(q,q')$ restricted
to values from $Reach^\sharp_P(q)$ only: $\sem{P}^{TInv}_\sharp(q,q')
= \{\langle \nu,\nu' \rangle \in \sem{P}^{+}_\sharp(q,q') ~|~ \nu \in
Reach^\sharp_P(q)\}$.

\begin{algorithm}[t!]
\begin{algorithmic}[0]
\State {\bf input} A~program $P = \langle \vec{x}, Q, q_{init}, \Delta
\rangle$, and distinct control states $q_{in},q_{out} \in Q$ 
\State {\bf output} An over-approximated transitive closure
$\sem{P}^+_\sharp(q_{in},q_{out})$
\end{algorithmic}
\begin{algorithmic}[1]
\Function{TransitiveRelation}{$P,q_{in},q_{out}$}
\State $\overline{P} = \langle \x, Q \cup \{\bar{q}_{in}, \bar{q}_{out}\}, 
\bar{q}_{in}, \Delta\cup\{\bar{q}_{in} \arrow{\mathcal{I}_\x}{} q_{in}, 
q_{out} \arrow{\mathcal{I}_\x}{} \bar{q}_{out}\} \rangle$ \label{alg:sum:copy}
\For{{\bf each} $q \in \overline{Q} \setminus \{  \bar{q}_{in}, \bar{q}_{out} \}$ 
with self-loops \label{alg:sum:forloop}
$q \arrow{R_1}{} q, \ldots, q \arrow{R_k}{} q \in \overline{\Delta}$}
\If{$k=0$}
\State $T \leftarrow \mathcal{I}_\x$
\Else\If{$k=1 ~\mbox{and}~ R_1$ is a finite monoid affine relation}
\State $H \leftarrow R_1$
\Else \State $H \leftarrow \Call{OctagonalHull}{R_1 \vee \ldots \vee R_k}$
\EndIf
\State $T \leftarrow \Call{ReflexiveTransitiveClosure}{H}$
\EndIf
\For{{\bf each} $q_1 \arrow{P}{} q$ and $q \arrow{Q}{} q_2$ such that $q\not\in\{q_1,q_2\}$}
\State $\overline{\Delta} \leftarrow \overline{\Delta} \cup \{q_1 \arrow{\exists \x_1\exists \x_2 . 
P(\x,\x_1) \wedge T(\x_1,\x_2) \wedge Q(\x_2,\x')}{} q_2\}$ \label{alg:sum:addtransition}
\EndFor
\State $\overline{Q} \leftarrow \overline{Q} \setminus \{q\}$  \label{alg:sum:removestate}
\State $\overline{\Delta} \leftarrow \overline{\Delta} \setminus \{q_1 \arrow{R}{} q_2 ~|~ q\in\{q_1,q_2\}\}$  \label{alg:sum:removetransitions}
\EndFor
\State\Return $\bigvee\{R ~|~ (\bar{q}_{in} \arrow{R}{} \bar{q}_{out}) \in \overline{\Delta}\}$ \label{alg:sum:return}
\EndFunction
\end{algorithmic}
\caption{Procedure Summary Algorithm}\label{alg:tc}
\end{algorithm}

Algorithm \ref{alg:tc} computes the over-approximated transitive
closures $\sem{P}^+_\sharp(q,q')$, that are the key of our method for
computing non-termination preconditions. The idea of this algorithm is
to eliminate control states which are neither initial or final, while
introducing new transitions labeled with compositions of relations
between the remaining states.\footnote{The algorithm resembles the
  schoolbook method for converting finite automata into regular
  expressions.} In the beginning (line \ref{alg:sum:copy}) we create a working copy
$\overline{P}$ of the program by adding two fresh control states
$\bar{q}_{in},\bar{q}_{out} \not\in Q$ and two copy transitions
$\bar{q}_{in} \arrow{\mathcal{I}_\x}{} q_{in}$ and $q_{out}
\arrow{\mathcal{I}_\x}{} \bar{q}_{out}$. This ensures that
$\bar{q}_{in}$ and $\bar{q}_{out}$ do not occur within loops in
$\overline{P}$. Then we iterate the following steps, until no more
states can be eliminated. For each control state with (possibly zero)
self-loops labeled with relations $R_1, \ldots, R_k$, we compute an
over-approximation of the reflexive and transitive closure $T=(R_1
\vee \ldots \vee R_k)^*$. Three situations may arise:
\begin{itemize}
\item if there is no such loop, i.e.\ $k=0$, $T$ is the identity
  relation.
\item if there is only one such loop labeled with a finite monoid
  affine relation $R_1$, $T=R_1^*$ can be computed using one of the
  techniques from \cite{finkel-leroux,boigelot-thesis,cav10}.
\item otherwise, we compute first the {\em octagonal hull} $H = (R_1
  \vee \ldots \vee R_k)^{oct}$, and then the reflexive and transitive
  closure of the octagonal hull $T = H^*$, using the algorithm
  described in \cite{cav10}. The octagonal hull of a set is the
  strongest octagonal constraint that defines an over-approximation of
  that set. In general, the octagonal hull of a Presburger-definable
  set can be computed using integer linear programming
  \cite{schrijver}.
\end{itemize}
Next, we compose the relation of each incoming transition $q_1
\arrow{R}{} q$ with $T$, and with the relation of each outgoing
transition $q \arrow{Q}{} q_2$. We replace the pair of incoming and
outgoing transitions with the transition $q_1 \arrow{P \circ T \circ
  Q}{} q_2$, which does not involve $q$ (line
\ref{alg:sum:addtransition}), and, finally, we eliminate $q$ and all
transitions involving it from the program (lines
\ref{alg:sum:removestate}-\ref{alg:sum:removetransitions}). The result
is the disjunction of all relations occurring on the remaining
transitions between the $q_{in}$ and $q_{out}$ states (line
\ref{alg:sum:return}), which defines $\sem{P}^+_\sharp(q_{in},
q_{out})$.

The argument for proving the soundness of Algorithm \ref{alg:tc} is
that the following invariant holds, at each iteration of the main loop
of the algorithm: after each elimination of a control state $q$ from a
program $P$ (line \ref{alg:sum:removestate}), the transitive closure
of the remaining program $P'$
is an over-approximation of the previous one, i.e.\ for all $q_1, q_2
\in Q \setminus \{q\}$, $\sem{P}^+(q_1,q_2) \subseteq
\sem{P'}^+(q_1,q_2)$. This is the case because the summary relation:
$$S_P^q(q_1,q_2) = \{\langle \nu_1,\nu_2 \rangle ~|~ \mbox{there is a
  run}~ \langle q_1,\nu_1 \rangle \arrow{}{} \ldots \arrow{}{} \langle
q,\nu \rangle \arrow{}{} \ldots \langle q_2,\nu_2 \rangle ~\mbox{in
  $P$}\}$$ induced by the set of runs between two configurations
$\langle q_1, \nu_1 \rangle$ and $\langle q_2, \nu_2 \rangle$, which
visits $q$, is over-approximated by the composition of $P$, $T$ and
$Q$ (line \ref{alg:sum:addtransition}):
$$S_P^q(q_1,q_2) \Rightarrow \exists \x_1 \exists \x_2 ~.~ P(\x,\x_1)
\wedge T(\x_1,\x_2) \wedge Q(\x_2,\x')$$ It is to be noticed that each
transition $q_1 \arrow{}{} q_2$ introduced at line
\ref{alg:sum:addtransition} in the algorithm corresponds to a~path
between $q_1$ and $q_2$ in the original control flow graph of the
program, which visits at least once the state $q$ removed at line
\ref{alg:sum:removestate}. A formal proof of soundness is given in
Lemma \ref{transitive-relation-computation}. 

\subsection{Flat Integer Programs}\label{sec:flat-integer-programs}

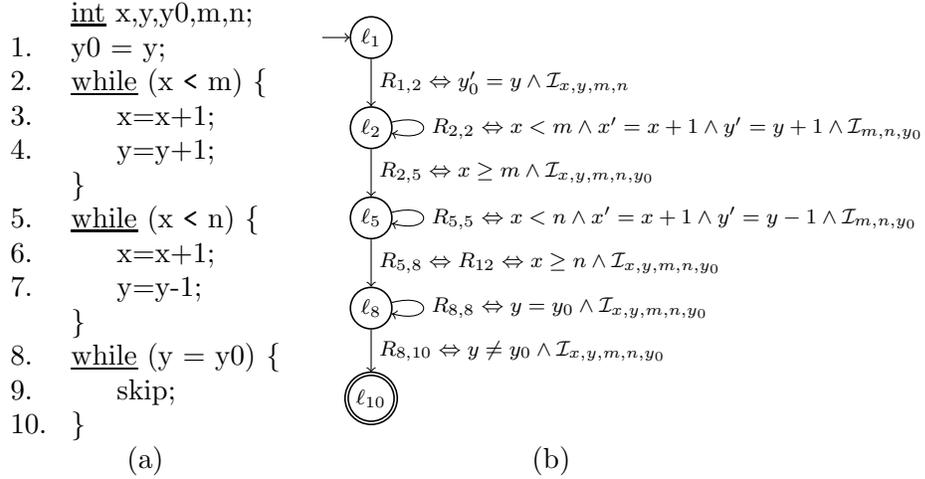
\begin{figure}[h]
\begin{tabular}{ccc}
\begin{minipage}{4cm}
\begin{tabbing} 
  xxxx\=xxx\=xxx\=xxx\=xxx\=xxx\=xxx\=\kill
     \> \key{int} x,y,y0,m,n; \\
  1. \> y0 = y; \\
  2. \> \key{while} (x \verb#<# m) \{ \\
  3. \>\>  x=x+1; \\
  4. \>\>  y=y+1; \\
     \> \} \\
  5. \> \key{while} (x \verb#<# n) \{ \\
  6. \>\>  x=x+1; \\
  7. \>\>  y=y-1; \\
     \> \} \\
  8. \> \key{while} (y = y0) \{ \\
  9. \>\>  skip; \\
  10.\> \}
\end{tabbing}
\end{minipage}
&
\mbox{\begin{minipage}{6.5cm}
\mbox{\scalebox{1.0}{\begin{tikzpicture}
\ttfamily
  \scriptsize

  \tikzset{
    sState/.style={draw=black,circle,inner sep=2pt,semithick},
    sInitial/.style={initial,initial text=}
  }

  \node[sState,sInitial] (l1) at (0mm,48mm) {$\ell_1$};
  \node[sState] (l2) at (0mm,36mm) {$\ell_2$};
  \node[sState] (l5) at (0mm,24mm) {$\ell_5$};
  \node[sState] (l9) at (0mm,12mm) {$\ell_8$};
  \node[sState,accepting] (l10) at (0mm,0mm) {$\ell_{10}$};

  \path[->]
    (l1) edge node [right] {$R_{1,2} \iff y_0'=y \wedge \mathcal{I}_{x,y,m,n}$} (l2)
    (l2) edge node [right] {$R_{2,5} \iff x\geq m \wedge \mathcal{I}_{x,y,m,n,y_0}$} (l5)
         edge [loop right,looseness=10] node [right] {$R_{2,2} \iff x<m \wedge x'=x+1 \wedge y'=y+1 \wedge \mathcal{I}_{m,n,y_0}$} (l2)
    (l5) edge node [right] {$R_{5,8} \iff R_{12} \iff x\geq n \wedge \mathcal{I}_{x,y,m,n,y_0}$} (l9)
         edge [loop right,looseness=10] node [right] {$R_{5,5} \iff x<n \wedge x'=x+1 \wedge y'=y-1 \wedge \mathcal{I}_{m,n,y_0}$} (l5)
    (l9) edge [loop right,looseness=10] node [right] {$R_{8,8} \iff y=y_0 \wedge \mathcal{I}_{x,y,m,n,y_0}$} (l9)
    (l9) edge node [right] {$R_{8,10} \iff y\neq y_0 \wedge \mathcal{I}_{x,y,m,n,y_0}$} (l10);
 
\end{tikzpicture}}}
\end{minipage}} 

\\
(a) & (b)
\end{tabular}
\caption{\label{fig:flat}A~flat integer program and its simplified control flow graph}
\end{figure}

In this section, we define a class of integer programs for which our
method computes precisely the weakest non-termination preconditions,
as formulas in Presburger arithmetic. As a consequence of the
decidability of the satisfiability problem for Presburger arithmetic
\cite{presburger29}, the universal termination problem is decidable
for this class. A recent result \cite{vmcai14,vmcai14-techreport}
shows that the {\em reachability problem}, i.e.\ the existence of a
finite run between two control states, in a flat program whose
transitions occurring within loops are labeled by octagonal
constraints, is NP-complete. As a byproduct, we show that the
non-termination problem, i.e.\ the existence of an infinite
computation, for these programs is NP-complete as well.
\begin{defi}
  Let $P = \langle \vec{x}, Q, q_{init}, \Delta \rangle$ be an integer
  program. For any elementary cycle $\pi ~:~
  q_1\arrow{R_1}{}q_2\arrow{R_2}{}\dots q_n \arrow{R_n}{} q_1$, let
  $\lambda(\pi)$ denote the formula $\exists \x_1,\ldots,\x_{n-1} ~.~
  R_1(\x,\x_1) \wedge \ldots \wedge R_n(\x_{n-1},\x)$. Then $P$ is
  said to be {\em flat} if and only if:
  \begin{enumerate}
  \item each control state $q \in Q$ belongs to at most one elementary
    cycle,
  \item for each elementary cycle $\pi$ in $P$, $\lambda(\pi)$ defines
    an octagonal, or a finite monoid affine relation. 
\end{enumerate}
\end{defi}
\begin{example}
Figure \ref{fig:flat} depicts a~flat integer programs $P$ and its
control flow graph. For simplicity, the elementary cycles have been
already reduced to one transition, by composition of all the relations
labeling the transitions within them. Since the labels of the
self-loops are octagonal constraints, we can compute their reflexive and
transitive closures precisely:
\[\begin{array}{lllllll}
R_{2,2}^* & \iff & \mathcal{I}_{x,y,m,n,y_0} \vee (x'-x=y'-y \wedge x'\geq x+1 \wedge m\geq x' \wedge \mathcal{I}_{m,n,y_0}) \\
R_{5,5}^* & \iff & \mathcal{I}_{x,y,m,n,y_0} \vee (x'-x=y-y' \wedge x'\geq x+1 \wedge n\geq x' \wedge \mathcal{I}_{m,n,y_0})
\end{array}\]
\[\begin{array}{llllllllllll}
\wnt(R_{2,2}) \iff \textbf{false} & \hspace{5mm}
\wnt(R_{5,5}) \iff \textbf{false} & \hspace{5mm}
\wnt(R_{8,8}) \iff y=y_0
\end{array}\]
Following the computation of Algorithm \ref{alg:wrs:proc}, the weakest
non-termination precondition of the integer program is:
\[
\wnt(P) \iff \begin{array}{ll}
\exists \x' ~.~ R_{1,2}(\x,\x') \wedge \wnt(R_{2,2})(\x') & \vee \\
\exists \x' ~.~ (R_{1,2} \circ R_{2,2}^* \circ R_{2,5})(\x,\x') \wedge \wnt(R_{5,5})(\x') & \vee \\
\exists \x' ~.~ (R_{1,2} \circ R_{2,2}^* \circ R_{2,5} \circ R_{5,5}^* \circ R_{5,8})(\x,\x') \wedge \wnt(R_{9,9})(\x')
\end{array}
\]
Since $\wnt(R_{2,2}) \iff \wnt(R_{5,5}) \iff \textbf{false}$, the
first two disjuncts are equivalent to \textbf{false}. The third
disjunct, and hence $\wnt(P)$, is equivalent to
\[
\wnt(P) \iff (n=2m-x \wedge m\geq x+1 \wedge n\geq m+1) \vee (m\leq x \wedge n\leq x)\eqno{\qEd}
\]
\end{example}

If $P = \langle \x,Q,q_{init},\Delta \rangle$ is a flat program, then
Algorithm \ref{alg:tc} can be shown to return the precise transitive
closures $\sem{P}^+(q,q')$, for any $q,q'\in Q$. Intuitively, this
is the case because during the state elimination process, at any step,
a state $q \in Q$ that is chosen to be removed can have at most one
self-loop (line 7 in Algorithm \ref{alg:tc}), which corresponds to the
(at most one) elementary cycle involving $q$ in $\Delta$. Since,
moreover the label of this cycle denotes an octagonal or finite monoid
affine relation, the transitive closure of this relation can be
computed as a Presburger formula, without loss of information, using
the algorithm from e.g.\ \cite{cav10}. As a direct consequence,
$\sem{P}^*(q,q')$ can also be computed without loss of precision, if
the program is flat.

\begin{lem}\label{transitive-relation-computation}
Let $P = \langle \vec{x}, Q, q_{init}, \Delta \rangle$ be an integer
program. Then, the result of Algorithm \ref{alg:tc} is a~Presburger
formula $\phi(\x,\x')$ that defines an over-approximation of
$\sem{P}^+(q_{in},q_{out})$. If, moreover, $P$ is flat, $\phi(\x,\x')$
defines precisely $\sem{P}^+$.
\end{lem}
\proof{ 

Let $\overline{P}_i = \langle \vec{x}, \overline{Q}_i,
\overline{q}_{in}, \overline{\Delta}_i \rangle$ be the program
$\overline{P}$ at the $i$-th iteration of the main loop of the
algorithm, $i \geq 0$, and $\overline{P}_0 = \overline{P}$. Since for
all $i \geq 0$, $\overline{Q}_{i+1} \subset \overline{Q}_i$ (line
\ref{alg:sum:removestate}) it is sufficient to prove that, for all $i
\geq 0$ we have $\sem{\overline{P}_{i}}^+(\bar{q}_{in}, \bar{q}_{out})
\subseteq \sem{\overline{P}_{i+1}}^+(\bar{q}_{in},
\bar{q}_{out})$. Moreover, $\overline{P}_0 = \overline{P}$ (line
\ref{alg:sum:copy}) and $\sem{P}^+ = \sem{\overline{P}}^+$ is an easy
exercise. Then we obtain that, for all $i \geq 0$, $\sem{P}^+(q_{in},
q_{out}) \subseteq \sem{\overline{P}_{i}}^+(\bar{q}_{in},
\bar{q}_{out})$. The algorithm is bound to terminate, by the fact that
the set of control states $Q$ is finite and the for loop at line
\ref{alg:sum:forloop} is executed once for each control state $q \in
\overline{Q} \setminus \{\bar{q}_{in}, \bar{q}_{out}\}$. Hence the
result is an over-approximation of $\sem{P}^+$.

Let $q \in \overline{Q}_{i-1}$ be a control state chosen at line
\ref{alg:sum:forloop}, $R_1,\dots,R_k$ be the labels of the self-loops
of $q$, and let $H$ be the relation computed by the algorithm. For
some $i > 0$, let $\pi$ be a~run between two configurations $\langle
\bar{q}_{in}, \nu' \rangle$ and $\langle \bar{q}_{out}, \nu'' \rangle$
in $\overline{P}_{i-1}$, for some valuations $\nu', \nu''\in\zed^\x$.
It is sufficient to show that each sub-run $\rho$ of $\pi$ of the form
$\langle q_0,\nu_0\rangle \arrow{}{} \dots \arrow{}{} \langle
q_n,\nu_n\rangle$, where $n\geq 2$, $q_1=\dots=q_{n-1}=q$, $q_0\neq
q$, and $q_n\neq q$, can be replaced with a sub-run $\rho' : \langle
q_0,\nu_0\rangle \arrow{}{} \langle q_n,\nu_n\rangle$ in
$\overline{P}_i$ of length $1$, thus obtaining a run $\pi'$ between
$\langle \bar{q}_{in}, \nu' \rangle$ and $\langle \bar{q}_{out}, \nu''
\rangle$ in $\overline{P}_i$. Consequently, we have:
$$\langle \nu',\nu'' \rangle \in
\sem{\overline{P}_{i-1}}^+(\bar{q}_{in}, \bar{q}_{out}) \Rightarrow
\langle\nu',\nu''\rangle\in\sem{\overline{P}_i}^+(\bar{q}_{in},
\bar{q}_{out})$$ and hence $\sem{P}^+(q_{in}, q_{out}) \subseteq 
\sem{\overline{P}_{i-1}}^+(\bar{q}_{in}, \bar{q}_{out}) \subseteq
\sem{\overline{P}_i}^+(\bar{q}_{in}, \bar{q}_{out})$.

Let us consider any sub-run $\rho$ of the above form. Since
$R_j\subseteq H$ for each $1\leq j\leq k$, we have that
$(\nu_\ell,\nu_{\ell+1}) \in H$ for each $1\leq \ell<n-1$, and hence
$(\nu_1,\nu_{n-1})\in H^{n-2}\subseteq H^*=T$. Let $q_0 \arrow{P}{} q$
and $q \arrow{Q}{} q_n$ be transitions in $\overline{\Delta}_{i-1}$
such that $(\nu_0,\nu_1)\in P$ and $(\nu_{n-1},\nu_n)\in Q$. Since
$(\nu_0,\nu_1) \models P$, $(\nu_1,\nu_{n-1}) \models T$ and
$(\nu_{n-1},\nu_n) \models Q$, we can choose $\rho'$ as the transition
labeled by $\exists \x_1\exists \x_2 . P(\x,\x_1) \wedge T(\x_1,\x_2)
\wedge Q(\x_2,\x')$, added at line \ref{alg:sum:addtransition}.

For the second part of the proof, suppose that the program $P$ is
flat. For some arbitrary $i \geq 0$ and two configurations
$\nu',\nu'' \in \zed^\x$, let $\pi$ be a run from $\langle
\bar{q}_{in},\nu'\rangle$ to $\langle \bar{q}_{out},\nu''\rangle$ in
$\overline{P}_{i}$. We show that there exists a~run in
$\overline{P}_{i-1}$ between the same configurations, proving thus
that $\sem{\overline{P}}^+_{i}(\bar{q}_{in}, \bar{q}_{out}) \subseteq
\sem{\overline{P}}^+_{i-1}(\bar{q}_{in}, \bar{q}_{out})$. By the
previous point, we obtain $\sem{\overline{P}}^+_{i}(\bar{q}_{in},
\bar{q}_{out}) = \sem{\overline{P}}^+_{i-1}(\bar{q}_{in},
\bar{q}_{out})$, and since the choice of $i \geq 0$ was arbitrary, we
conclude that $\sem{\overline{P}}^+_{i}(\bar{q}_{in}, \bar{q}_{out}) =
\sem{P}^+(q_{in}, q_{out})$.

Let $\langle q_1,\nu_1\rangle \arrow{}{} \langle q_2,\nu_2\rangle$ be
a step of $\pi$ such that $(\nu_1,\nu_2) \models V$ for some
transition
$t=(q_1\arrow{V}{}q_2)\in(\overline{\Delta}_{i}\setminus\overline{\Delta}_{i-1})$,
and let $t_1=q_1 \arrow{P}{} q$ and $t_2=q \arrow{Q}{} q_2$ be the
transitions in $\overline{\Delta}_{i-1}$ used to construct $t$. Since
$P$ is flat, there is at most 1 self-loop involving the control state
$q$. If there is no such self-loop, the algorithm computes $T =
\mathcal{I}_\x$, hence $V(\x,\x') \iff \exists \z . P(\x,\z) \wedge
Q(\z,\x')$. Consequently, there exists a valuation $\eta\in\zed^\x$
such that $(\nu_1,\eta) \models P$, $(\eta,\nu_2) \models Q$ and thus,
there is a~run $\langle q_1,\nu_1\rangle \arrow{}{} \langle
q,\eta\rangle \arrow{}{} \langle q_2,\nu_2\rangle$ in
$\overline{P}_{i-1}$. If there is one self-loop, then the algorithm
computes precisely the reflexive and transitive closure $T = R_1^*$
and hence, $V(\x,\x') \iff \exists \z,\z' . P(\x,\z) \wedge
R_1^*(\z,\z') \wedge Q(\z',\x')$. Since $(\nu_1,\nu_2) \models V$,
there exists $n \geq 0$, such that $(\nu_1,\nu_2) \models \exists
\z,\z' . P(\x,\z) \wedge R_1^n(\z,\z') \wedge Q(\z',\x')$. If $n = 0$,
$R^0=\mathcal{I}_\z$ and we obtain a run in $\overline{P}_{i-1}$
similarly as in the case with no self-loop. If $n \geq 1$, there exist
valuations $\eta_0,\dots,\eta_n\in \zed^\x$ such that
$(\eta_\ell,\eta_{\ell+1})\in R_1$ for each $0\leq\ell<n$,
$(\nu_1,\eta_0)\in P$, and $(\eta_n,\nu_2)\in Q$. Hence we obtain the
run $\langle q_1,\nu_1\rangle \arrow{}{} \langle q,\eta_0\rangle
\arrow{}{} \dots \arrow{}{} \langle q,\eta_n\rangle \arrow{}{} \langle
q_2,\nu_2\rangle$ in $\overline{P}_{i-1}$. We obtain thus:
$$\langle\nu',\nu''\rangle\in\sem{\overline{P}_{i}}^+(\bar{q}_{in},\bar{q}_{out})
\Rightarrow
\langle\nu',\nu''\rangle\in\sem{\overline{P}_{i-1}}^+(\bar{q}_{in},\bar{q}_{out})$$
and consequently,
$\sem{\overline{P}_{i}}^+(\bar{q}_{in},\bar{q}_{out}) \subseteq
\sem{\overline{P}_{i-1}}^+(\bar{q}_{in}, \bar{q}_{out})$.

Since the transitive closure of octagonal and finite monoid affine
relations is Presburger definable (see e.g.\ \cite{cav10}), Presburger
arithmetic is closed under existential quantification, and since the
octagonal hull of a Presburger formula can be computed using integer
linear programming \cite{schrijver}, it follows that the algorithm
manipulates and returns only Presburger formulas.  \qed}

Moreover, Algorithm \ref{alg:wrs:proc} will also compute the weakest
non-termination precondition for flat programs. Since every state
occurs within at most one elementary cycle, the test on line 4 of the
algorithm will succeed for every state on a loop, and since the
formula defining the composition $R$ of all relations along the cycle
is equivalent to an octagonal or a finite monoid affine relation, the
test on line 6 will also succeed. In this case,
$\mbox{\textsc{WNT}}(R)$ is bound to return the weakest
non-termination precondition of $R$, thus the result of Algorithm
\ref{alg:wrs:proc} is the weakest non-termination precondition of the
entire program.

\begin{lem}\label{wrs-computation}
Let $P = \langle \vec{x}, Q, q_{init}, \Delta \rangle$ be an integer
program. Then, the result of Algorithm \ref{alg:wrs:proc} is
a~Presburger formula $\phi(\x,\x')$ that defines an over-approximation
of $\sem{P}^{wnt}(q_{init})$. If, moreover, $P$ is flat,
$\phi(\x,\x')$ defines precisely $\sem{P}^{wnt}(q_{init})$.
\end{lem}
\proof{ 
Consider the iteration of the for-loop during which the control state
$q\in Q$ is chosen. First, suppose that the test at line
\ref{alg:wrs:proc:N1} fails. In this case the algorithm enters line
\ref{alg:wrs:proc:N2}, and the correctness of the assignment at this
line follows from Theorem \ref{theorem:wnt:proc}. Second, suppose that
the test at \ref{alg:wrs:proc:N1} succeeds. In this case, there is a
unique elementary cycle of the form
$q\arrow{R_1}{}\dots\arrow{R_n}{}q$, where $n\geq 1$. Let
$R\stackrel{def}{=} R_1\circ\dots\circ R_n$. Then, it follows from the
definition of $\sem{P}^{TInv}$ that:
\[ \begin{array}{cl}
 & (\sem{P}^*(q_{init},q))^{-1}(\sem{P}^{TInv}(q,q)) \vspace{1mm}\\
 = & \left\{ \nu_0\in\zed^\x ~|~ 
   \begin{array}{l}
     \exists \textrm{ valuations } \{\nu_i\in\zed^\x\}_{i\geq 1} \textrm{ and runs }
     \pi_0 = \langle q_{init},\nu_0 \rangle \arrow{}{}^* \langle q,\nu_1 \rangle, \\
     \pi_i = \langle q,\nu_i \rangle \arrow{}{}^+ \langle q,\nu_{i+1} \rangle \textrm{ for each } i\geq 1
   \end{array} \right\} \\
 = & \left\{ \nu_0\in\zed^\x ~|~ 
   \begin{array}{l}
     \exists \textrm{ valuations } \{\nu_i\in\zed^\x\}_{i\geq 1} \textrm{ and run }
     \pi_0 = \langle q_{init},\nu_0 \rangle \arrow{}{}^* \langle q,\nu_1 \rangle \\
     \textrm{such that } (\nu_i,\nu_{i+1})\in R^+ \textrm{ for each } i\geq 1
   \end{array} \right\} \\
 = & \left\{ \nu_0\in\zed^\x ~|~ 
   \begin{array}{l}
     \exists \textrm{ valuations } \{\nu_i\in\zed^\x\}_{i\geq 1} \textrm{ and run }
     \pi_0 = \langle q_{init},\nu_0 \rangle \arrow{}{}^* \langle q,\nu_1 \rangle \\
     \textrm{such that } (\nu_i,\nu_{i+1})\in R \textrm{ for each } i\geq 1
   \end{array} \right\}
\end{array} \]
\[ \begin{array}{cl}
 = & \left\{ \nu_0\in\zed^\x ~|~ 
   \begin{array}{l}
     \exists \textrm{ valuation } \nu_1\in\zed^\x \textrm{ and run }
     \pi_0 = \langle q_{init},\nu_0 \rangle \arrow{}{}^* \langle q,\nu_1 \rangle \\
     \textrm{such that } \nu_1 \in \wnt(R) \textrm{ for each } i\geq 1
   \end{array} \right\} \vspace{1mm}\\
 = & (\sem{P}^*(q_{init},q))^{-1}(\wnt(R))
\end{array}\]
Then, the correctness of line \ref{alg:wrs:proc:N1} follows from Theorem \ref{theorem:wnt:proc}. Consequently, the algorithm always returns an over-approximation of $\sem{P}^{wnt}(q_{init})$. 

Next, suppose that $P$ is flat. Moreover, line \ref{alg:wrs:proc:N2} is reached if and only if there is no cycle that involves $q$, in which case $\sem{P}^+(q,q)=\emptyset$. Consequently, $\sem{P}^{TInv}(q,q)=\emptyset$ and hence, the algorithm can always choose $R'_1\iff\textbf{false}$ before executing line \ref{alg:wrs:proc:N2}. Previously, we argued that \[(\sem{P}^*(q_{init},q))^{-1}(\sem{P}^{TInv}(q,q)) = (\sem{P}^*(q_{init},q))^{-1}(\wnt(R))\] Since $P$ is flat, $\sem{P}^*(q_{init},q)$ can be computed precisely as a Presburger formula, by Lemma \ref{transitive-relation-computation}. Moreover, $R$ is an octagonal or a~finite monoid affine relations and hence, $\wnt(R)$ can be computed precisely as a Presburger formula too, by Theorem \ref{th:oct:wrs} and \ref{theorem:term:monoids}. Hence, the algorithm returns a Presburger formula that precisely defines $\sem{P}^{wnt}(q_{init})$.
\qed}

If we restrict the class of flat integer programs further, by
considering that only octagonal constraints appear as labels within
the loops of the program, we can characterize the complexity class for
the problem asking for the existence of an infinite run, within this
class of programs. The result is based on a characterization of the
{\em reachability problem} in this class of programs. Given a program
$P = \langle \x,Q,q_{init},\Delta \rangle$ and a control state $q \in
Q$, the reachability problem asks for the existence of a run of $P$ from
$q_{init}$ to $q$.

\begin{thm}[\cite{vmcai14}]
  The reachability problem for the class of programs:
  $$\mathcal{P}_{OCT} = \left\{P ~\mbox{flat program} ~|~ \begin{array}{l}
    \mbox{if $q \arrow{R}{} q'$ is in a cycle, $R$ is an octagonal
      constraint} \\ \mbox{otherwise, $R$ is a quantifier-free
      Presburger formula} \end{array} \right\}
    $$ is NP-complete.
\end{thm}
This result can be used in conjunction with Theorem \ref{th:oct:wrs}
to obtain the following: 

\begin{thm}
  The problem asking for the existence of an infinite run is
  NP-complete for the class of programs $\mathcal{P}_{OCT}$.
\end{thm}
\proof{Let $P = \langle \x, Q, q_{init}, \Delta \rangle$ be an
  instance of the $\mathcal{P}_{OCT}$ class. Since $P$ is a flat
  program, each strongly connected component consists of at most
  one cycle, which is elementary. 
  Let $C_1, \ldots, C_k$ be the non-trivial elementary
  cycles of $P$, and let $q_1, \ldots, q_k$ be arbitrary control
  states belonging to each of these cycles, respectively. Let $R_i$ be
  the composition of all octagonal relations on $C_i$ starting from
  $q_i$, for all $i = 1,\ldots,k$, respectively. Since all of these
  relations are defined by octagonal constraints, their composition
  can be computed in PTIME, according to Corollary
  \ref{oct:consistency:test}. Since PTIME $\subseteq$ PSPACE, the
  sizes of $R_1,\ldots,R_k$ are at most polynomial in the size of
  $P$. Then one uses Algorithm \ref{alg:recurset} to compute
  $\wnt(R_1), \ldots, \wnt(R_k)$ in PTIME, respectively (Theorem
  \ref{th:oct:wrs}). Clearly, the sizes of $\wnt(R_1), \ldots,
  \wnt(R_k)$ are also polynomial in the size of $P$. Finally, we
  construct $P' = \langle \x, Q \cup \{q_{nt}\}, q_{init}, \Delta'
  \rangle$, where $q_{nt} \not\in Q$ is a fresh control state, and:
  $$\Delta' = \Delta \cup \{q_i \arrow{\wnt(R_i)}{} q_{nt} ~|~ i =
  1,\ldots,k\}$$ The size of $P'$ is bounded by a polynomial in the
  size of $P$, and, moreover, $P$ has an infinite run if and only if
  the control state $q_{nt}$ is reachable by a finite run of
  $P'$. Hence the existence of an infinite run is in NP. 

  To show NP-hardness, let $\varphi(\x)$ be an arbitrary
  quantifier-free Presburger formula, and consider the following
  integer program:
  \begin{equation}\label{one-loop-cm}
    q_{init} \arrow{\varphi(\vec{x'})}{}~  
    \stackrel{\stackrel{\true}{\curvearrowright}}{q}    
  \end{equation}  
  Clearly, the program (\ref{one-loop-cm}) has an infinite run if and
  only if $\varphi(\x)$ is satisfiable. However, this is an
  NP-complete problem, since $\varphi$ is an arbitrary quantifier-free
  Presburger formula. \qed}

\section{Experiments}
\label{sec:experiments}

We have validated the methods described in this paper by automatically
finding preconditions for termination of all the octagonal running
examples, and of several integer programs synthesized from
(i)~programs with lists obtained using the translation scheme from
\cite{cav06} which generates an integer program from a program
manipulating dynamically allocated single-selector linked lists,
(ii)~VHDL designs such as hardware counter and synchronous LIFO
\cite{smrcka-vojnar08}, (iii)~small C~programs with challenging loops
and (iv)~small recursive Java programs from
\cite{termination-competition} translated to non-recursive programs
using the procedure summarization method described in
\cite{cook-podelski-rybalchenko-fmsd09}.

We have computed the weakest non-termination preconditions reported in
Table \ref{fig:results:termination} using the methods from Section
\ref{sec:termination:octagons} and \ref{sec:intprograms} which we
implemented in the \textsc{Flata} tool \cite{flata}. By computing
octagonal abstractions of disjuncts of a~transition invariant, we have
verified universal termination of the \textsc{ListCounter} and
\textsc{ListReversal} programs. Next, we have verified the
\textsc{Counter} and \textsc{SynLifo} programs by computing the
precise transition invariant and then the weakest non-termination
precondition, which was empty in both cases. Thus, these models have
infinite runs for any input values, which is to be expected as they
encode the behavior of synchronous reactive circuits. Similarly, we
have computed the weakest non-termination preconditions for numerical
programs \textsc{anubhav}, \textsc{cousot}, \textsc{leq}, and
\textsc{plus}.

\begin{table}[t]
\caption{Weakest Non-termination Preconditions for Integer Programs.}
\begin{scriptsize}
\begin{center}
\begin{minipage}[t]{.5\textwidth}
\scalebox{0.95}{\begin{tabular}[t]{|c|ccc|c|c|c|}
\hline
\multicolumn{1}{|c|}{\multirow{2}{*}{Model}} & 
\multicolumn{3}{|c}{Size} & 
\multicolumn{1}{|c|}{\multirow{2}{*}{Time [s]}} &
\multicolumn{1}{|c|}{\multirow{2}{*}{\shortstack{Weakest \\ Non-termination Preconditions}}}
\\
& \multicolumn{1}{|c}{$\card{\x}$} & \multicolumn{1}{c}{$\card{Q}$} & \multicolumn{1}{c|}{$\card{\Delta}$} &&
\\
\hline
\multicolumn{6}{l}{\textbf{(i) Examples from L2CA \cite{cav06}}} \\
\hline
listcounter & 4 & 31 & 35 &
  1.2 & $false$
  \\
listreversal & 7 & 97 & 107 &
  32.6 & $false$
  \\
\hline
\multicolumn{6}{l}{\textbf{(ii) VHDL models from \cite{smrcka-vojnar08}}} \\
\hline
counter & 2 & 6 & 13 &
  0.8 & $true$
  \\
register & 2 & 10 & 49 &
  1.4 & $true$
  \\
synlifo & 3 & 43 & 1006 &
  1016.4 & $true$
  \\
\hline
\multicolumn{6}{l}{\textbf{(iii) Examples from \cite{jhala-mcmillan06}}} \\
\hline
anubhav  & 29 & 20 & 25 &
  3.2 & $i<0$
  \\
cousot & 29 & 31 & 34 &
  4.0 & $true$
  \\
\hline
\multicolumn{6}{l}{\textbf{(iv) Examples from \cite{termination-competition}}} \\
\hline
leq  & 3 & 5 & 6 &
  0.6 & $false$
  \\
leq.modif  & 3 & 5 & 6 &
  2.4 & $x<0 \wedge y<0$
  \\
plus & 3 & 7 & 9 &
  0.7 & $false$
  \\
plus.modif & 3 & 7 & 9 &
  0.9 & $x<0 \vee y<0$
  \\
\hline
\end{tabular}}
\end{minipage}%
\end{center}
\end{scriptsize}
\label{fig:results:termination}
\end{table}


Second, we have compared (Table \ref{fig:comparison:poly:bounded}) our method for termination of polynomially bounded linear affine loops from Section~\ref{sec:termination:affine} with the examples given in \cite{byron}, and found the same termination preconditions as they do, with one exception, in which we can prove universal termination in integer input values (row 3 of Table \ref{fig:comparison:poly:bounded}).

\begin{table}[t]
\caption{Termination preconditions for several program fragments from \cite{byron}}
\begin{center}
\begin{tabular}{|c|c|c|}
\hline
{\scriptsize \textsc{Program}} & 
{\scriptsize \textsc{Cook et al. \cite{byron}}} & 
{\scriptsize \textsc{Linear Affine Loops}} \\
\hline
\hline
\begin{minipage}{3.5cm}
{\scriptsize\begin{tabbing}
if \= (lvar $\geq$ 0) \\
 \> while \= (lvar $<$ $2^{30}$) \\
 \> \> lvar = lvar $<\!\!<$ 1; 
\end{tabbing}}
\end{minipage}
& {\scriptsize $lvar > 0 \vee lvar < 0 \vee lvar \geq 2^{30}$} & 
${\scriptstyle \neg(lvar = 0) \vee lvar \geq 2^{30}}$
\\
\hline
\begin{minipage}{3.5cm}
{\scriptsize\begin{tabbing}
while \= (x $\geq$ N) \\
\> x = -2*x + 10;
\end{tabbing}}
\end{minipage}
& {\scriptsize $x > 5 \vee x + y \geq 0$ } & 
$x \neq \frac{10}{3} \iff \mbox{true}$
\\
\hline
\begin{minipage}{3.5cm}
{\scriptsize\begin{tabbing}
//@ requires $n > 200$ \\
x = 0; \\
while \= (1) \\
\> if \= (x $<$ n) \{ x=x+y; \\
\> \> if (x $\geq$ 200) break; \}
\end{tabbing}}
\end{minipage}
 & {\scriptsize $y > 0$} & 
${\scriptstyle y > 0}$
\\
\hline
\end{tabular}
\label{fig:comparison:poly:bounded}
\end{center}
\end{table}

\section{Conclusion}
\label{sec:conclusion}

We have presented several methods for deciding conditional termination
of several classes of program loops manipulating integer
variables. The universal termination problem has been found to be
decidable for octagonal relations and linear affine loops with the
finite monoid property. For the class of polynomially bounded linear
affine loops, we give sufficient termination conditions. Further, we
extend the computation of weakest non-termination preconditions from
simple loops to general programs, and define a class of programs,
called flat, for which this computation yields precise
results. Finally, we have implemented our method in the \textsc{Flata}
tool \cite{flata} and performed a number of preliminary experiments.

\noindent \\
{\bf Acknowledgments} The authors
wish to thank the anonymous reviewers for their important contribution
to improving the quality of this paper.

\bibliographystyle{plain}
\bibliography{termination}

\end{document}

%% file: commands.tex
\newtheorem{example}{Example}

\renewcommand{\iff}{\Leftrightarrow}

\renewcommand{\Call}[2]{\ensuremath{\textsc{#1}} ({#2})}

\DeclareMathOperator{\pre}{pre}

\DeclareMathOperator{\wrs}{wrs}
\DeclareMathOperator{\wnt}{wnt}

\DeclareMathOperator{\true}{\mathbf{true}}
\DeclareMathOperator{\false}{\mathbf{false}}
\DeclareMathOperator{\gfp}{gfp}
\DeclareMathOperator{\mmod}{mod}

\newcommand{\maxcoef}[1]{\mu(#1)}

\newcommand{\zedToX}[0]{\zed^\x}
\newcommand{\zedToY}[0]{\zed^\y}
\newcommand{\relRFc}[0]{R(\x,\x') \wedge \exists \x' . R^{N^2}(\x,\x')}
\newcommand{\relRFcBr}[0]{R(\x,\x') \wedge (\exists \x' . R^{N^2}(\x,\x'))}
\newcommand{\relRFcOct}[0]{R(\x,\x') \wedge \exists \x' . R^{4N^2}(\x,\x')}
\newcommand{\relRFcOctBr}[0]{R(\x,\x') \wedge (\exists \x' . R^{4N^2}(\x,\x'))}
\newcommand{\relRFcDbOct}[0]{\overline{R}(\y,\y') \wedge \exists \y' . \overline{R}^{4N^2}(\y,\y')}
\newcommand{\relRFd}[0]{R(\x,\x') \wedge \exists \x' . R^{5^{2N}}(\x,\x')}
\newcommand{\relRFdBr}[0]{R(\x,\x') \wedge (\exists \x' . R^{5^{2N}}(\x,\x'))}
\newcommand{\relRFm}[0]{R(\x,\x') \wedge \exists \x' . R^m(\x,\x')}

\newcommand{\x}[0]{\vec{x}}
\newcommand{\xk}[1]{\vec{x}^{(#1)}}

\newcommand{\xxki}[2]{x^{(#1)}_{#2}}
\newcommand{\y}[0]{\vec{y}}
\newcommand{\yk}[1]{\vec{y}^{(#1)}}

\newcommand{\yyki}[2]{y^{(#1)}_{#2}}
\newcommand{\z}[0]{\vec{z}}

\newcommand{\pl}[0]{+\!}
\newcommand{\mi}[0]{-\!}

\newcommand{\dbc}[1]{\Delta[{#1}]}
\newcommand{\oct}[1]{\Omega[{#1}]}
\newcommand{\botdb}[1]{\bot\!\!\!\!\bot^{\!\!{#1}}}

\newcommand{\topleft}[1] {{{}^{\scriptscriptstyle{\blacksquare}} \! {#1}}}
\newcommand{\botleft}[1] {{{}_{\scriptscriptstyle{\blacksquare}} \! {#1}}}
\newcommand{\topright}[1] {{{#1}^{\scriptscriptstyle{\blacksquare}}}}
\newcommand{\botright}[1] {{{#1}_{\scriptscriptstyle{\blacksquare}}}}

\newcommand{\key}[1]{\underline{#1}}

\newcommand{\rbr}{{\bf ]\!]}}
\newcommand{\lbr}{{\bf [\![}}
\newcommand{\sem}[1]{\lbr #1 \rbr}

\renewcommand{\vec}[1]{{\bf {#1}}}

\newcommand{\card}[1]{{|\!|{#1}|\!|}}

\newcommand{\arrow}[2]{\xrightarrow{{\scriptscriptstyle #1}}}

\newcommand{\nat}{{\bf \mathbb{N}}}
\newcommand{\zed}{{\bf \mathbb{Z}}}

\newcommand{\rat}{{\bf \mathbb{Q}}}
\newcommand{\real}{{\bf \mathbb{R}}}
\newcommand{\complex}{{\bf \mathbb{C}}}

\newif\ifLongVersion\LongVersionfalse


%% file: tikzmacros-period.tex
\newcommand{\gridName}[2]{n#1n#2}

\tikzset{
  nborder/.style={circle,inner sep=0.75mm,minimum size=0mm},
  ngrid/.style={nborder,fill=black!30,draw=black},
  matchingsep/.style={shape=circle,fill=red!10,inner sep=0mm,minimum size=0mm},
  matching/.style  = {matchingsep,midway,left}
}

\tikzset{
  zrBorder/.style={circle,inner sep=0.5mm,minimum size=0mm},
  zrNode/.style={zrBorder,fill=black!40} 
}



%% file: tikzmacros-termination.tex
\tikzset{
  nborder/.style={circle,inner sep=0.5mm,minimum size=0mm},
  ngrid/.style={nborder,fill=black!40},
  matchingsep/.style={shape=circle,fill=red!10,inner sep=0mm,minimum size=0mm},
  matching/.style  = {matchingsep,midway,left}
}

\newcommand{\TermGridGenNoCap}[8]
{
  \begin{scope}
    \def\xIndent{#7}
    \def\yIndent{#8}
    \def\yBase{#4+(#6-1)*#5}
    \foreach \x in {1,...,#3} {
      \pgfmathsetmacro{\aa}{\xIndent+#1+(\x-1)*#2}
      \foreach \y in {1,...,#6} {
        \pgfmathsetmacro{\bb}{\yBase-(\y-1)*#5}
        \def\nName{\gridName{\x}{\y}}
        \node (\nName) at (\aa cm,\bb cm) [ngrid] {};
      }
    }
  \end{scope}
}

\newcommand{\TermGridGenNoCapBoxed}[8]
{
    \TermGridGenNoCap{#1}{#2}{#3}{#4}{#5}{#6}{#7}{#8}

    \pgfmathsetmacro{\cXleft}{#1+(#7)-0.2}
    \pgfmathsetmacro{\cXright}{#1+(#2)*(#3-1)+#7+0.2}
    \pgfmathsetmacro{\cYbot}{#4-0.2}
    \pgfmathsetmacro{\cYtop}{#4+(#5)*(#6-1)+0.2}
    \draw [rounded corners=4pt,dotted,very thick] (\cXleft cm,\cYbot cm) rectangle (\cXright cm,\cYtop);
}

\newcommand{\TermGridGen}[9]
{

  \begin{scope}

    \def\xIndent{#7}
    \def\yIndent{#8}

    \pgfmathsetmacro{\aa}{#1}
    \foreach \y in {1,...,#6} {
      \pgfmathsetmacro{\bb}{#4+(\y-1)*#5}
      \def\nName{\gridName{0}{\y}}
      \pgfmathtruncatemacro\inx{#6-\y+1}
      \node (\nName) at (\aa cm,\bb cm) [nborder] {$x_{\inx}$};
    }
    \pgfmathsetmacro{\bb}{#4+(#6-1)*#5+\yIndent}
    \foreach \x in {1,...,#3} {
      \pgfmathsetmacro{\aa}{\xIndent+#1+(\x-1)*#2}
      \def\nName{\gridName{\x}{0}}

      \pgfmathtruncatemacro\inxvv{mod(\x-1,#3-1)}
      \pgfmathtruncatemacro\inxww{\x-1}
      \def\inx{\ifthenelse{\equal{#9}{1}}{\inxvv}{\inxww}}

      \node (\nName) at (\aa cm,\bb cm) [nborder] {$\vec{x}^{(\inx)}$};
    }

    \TermGridGenNoCap{#1}{#2}{#3}{#4}{#5}{#6}{#7}{#8}

  \end{scope}
}
\newcommand{\TermGridGenDifferentCap}[9]
{

  \begin{scope}

    \def\xIndent{#7}
    \def\yIndent{#8}

    \pgfmathsetmacro{\aa}{#1}
    \pgfmathsetmacro{\aap}{#1+(#2)*(#3-1)+2*#7}
    \foreach \y in {1,...,#6} {
      \pgfmathsetmacro{\bb}{#4+(\y-1)*#5}
      \def\nName{\gridName{0}{\y}}
      \pgfmathtruncatemacro\inx{#6-\y+1}
      \node (\nName) at (\aa cm,\bb cm) [nborder] {$x_{\inx}$};
      \node (\nName) at (\aap cm,\bb cm) [nborder] {$x'_{\inx}$};
    }

    \TermGridGenNoCap{#1}{#2}{#3}{#4}{#5}{#6}{#7}{#8}

  \end{scope}
}

\newcommand{\TermGridEdgeC}[6]
{
  \draw [->,thick] (\gridName{#1}{#2}) -- (\gridName{#3}{#4})
     node[midway,#6] {#5}; 
}

\newcommand{\TermGridBind}[5]
{
  \draw [->,bend left=20,relative,dotted,very thick,draw=red] (\gridName{#1}{#2}) to (\gridName{#3}{#4}) 
     node[matching] {#5};
}

\tikzset{
  zrBorder/.style={circle,inner sep=0.5mm,minimum size=0mm},
  zrNode/.style={zrBorder,fill=black!40} 
}

\newcommand{\TermZrName}[2]{n#1n#2}
\newcommand{\TermZrNameP}[2]{np#1n#2}
\newcommand{\TermZrNameC}[2]{nc#1n#2}

\newcommand{\TermZrCapState}[6]
{
  \pgfmathsetmacro{\capStart}{#1*(#4+#5)+(#4/2)} 
  \pgfmathsetmacro{\capStep}{#4+#5}
  \pgfmathtruncatemacro\aSize{#2-#1+1}
  \foreach \y in {1,...,\aSize} {
     \pgfmathsetmacro{\capPos}{\capStart+(\y-1)*\capStep}
     \node at (\capPos,-0.6	) {\aCaption(\y)};
  }
}
\newcommand{\TermZrCapGraph}[6]
{
  \pgfmathsetmacro{\capStart}{(#1-1)*(#4+#5)+#4+(#5/2)} 
  \pgfmathsetmacro{\capStep}{#4+#5}
  \pgfmathsetmacro{\capHeight}{(#3-1)*#6+0.5}
  \pgfmathtruncatemacro\aSize{#2-#1+1}
  \foreach \y in {1,...,\aSize} {
     \pgfmathsetmacro{\capPos}{\capStart+(\y-1)*\capStep}
     \node at (\capPos,\capHeight) {\aCaption(\y)};
  }
}

\newcommand{\TermZrBase}[9]
{
  \begin{scope}
    \pgfmathtruncatemacro\xStart{#1-1}
    \foreach \x in {#1,...,#2} {
      \pgfmathsetmacro{\aa}{#4+(\x-1)*(#4+#5)}
      \pgfmathsetmacro{\aaP}{\aa+#5}
      \foreach \y in {1,...,#3} {
        \pgfmathsetmacro{\bb}{(#3-\y)*#6}
        \node (\TermZrName{\x}{\y}) at (\aa cm,\bb cm) [zrNode] {};  
        \node (\TermZrNameP{\x}{\y}) at (\aaP cm,\bb cm) [zrNode] {}; 
      }
      \pgfmathsetmacro{\cXleft}{\aa-0.1}
      \pgfmathsetmacro{\cXright}{\aaP+0.1}
      \pgfmathsetmacro{\cYbot}{-0.2}
      \pgfmathsetmacro{\cYtop}{(#3-1)*#6+0.2}
      \draw [rounded corners=4pt,dotted,very thick] (\cXleft cm,\cYbot cm) rectangle (\cXright cm,\cYtop);
    }

    \def\TermZrXGap{0.2}
    \def\TermZrYExtra{0.3}
    \pgfmathsetmacro{\yBase}{(#3-1)*#6} 
    \foreach \x in {\xStart,...,#2} {
      \pgfmathsetmacro{\aaL}{(\x)*(#4+#5)+\TermZrXGap} 
      \pgfmathsetmacro{\aaR}{\aaL+#4-(2*\TermZrXGap)} 
      \pgfmathsetmacro{\bbB}{0-\TermZrYExtra} 
      \pgfmathsetmacro{\bbT}{\yBase+\TermZrYExtra} 
      \draw [rounded corners=4pt] (\aaL cm,\bbB cm) rectangle (\aaR cm,\bbT cm);
    }
    \foreach \x in {\xStart,...,#2} {
      \pgfmathsetmacro{\aaC}{\x*(#4+#5)+(#4/2)} 
      \foreach \y in {1,...,#3} {
        \pgfmathsetmacro{\bb}{(#3-\y)*#6}
        \node (\TermZrNameC{\x}{\y}) at (\aaC cm,\bb cm) {}; 
      }
    }
    \foreach \x in {1,...,#9} {
      \pgfmathsetmacro{\aaL}{#7*(#4+#5)+(\x-1)*(#4+#5)*#8+(\TermZrXGap/2)} 
      \pgfmathsetmacro{\aaR}{\aaL+(#4+#5)*#8} 
      \pgfmathsetmacro{\bbB}{0-\TermZrYExtra-0.6} 
      \pgfmathsetmacro{\bbT}{\yBase+\TermZrYExtra+0.6} 
      \draw [rounded corners=4pt,dotted,very thick] (\aaL cm,\bbB cm) rectangle (\aaR cm,\bbT cm);
      \pgfmathsetmacro{\capX}{(\aaL+\aaR)/2}
      \pgfmathsetmacro{\capY}{\bbT+0.3}
      \node at (\capX,\capY) {$\lambda$};
    }

  \end{scope}
}

\newcommand{\TermZrStateElem}[3]
{
    \path let \p1 = (\TermZrNameC{#1}{#2}) in node at (\x1,\y1) [nborder] {$#3$};
}

\newcommand{\TermZrEdgeFWLab}[4]
{
  \draw [->,thick] (\TermZrName{#1}{#2}) -- (\TermZrNameP{#1}{#3}) node[midway,above] {#4};
}

\newcommand{\TermZrEdgeBWLab}[4]
{
  \draw [->,thick] (\TermZrNameP{#1}{#2}) -- (\TermZrName{#1}{#3}) node[midway,above] {#4};
}
